\documentclass{aa}
\usepackage[varg]{txfonts}
\usepackage{graphicx}
\usepackage{booktabs}
\usepackage[colorlinks=true,linkcolor=blue,citecolor=blue,urlcolor=blue]{hyperref}

\newcommand{\orcidicon}[1]{\unskip\protect\href{https://orcid.org/#1}{\protect\includegraphics[width=8pt,clip]{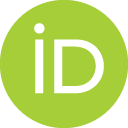}}}

\defcitealias{baker2025doubleA}{JWST GS-z5-Q1}
\defcitealias{baker2025doubleB}{JWST COS-z5-Q1}
\defcitealias{alberts2024highredshift}{JADES 35453}
\defcitealias{delaVega25searchingA}{JADES 1025958}
\defcitealias{delaVega25searchingB}{JADES 1029481}

\begin{document} 

\title{Environmental quenching of high-redshift galaxies: Interpreting JWST observations with simulations}

\titlerunning{high-$z$ quenched low-mass galaxies}

\author{Aleyna Döven\inst{1}\thanks{adoeven@astro.uni-bonn.de}\orcidicon{0009-0001-6397-5695}
      \and Mohammadreza Ayromlou\inst{1}\orcidicon{0000-0003-3783-2321}
      \and Cristiano Porciani\inst{1}\orcidicon{0000-0002-7797-2508}
 }

\institute{Argelander-Institute f\"ur Astronomie, Auf dem H\"ugel 71, D-53121 Bonn, Germany \label{1}}

\date{}
 
\abstract
{
Recent observations of the high-redshift Universe ($z>3$), particularly with the \textit{James Webb} Space Telescope (JWST), have revealed a population of quenched galaxies that challenges current galaxy formation models, which systematically underpredict their abundance in most cases. This discrepancy has been extensively studied for massive systems, motivating revisions to internal quenching mechanisms such as active galactic nucleus feedback. However, the origin of quenching in lower-mass galaxies at high-$z$ has received far less attention, largely due to previous observational limitations. JWST has now identified low-mass quenched galaxies (${M_{\star}}<10^{10}{\rm M_{\odot}}$) where the internal feedback could not have quenched galaxies based on our understanding of low-$z$ analogs.

Given this emerging observational evidence, we investigated the viability of environmental quenching as the primary mechanism suppressing star formation in low-mass galaxies at $z>3$. We analyzed several cosmological simulations, including \textsc{L-Galaxies}, \textsc{IllustrisTNG}, \textsc{Simba}, and the \textsc{TNG-Cluster} zoom simulations, which jointly comprise more than half a million galaxies at $z\sim 5$. Across all simulations, quenched systems are overwhelmingly satellites, despite satellites representing less than 10\% of the total galaxy population. Satellite quenching increases with host halo mass and decreases with both stellar mass and halocentric distance, showing strong correlations with enhanced ram-pressure exposure and gas depletion.

The simulations, particularly \textsc{L-Galaxies}, produce low-mass quenched galaxies broadly consistent with those observed by JWST. Our results suggest that the recently discovered high-redshift quenched low-mass galaxies are possibly environmentally quenched systems residing in the vicinity of massive halos. According to the simulations, these galaxies are often only temporarily quenched: nearly 90\% of them merge within a few hundred megayears, and a small fraction rejuvenate and resume star formation. Extended samples from future observations will enable robust tests of the environmental origin of galaxy quenching in the early Universe.
}


\maketitle

\section{Introduction}
\label{sec:Introduction}

A major challenge in the study of galaxy formation and evolution is to understand the mechanisms responsible for galaxy quenching, the transformation of galaxies from actively star-forming systems to quiescent ones \citep{blanton2003broadband, Kauffmann2003Stellar, Baldry2004Quantifying, delucia2024tracing}. In principle, the suppression of star formation is crucial for explaining the observed division of galaxies into star-forming and passive populations, both in the local Universe and at high redshifts.

Galaxies are broadly classified as central or satellite systems \citep{Zheng2005theoretical, weinmann2006properties, Blanton2007what}. Central galaxies, which reside near the gravitational potential minimum of their host halo, are expected to quench primarily through internal mechanisms such as stellar feedback and active galactic nucleus (AGN) activity \citep{Kauffmann2003dependence, croton2006many, henriques2017galaxy, Fu2025role}. Stellar feedback can regulate star formation in low- and intermediate-mass systems but is generally insufficient to fully quench high-mass galaxies \citep{Ciotti1991winds, Springel2003}. In massive systems, AGN-driven heating and gas ejection are therefore required to efficiently suppress star formation and maintain long-term quiescence \citep{bower2006breaking, croton2006many, weinberger18}.

In contrast to centrals, satellite galaxies orbit within the dark matter halo of a more massive central galaxy. After infall into a denser environment, they are subject to a range of environmental processes, including ram-pressure stripping by the ambient medium \citep{gunn1972on, moore1996galaxy, ayromlou2019new}, removal of their hot gas reservoirs through strangulation \citep{peng2015strangulation}, and tidal interactions with the host potential or with neighboring galaxies \citep{guo2011dwarf}. These mechanisms can efficiently deplete, heat, or prevent the replenishment of a satellite's gas supply, particularly in dense environments where low-mass satellites are most strongly affected \citep{gunn1972on, Peng2010mass, Ayromlou2023physical}. Although environmental quenching is most efficient for satellite galaxies, central galaxies, particularly those located in the vicinity of high-mass halos, can also lose gas and quench through ram-pressure stripping by the surrounding hot gas \citep{Wetzel2012Galaxy, Ayromlou2020Comparing}.

Due to observational limitations, quenching has historically been investigated primarily at low redshifts, where large spectroscopic surveys have provided statistically robust galaxy samples. These studies established key correlations between galaxy quenching and stellar mass, halo mass, and the local background environment \citep[LBE;][]{Bell2003Optical, Kauffmann2003dependence, Peng2010mass}. Consequently, most theoretical models and simulations have been calibrated to reproduce the properties of galaxies in the nearby Universe, including the galaxy stellar mass function (SMF) and star formation rates (SFRs) at $z \lesssim 1$ \citep{vogelsberger2014Introducing, henriques2015galaxy,Davies2020Galaxy,Schaye2023FLAMINGO}.

This picture has changed dramatically with the emergence of deep high-redshift observations, particularly from the \textit{James Webb} Space Telescope (JWST). Recent studies have revealed a large population of high-mass quenched galaxies at $z \gtrsim 2$ \citep{carnall2023surprising, Valentino2023atlas, alberts2024highredshift, Nanayakkara2024population}, with spectroscopically confirmed systems extending as far back as $z \sim 7$--8 \citep{Carnall2023massive, Carnall2024JWST, Weibel2025RUBIES}. These galaxies are compact, gas-poor, and contain old stars, implying rapid and efficient quenching within the first few billion years of cosmic history. Similar studies have also been carried out with the Atacama Large Millimeter/submillimeter Array and the \textit{Hubble} Space Telescope \citep{schreiber2018jekyllhyde, whitaker2013quiescent, straatman2014substantial}.

Growing evidence now points to the presence of quenched systems in the low-mass regime at early times. Several studies have reported quiescent galaxy candidates with ${M_{\star}} < 10^{10}\,{\rm M_{\odot}}$ at $z \geq 4$ based on photometric analyses \citep{Valentino2023atlas, alberts2024highredshift, delaVega25searching, Merlin2025Witnessing}, and recent spectroscopic work has begun to provide further support for the existence of such systems \citep{baker2025double}. This population is particularly important because it is difficult to explain with the same internal-feedback channels typically invoked for high-mass galaxies, making it a key probe of alternative pathways such as environmental quenching.

On the theoretical side, reproducing the abundance and properties of high-$z$ quenched galaxies remains a major challenge for most numerical simulations. Models generally underpredict the number density of high-mass quiescent systems at $z \gtrsim 2$ \citep{Hartley2023first, Valentino2023atlas, delucia2024tracing, lagos2024quenching, weller2025quenched}. Using the \textsc{L-Galaxies} semi-analytic model, \citet{vani2025probing} showed that although the quiescent SMF is reasonably reproduced at low-$z$, discrepancies grow rapidly toward earlier epochs: high-mass quenched galaxies $({M_{\star}} \gtrsim 10^{10.5} \, \rm M_\odot)$ are strongly underabundant by $z \sim 3$ and are largely absent beyond $z \sim 3$--3.5, motivating a reassessment of quenching models. Similarly, \citet{lagos2025diverse} found a comparable deficit of high-mass quenched galaxies across several other semi-analytic models and hydrodynamical simulations.

Internal physical processes such as stellar and AGN feedback are likely key to resolving the deficit of high-mass quiescent galaxies \citep{Hartley2023first, lagos2024quenching, vani2025probing} in the early Universe, as suggested by several studies. On the other hand, these internal processes are less effective in low-mass systems. This raises a timely question: could environmental quenching already be active in the early Universe, especially for low-mass galaxies that are more susceptible to gas stripping due to their weaker self-gravity? This is the topic of this paper.

We investigated the role of environment in galaxy evolution at high redshifts ($z > 3$) using several simulations in comparison with the most recent high-redshift observational data \citep{alberts2024highredshift,delaVega25searching,baker2025double}. In particular, we examined whether environmental quenching mechanisms traditionally invoked at low redshifts left detectable imprints in the early Universe.

The structure of this paper is as follows. Section~\ref{sec:Methodology} describes the simulations, observations, and analysis methods. In Section~\ref{sec:Doubletrouble} we examine the predictions of simulations against the recently observed low-mass high-$z$ quenched galaxies. Section~\ref{sec:Results} is dedicated to uncovering the physical drivers of the quenching of low-mass galaxies at high redshifts, investigating the role of environment. Finally, in Section~\ref{sec:Summary} we summarize our findings and present our conclusions.

\section{Methodology}
\label{sec:Methodology}

\subsection{Galaxy formation simulations}
In this subsection we briefly describe the galaxy formation simulations used in this work. For a full model description, we refer the reader to the model paper of each simulation: \textsc{L-Galaxies} \citep{ayromlou2021galaxy}, \textsc{IllustrisTNG} \citep{weinberger17,pillepich2018Simulating}, and \textsc{Simba} \citep{Dave2019SIMBA}. The number of galaxies analyzed from each simulation is provided in Appendix~\ref{tab:quenched_comparison}. Across all simulations, we imposed a stellar mass cut of ${M_{\star}} \geq 10^9{\rm M_{\odot}}$ to ensure that we analyzed well-resolved galaxies given the resolution limits of the simulations.

\subsubsection{\textsc{L-Galaxies}}
The \textsc{L-Galaxies} semi-analytic model \citep{ayromlou2021galaxy} is built upon the dark matter merger trees of the Millennium \citep{springel2005cosmological} N-body simulation. The simulations have been rescaled to a Planck cosmology \citep{angulo2010one, angulo2015cosmological} with parameters: $\sigma_8 = 0.829$, $H_0 = 67.3\ \text{km,s}^{-1}\mathrm{Mpc}^{-1}$, $\Omega_\Lambda = 0.685$, $\Omega_m = 0.315$, $\Omega_b = 0.0487$ ($f_b = 0.155$), and $n_s = 0.96$ \citep{planck2015_xiii}.

Dark matter halos and subhalos are identified using the friends-of-friends \citep[FOF;][]{Davis1985TheEvolution} and \textsc{SUBFIND} \citep{springel2001populating} algorithms, respectively. Galaxies are then associated with these subhalos, such that each FOF halo hosts one central galaxy, while all other galaxies within the FOF group are classified as satellites. 
The central galaxy resides in the primary subhalo of the FOF halo, which is often the most massive subhalo of the halo. Satellite galaxies are associated with secondary subhalos or, in cases where the subhalo is no longer resolved, are tracked as orphan satellites. We note that satellites can reside at distances beyond the halo boundary, $R_{\rm 200c}$, since FOF groups often extend to scales larger than $R_{\rm 200c}$. The central galaxy, satellites with subhalos, and orphan satellites are also referred to as Type 0, Type 1, and Type 2 galaxies, respectively. Throughout this work, when we refer to ``satellites,'' we include both Type 1 and Type 2 galaxies, unless otherwise specified.

\textsc{L-Galaxies} models galaxy evolution by solving a set of coupled equations that describe key baryonic processes, including gas accretion and cooling, star formation, and chemical enrichment. Feedback from stars and supermassive black hole (supermassive black hole or AGN feedback) plays a central role in regulating and quenching star formation by heating or ejecting gas from galaxies and their halos. In addition to these internal processes, the model includes environmental mechanisms such as ram-pressure stripping and tidal stripping. \textsc{L-Galaxies} is unique among semi-analytic models in modeling gas stripping for both satellite and central galaxies \citep{ayromlou2019new}, enabling a more consistent treatment of quenching across environments, which is the focus of this study.

\subsubsection{\textsc{IllustrisTNG}}
\label{sec:TheIllustrisTNG}
The \textsc{IllustrisTNG} simulation suite \citep[TNG;][]{marinacci2018first, naiman2018first, nelson2018first, pillepich2018First, springel2018first} follows the coupled evolution of dark matter and baryonic matter from the early Universe to the present day, self-consistently modeling gravitational interactions and (magneto-)hydrodynamical processes using the moving-mesh code \textsc{AREPO} \citep{springel2010pur}. \textsc{IllustrisTNG} also adopts a Planck $\Lambda$ cold dark matter cosmology \citep{planck2015_xiii}. Sub-grid physics models are employed to capture unresolved baryonic physics, including star formation, cooling, stellar feedback, and black hole growth and feedback \citep{weinberger17,pillepich2018Simulating}. Similar to \textsc{L-Galaxies}, \textsc{IllustrisTNG} uses FOF and \textsc{SUBFIND} algorithms to identify halos, subhalos, and galaxies therein.

In this work, we used the galaxy catalogs from the \textsc{TNG300-1} simulation, which provides the largest volume in the  \textsc{TNG} suite with a box side length of 302.6 $\mathrm{cMpc}$. The simulation contains $2500^3$ gas cells and dark matter particles, with masses of $1.1 \times 10^7 {\rm M_{\odot}}$ and $5.9 \times 10^7 {\rm M_{\odot}}$, respectively. In \textsc{TNG300}, SFRs below $\log_{10}(\rm SFR [M_{\odot}/yr]) \sim -3$ are not reliably resolved and are reported as zero. So in this work, we manually set specific star formation rate (sSFR) values of $\log_{10} ({\rm sSFR} [\rm{yr}^{-1}]) \leq -13$ to this value.

\subsubsection{\textsc{TNG-Cluster}}
\label{sec:TNG-Cluster}

For the analysis, we additionally used the \textsc{TNG-Cluster} simulation, a suite of 352 high-resolution cosmological zoom-in simulations of galaxy clusters, with ${M_{200c}} \gtrsim 10^{14.4}{\rm M_{\odot}}$ at $z=0$, whose first results have recently been presented \citep{Ayromlou2024atlas,Lehle2024Heart, Lee2024Radio, Nelson2024Introducing, Rohr2024hot, Truong2024xray}. \textsc{TNG-Cluster} adopts the physics of the original \textsc{TNG} simulation and the resolution of \textsc{TNG300}.

At high redshifts, the simulated zoom regions correspond to protocluster environments tracing the progenitors of present-day massive clusters. Because the \textsc{TNG-Cluster} sample was selected based on its final ($z=0$) halo masses, the resulting high-redshift protocluster population is biased toward regions that evolve into extreme overdensities and may not be representative of the general high-redshift galaxy population. We therefore used \textsc{TNG-Cluster} mainly to enlarge our \textsc{TNG300} sample when comparing with observations (Sect.~\ref{sec:Doubletrouble}), and we did not rely on it for the statistical analysis in this work.

\subsubsection{\textsc{Simba}}
\label{sec:simba}

\textsc{Simba} \citep{Dave2019SIMBA} is a cosmological hydrodynamical simulation based on the meshless finite-mass implementation of the \textsc{GIZMO} code \citep{hopkins15, hopkins18}, with gravitational forces computed using the \textsc{GADGET-3} solver \citep{Springel2005Gadget}. The simulation volume is 147.7 $\mathrm{cMpc}$ and contains $1024^3$ dark matter particles and gas resolution elements, with masses of $9.7 \times 10^7{\rm M_{\odot}}$ and $1.8 \times 10^7{\rm M_{\odot}}$, respectively. The \textsc{Simba} simulations include non-equilibrium radiative cooling for primordial and metal-enriched gas, as well as an $\mathrm{H}_2$-based star formation prescription tied to the local molecular gas fraction \citep{Kennicutt1998global}. Dark matter halos are identified on-the-fly using a standard 3D FOF algorithm with a linking length of 0.2 times the mean inter-particle separation. Galaxies are identified separately using another FOF algorithm applied to star particles, black holes, and dense gas, with a different linking length. The CAESAR package is then employed in post-processing to cross-match galaxies and halos and to produce the data catalogs.

\subsection{Observations}
\label{subsec:Observations}
Although most observed high-redshift quenched galaxies are high-mass systems with ${M_{\star}} \gtrsim 10^{10}{\rm M_{\odot}}$ (e.g., \citealt{Valentino2023atlas}), recent JWST observations have revealed a growing population of low-mass quenched galaxies at $z > 4$. In this work, we focused on these emerging galaxies and adopted the observed galaxies reported by \citet{delaVega25searching} and \citet{alberts2024highredshift} (photometric candidates), together with the spectroscopically confirmed systems presented by \citet{baker2025double}.

Deep JWST imaging in the JADES (JWST Advanced Deep Extragalactic Survey) fields enables the identification of galaxies with suppressed star formation at $z \gtrsim 4$ and at stellar masses well below $10^{10}{\rm M_{\odot}}$. Using observed-frame colors and spectral energy distribution fitting, \citet{delaVega25searching} reported several low-mass quiescent candidates in the JADES fields, including JADES-1025958 and JADES-1029481, with stellar masses of ${{M_{\star}}} \sim 10^{9.4}$--$10^{9.9}{\rm M_{\odot}}$ at $z \sim 5.5$. Similarly, \citet{alberts2024highredshift} identified additional low-mass quiescent and post-starburst candidates, such as JADES-35453, with ${M_{\star}} \sim 10^{9.9}{\rm M_{\odot}}$ at $z \sim 5.3$, by exploiting ultra-deep NIRCam (Near-Infrared Camera) and MIRI (Mid-Infrared Instrument) imaging to distinguish old stellar populations from dust-obscured star-forming galaxies. Finally, \citet{baker2025double} presented the spectroscopic confirmation of two low-mass quenched galaxies at $z \geq 5$, referred to as the ``Double Trouble'' galaxies, with stellar masses of ${M_{\star}} \sim 10^{9.62}{\rm M_{\odot}}$ and ${M_{\star}} \sim 10^{9.55}{\rm M_{\odot}}$ at $z=5.39$ and $z=5.11$, respectively. The properties of these observed low-mass quenched galaxies, together with the simulated analogs identified in this work, are summarized in Table~\ref{tab:pc_level_summary}.

\begin{figure*}
    \centering
    \includegraphics[width=0.6\textwidth]{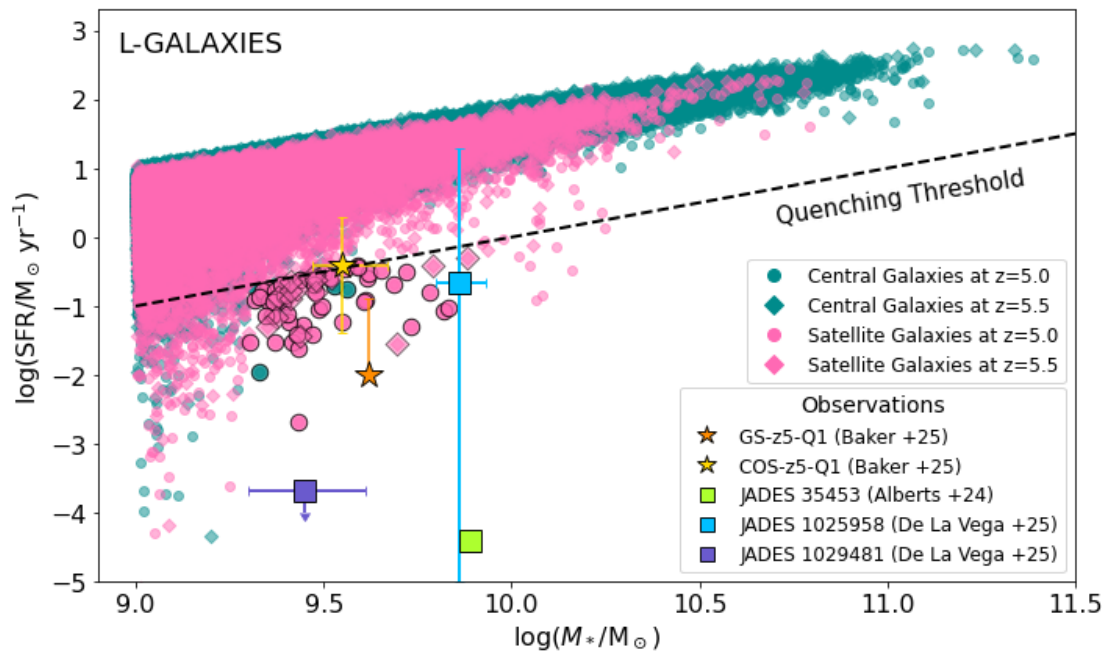}
    \includegraphics[width=0.48\textwidth]{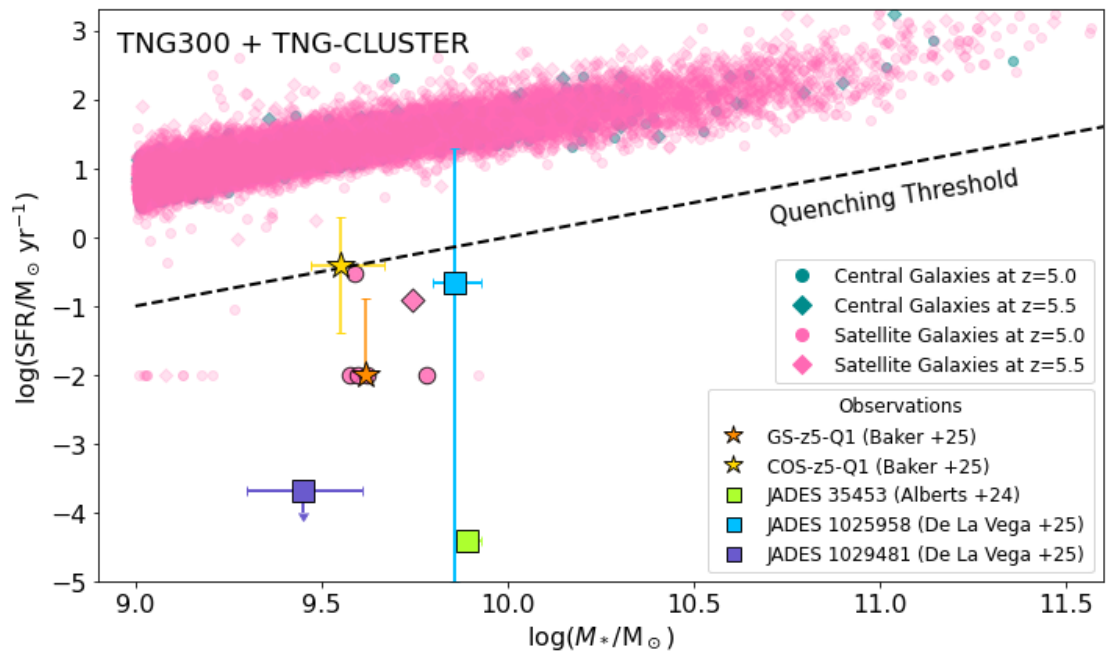}
    \includegraphics[width=0.48\textwidth]{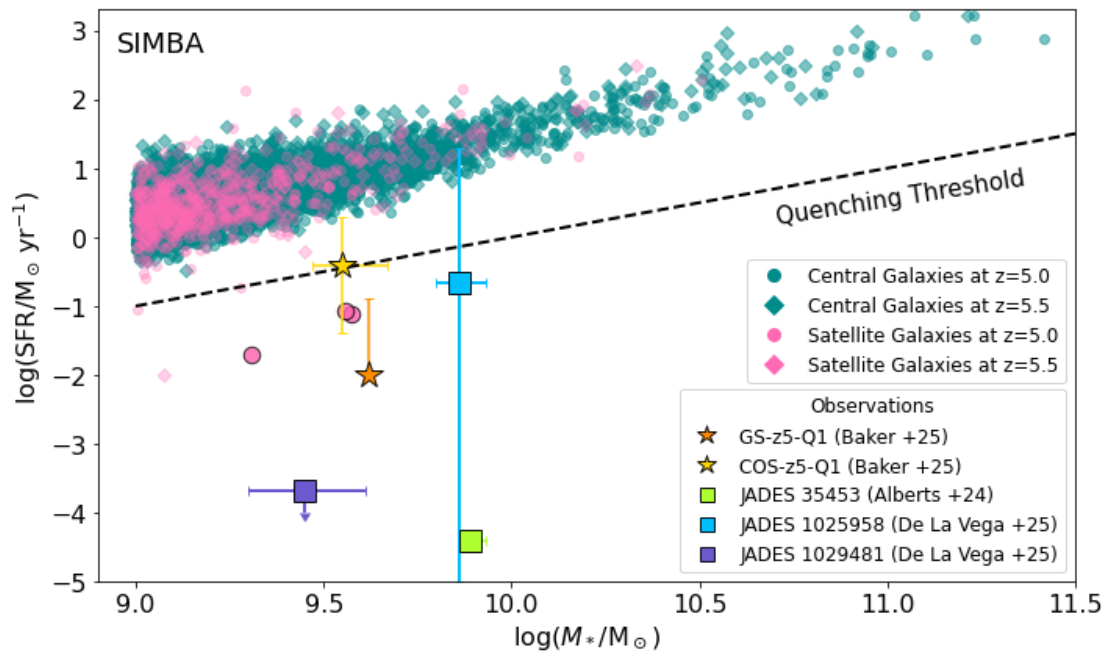}
    \caption{SFR as a function of stellar mass. The two observed spectroscopically confirmed ``Double Trouble'' galaxies (stars) are shown together with three additional observed quenched candidates (squares) from the literature, and compared to simulated satellite (pink) and central (dark cyan) galaxies from multiple snapshots of the \textsc{L-Galaxies} (\textit{top}), \textsc{TNG} (\textit{bottom left}), and \textsc{Simba} (\textit{bottom right}) simulations. Simulated quenched galaxies with $\log_{10}({M_{\star}}/{\rm M_{\odot}}) = 9.6 \pm 0.3$ are highlighted with larger symbols to emphasize the closest analogs to the observed systems. The downward arrow for JADES-1029481 indicates an upper limit on its SFR. All simulations, particularly \textsc{L-Galaxies}, produce low-mass quenched galaxies with properties broadly comparable to the observed systems; most of these analogs are satellites.}
    \label{fig:doubletroublesfr} 
\end{figure*}

\subsection{Analysis at high redshifts}
\label{sec:analysisathighredshift}

We explored the redshift evolution of galaxy quenching up to $z=8$ (e.g., see Sect.~\ref{sec:quenchedfraction}). For conciseness and clarity, most of our analysis focused on $z\sim 4, 5$, and $5.5$. The latter two ($z=5-5.5$) match the redshifts of the observed low-mass quenched galaxies \citep{alberts2024highredshift,delaVega25searching,baker2025double}, and $z=4$ is chosen as a representative redshift approximating the average redshift of observed high-$z$ quenched galaxies (e.g., \citealt{Valentino2023atlas}). We classified a galaxy as quenched based on an sSFR threshold of $0.2/t_{H(z)}$, below which the galaxy is considered quenched \citep[e.g., see][]{Pacifici2016evolution,vani2025probing}. Here $t_{H(z)}$ is the Hubble time at redshift $z$, indicating no star formation on cosmologically relevant timescales. For \textsc{L-Galaxies}, stellar masses and SFRs correspond to the whole galaxy, as modeled by the semi-analytical model. For \textsc{Simba}, we also used the total galaxy stellar mass and SFR, whereas for \textsc{TNG}, these quantities are measured within twice the stellar half-mass radius. These choices do not affect any of our conclusions.

Throughout the study we imposed a minimum of five galaxies per x-axis bin for all relevant plots, unless stated otherwise. This choice ensures statistical robustness and minimizes noise in the measurements. For the ram pressure analysis, Type 2 galaxies (satellites without a dark matter subhalo) are excluded due to the unreliability of their LBE estimates, which are essential for a robust assessment of ram pressure stripping \citep[see][]{ayromlou2019new}. As a result, only galaxies with well-defined baryonic and orbital properties are included in this part of the analysis. All ram pressure values smaller than $10^{-18} \ \mathrm{dyn/cm^{-2}}$ are mapped to this value.

\begin{figure*}
    \centering
    \includegraphics[width=0.48\textwidth]{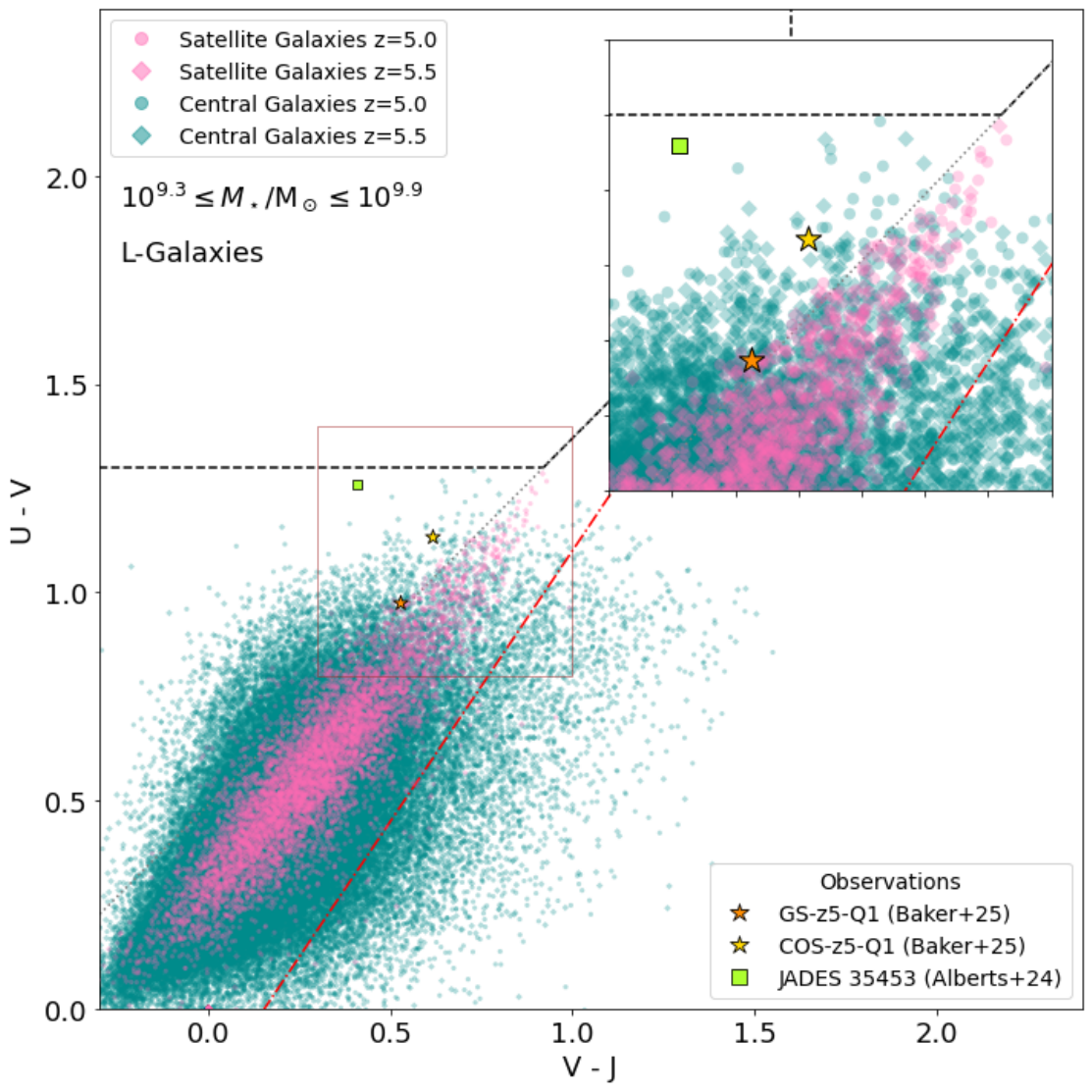}
    \includegraphics[width=0.48\textwidth]{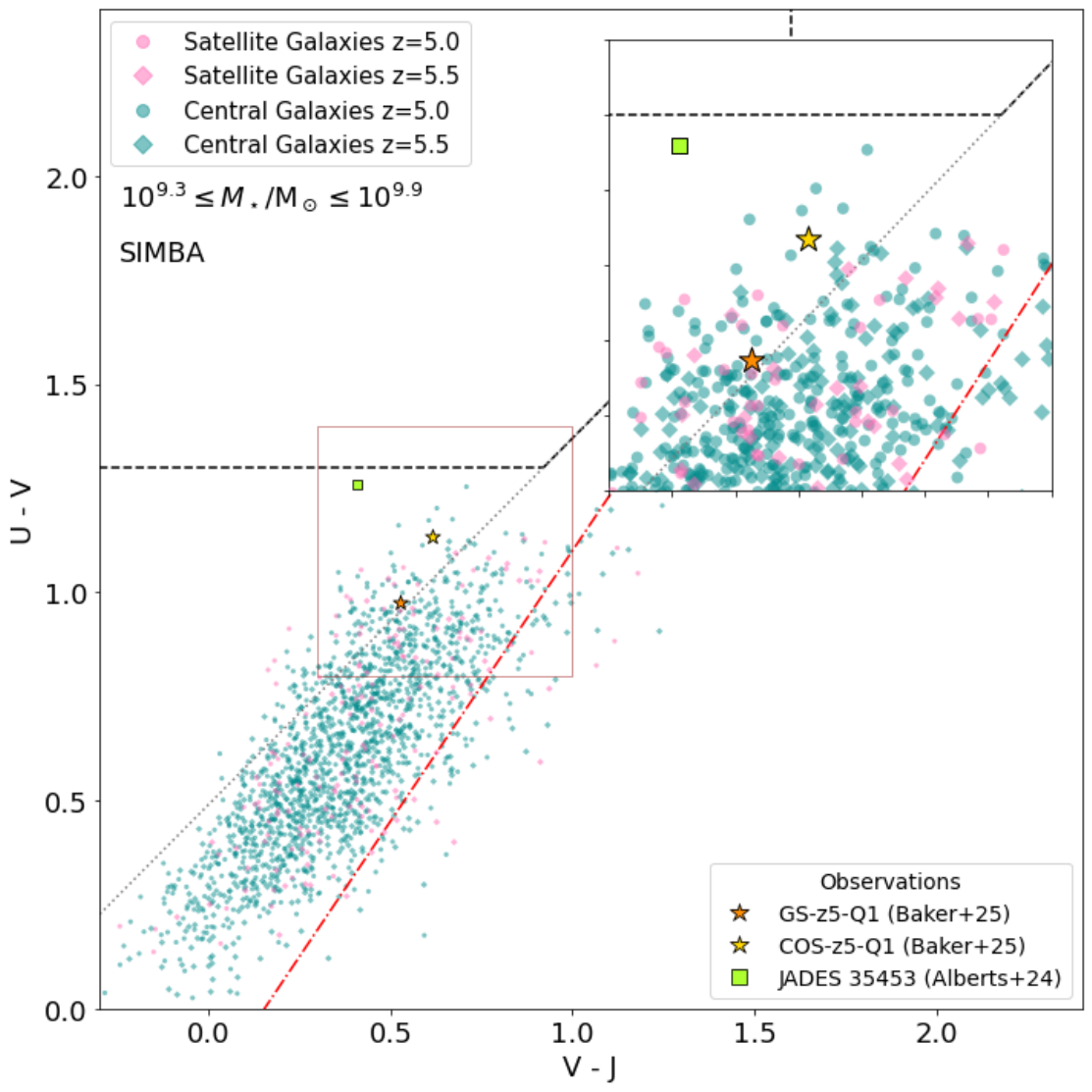}
    \caption{UVJ diagram for the observed galaxies shown together with satellite (pink) and central (dark cyan) galaxy populations from multiple snapshots of \textsc{L-Galaxies} (\textit{left}) and \textsc{Simba} (\textit{right}). Only galaxies with $\log_{10}({M_{\star}}/{\rm M_{\odot}}) = 9.6 \pm 0.3$ are included. The extended UVJ quiescent selection proposed by \citet{Baker2025abundance} is shown by the dash-dotted red line. For comparison, the classical UVJ selection boundaries from \citet{schreiber2015herschel} and \citet{Belli2019MOSFIRE} are overplotted as dashed black and dotted gray lines, respectively. Spectroscopically confirmed and photometrically selected quenched JWST galaxies are indicated with star and square markers.}
    \label{fig:doubletroubleuvj}
\end{figure*}

\section{JWST low-mass quenched galaxy analogs}
\label{sec:Doubletrouble}

\subsection{Quenching: Simulations versus observations}
\label{sec:comparison}

Figure~\ref{fig:doubletroublesfr} shows the SFR as a function of stellar mass in the models and data. All three simulations produce some quenched galaxies with properties broadly comparable to the observed systems. Among the models, \textsc{L-Galaxies} yields the largest number of low-mass quenched galaxies, reflecting both its physical prescriptions and its larger volume. \textsc{Simba} also produces a small number of galaxies similar to the observed ones. \textsc{TNG300}, despite its volume being nearly ten times larger than \textsc{Simba}, does not predict any quenched galaxies in the relevant stellar mass range. To improve statistics for \textsc{TNG300}, we additionally included \textsc{TNG-Cluster} (see Sect.~\ref{sec:TNG-Cluster}). Owing to its focus on protocluster regions, \textsc{TNG-Cluster} produces a few low-mass quenched satellites comparable to those observed. This indicates that extending the sample to include the most extreme environments increases the likelihood of finding quenched satellites. Most simulated analogs of the observed systems are satellites, suggesting that the high-$z$ quenched low-mass JWST galaxies could be satellites affected by environmental quenching.

In Fig.~\ref{fig:doubletroubleuvj} we further examine the rest-frame UVJ colors of the spectroscopically confirmed quenched galaxies in comparison with the models for the observed mass range. The same UVJ selection criteria adopted by \citet{baker2025double} are applied to both the observed and simulated galaxies. In particular, the extended UVJ quiescent selection proposed by \citet{Baker2025abundance} and the classical UVJ selection boundaries from \citet{schreiber2015herschel} and \citet{Belli2019MOSFIRE} are shown\footnote{Because UVJ measurements are not available in \textsc{TNG} public catalogs at these redshifts, they are not shown here.}. Simulated galaxies with UVJ colors consistent with the observed quenched systems are found among both centrals and satellites, although centrals dominate. This contrasts with selections based on sSFR (Fig.~\ref{fig:doubletroublesfr}), which instead identify satellites as the dominant quenched population, highlighting the strong dependence of inferred quenching pathways on the adopted classification criterion.

We note, however, that the simulated UVJ colors are corrected for dust attenuation in post-processing and therefore depend strongly on the adopted dust prescription. Because observational constraints on dust properties at these redshifts are limited, the simulated colors should be interpreted with caution, whereas SFRs are intrinsic model predictions and more reliably comparable to observations.

\subsection{The fate of passive galaxies}
\label{sec:fate}
\begin{figure*}
    \centering
    \includegraphics[width=0.6\textwidth]{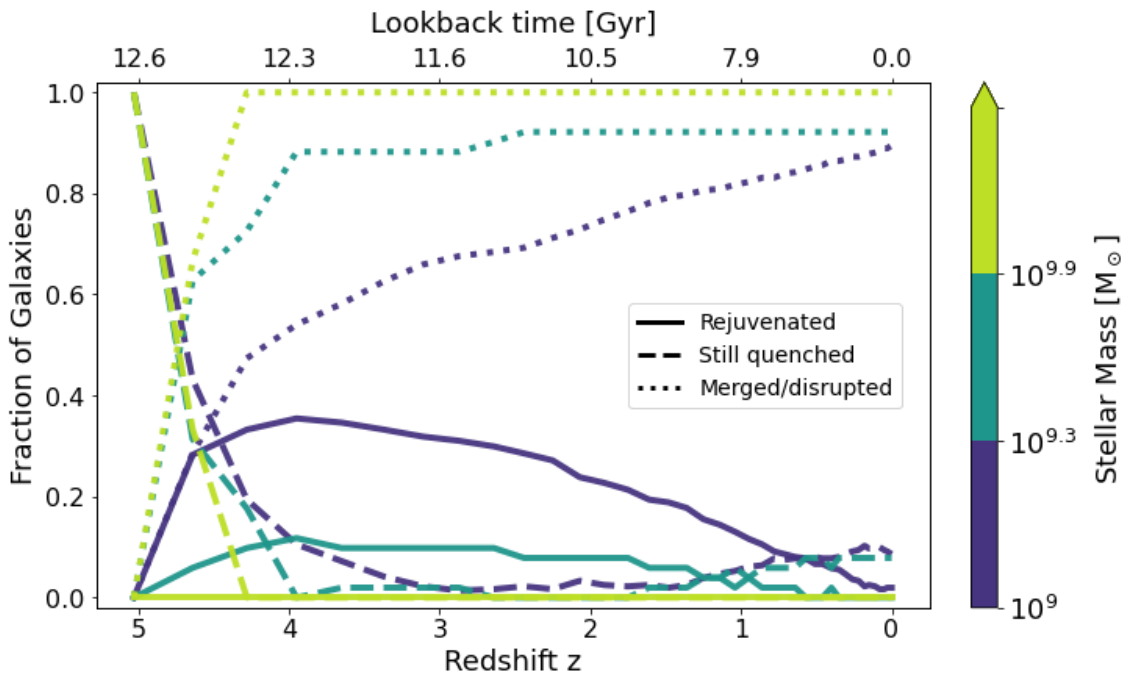}
    \includegraphics[width=0.48\textwidth]{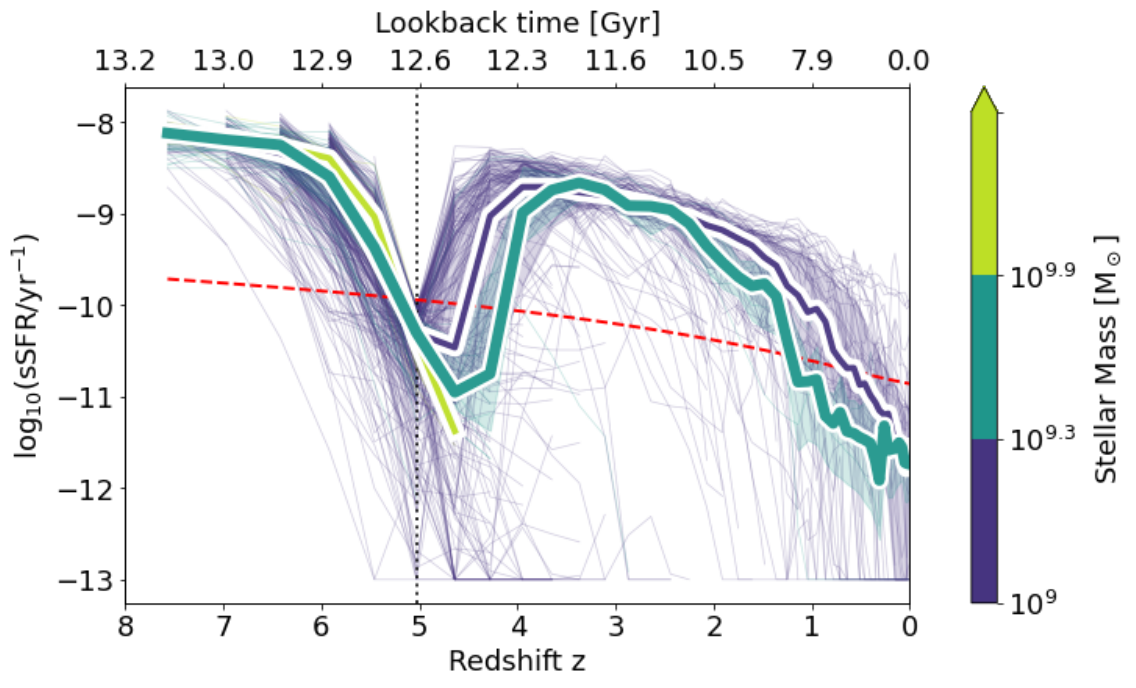}
    \includegraphics[width=0.48\textwidth]{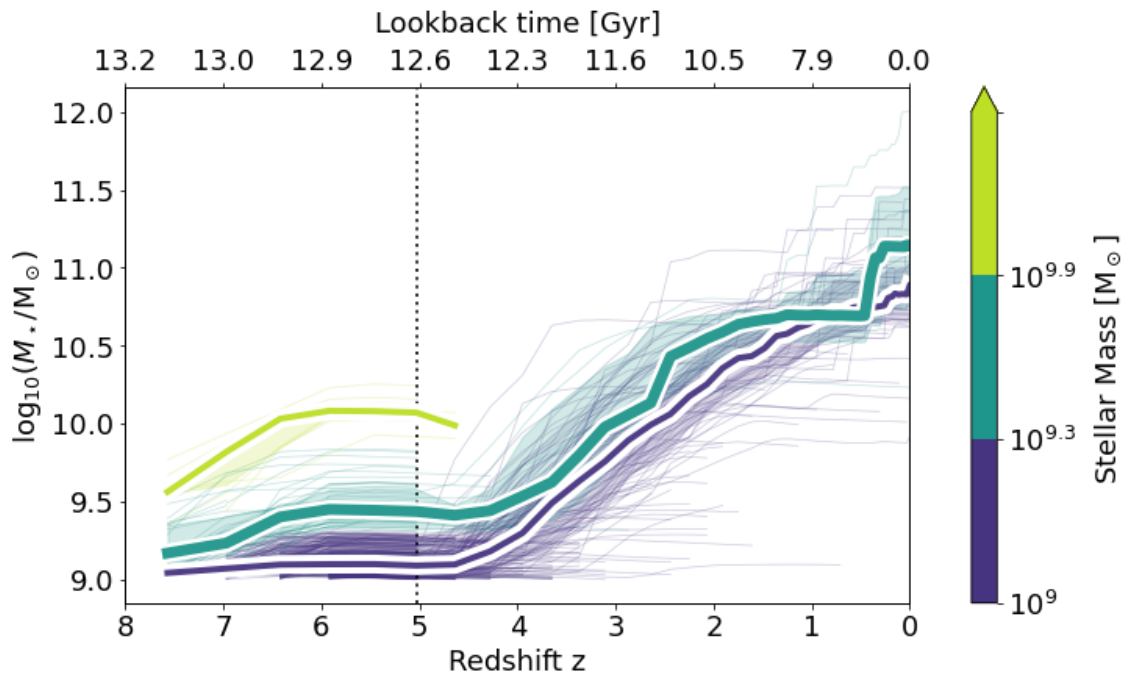}
    
    \caption{\textit{Top}: Fate of galaxies quenched at $z=5$ up to $z=0$. Solid lines show the rejuvenated fraction, defined at each redshift as the fraction of the initially quenched population that is still present and star-forming at that time. This quantity is instantaneous rather than cumulative and therefore does not necessarily increase with time. Dashed lines show the fraction that remains quenched, while dotted lines show the fraction of galaxies that have merged or been disrupted by that redshift. All fractions are normalized by the corresponding initially quenched population at $z=5$. \textit{Bottom}: Redshift evolution of surviving galaxies quenched at $z=5$ in \textsc{L-Galaxies}, shown in terms of sSFR (\textit{left}) and stellar mass (\textit{right}). Colors indicate the stellar mass bins at $z=5$; the thicker line corresponds to the intermediate bin, $9.3 < \log_{10}(M_{\star}/{\rm M_{\odot}}) < 9.9$, which matches the stellar-mass range of the JWST-observed quenched galaxies in Fig.~\ref{fig:doubletroublesfr}. No minimum galaxy-number threshold is applied to the stellar-mass bins, in order to show all available trends. sSFR values below $10^{-13}\mathrm{yr}^{-1}$ are set to this floor value. The quenching sSFR threshold, defined as $0.2/t_{H(z)}$, is shown as a dashed red line. For completeness, the gas-mass evolution of $z=5$ quenched galaxies and the corresponding trends for $z=4$ quenched galaxies are shown in Appendix~\ref{sec:analogs}.}
    
    \label{fig:rejuv_plots}
\end{figure*}

We traced the main progenitor branches of \textsc{L-Galaxies} analogs of JWST-observed low-mass quenched galaxies to determine their subsequent evolution and fate. We uncovered whether these systems remain quenched, rejuvenate, merge, or become tidally disrupted. Galaxies were selected to be quenched at $z=5$ and divided into three stellar mass bins; the intermediate bin matches the observed systems.

The top panel of Fig.~\ref{fig:rejuv_plots} shows the fraction of galaxies in different evolutionary states and reveals a similar trend across all stellar mass bins. For galaxies quenched at $z=5$, the dominant outcome is merging or disruption. Quantitatively, approximately $\sim90\%$ of these galaxies have merged or undergone disruption by $z=0$. This fraction increases monotonically with time because once a galaxy merges or gets disrupted, it can no longer be found in the simulation. Therefore, this fraction can only increase with time. Only about $10\%$ across all stellar mass bins combined end up being passive up to the present day. Persistent, uninterrupted quiescence from $z=5$ to $z=0$ is therefore rare. The rejuvenated fraction, on the other hand, does not monotonically increase with time. It measures the fraction of the initially quenched population that is still present and star-forming at a given redshift, rather than the fraction that has ever rejuvenated since $z=5$. Among the galaxies that do not merge, the majority rejuvenate rapidly after their initial quenching phase. Following this rejuvenation episode, they either merge at later times or eventually become quenched again by $z=0$. It is important to clarify that a galaxy is classified as merged or disrupted when its main branch can no longer be traced in the merger tree. This typically occurs when the galaxy merges with a more massive system and ceases to exist as a distinct object.\footnote{Disrupted galaxies could either be completely disrupted or just have fallen below the simulation resolution.}

A crucial distinction emerges when separating galaxies into central and satellite populations. The vast majority of galaxies quenched at $z=5$ are satellites (see Sect.~\ref{sec:quenchedfraction} and Table~\ref{tab:pc_level_summary}). Rejuvenation in satellites, while it does occur, is rare, with fewer than $10\%$ of systems showing signs of renewed star formation even 200 Myr after their initial quenching. The remaining few quenched galaxies that are centrals, however, are significantly more likely to rejuvenate than their satellite counterparts (see the bottom panel of Fig.~\ref{fig:gasfractionrejuvlgal5}).

The bottom panels of Fig.~\ref{fig:rejuv_plots} show the redshift evolution of sSFR and stellar mass for a specific subset of galaxies: those that have not merged at each redshift, i.e., survivor galaxies. These galaxies, therefore, do not represent the full population of galaxies that were quenched at $z=5$, as the majority have already merged or been disrupted by earlier epochs. In both panels, the most massive galaxies ($\log_{10} {M_{\star}}/{\rm M_{\odot}} >9.9$) exhibit striking behavior, with the vast majority of their members either merging or being disrupted. However, this trend should be interpreted with caution, as it is driven by small-number statistics: only nine galaxies populate this bin at $z=5$, all of which are satellites whose dark matter subhalos have already been disrupted. For the remaining systems, the evolution of stellar mass and sSFR offers deeper insight into their post-quenching behavior. The thicker line in each panel corresponds to the intermediate stellar mass bin, the population directly comparable to the JWST-observed galaxies at $z=5$ (see Table~\ref{tab:pc_level_summary}).

In the sSFR evolution (bottom left), although the galaxies are chosen to be quenched at $z=5$, their median sSFR increases shortly afterward, indicating rejuvenation episodes. This temporary rise is followed by a gradual decline toward lower redshifts, with many systems eventually reaching strongly passive states. This non-monotonic evolution demonstrates that quenching at high redshifts is not necessarily a permanent condition. Instead, galaxies can rejuvenate, become quenched again, or merge with other, often larger, systems.

The right panel sheds more light on this picture: galaxies continue to grow in stellar mass after being quenched at $z=5$. While the increase is modest until $z=4$, it becomes more pronounced toward lower redshifts. This mass growth likely reflects a combination of rejuvenated star formation episodes, mergers, and merger-driven accretion — all of which contribute to their late-time evolution.

\section{Main drivers of quenching for low-mass galaxies at high redshifts}
\label{sec:Results}

Having established that simulations can produce analogs of the recently observed low-mass quenched galaxies, we next placed these systems in a broader context by examining the full galaxy population at $z=4$, with the aim of quantifying the dominant quenching channel for low-mass galaxies\footnote{We also repeated the analysis at $z=5$, where the low-mass quenched systems are observed by JWST, and find consistent trends and conclusions, albeit with lower number statistics.}.

\begin{figure}
  \centering
  \includegraphics[width=\columnwidth]{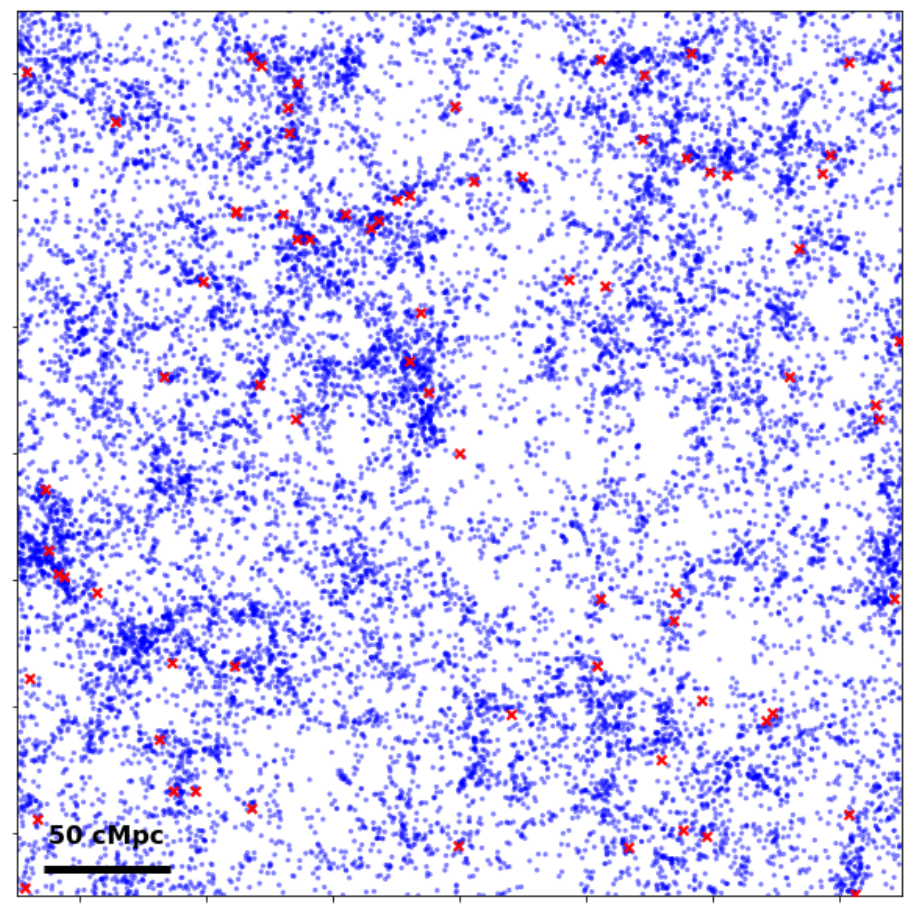} 
  \caption{Overview of the cosmic web at $z=4$ in a $350 \times 350\,\mathrm{cMpc}^2$ slice of \textsc{L-Galaxies}, projected over $75\,\mathrm{cMpc}$ along the line of sight. Blue and red points denote star-forming and quenched galaxies, respectively; quenched galaxies are shown with slightly larger symbols.}
  \label{fig:quenchingscatterplot}
\end{figure}

Figure~\ref{fig:quenchingscatterplot} presents an overview of the cosmic web at $z=4$ using a projected slice of the \textsc{L-Galaxies} simulation. It illustrates the spatial distribution of star-forming and quenched galaxies: star-forming systems, which constitute the majority of the population, are found throughout all environments (nodes, filaments, and voids), whereas the sparse quenched population is predominantly located in denser regions. In this section we first provide a global characterization of galaxy properties and then focus on the role of environment at high redshifts.

\subsection{Galaxy population overview}
\label{sec:galaxypopulationoverview}

\subsubsection{The stellar mass function}
\label{sec:StellarMassFunction}

\begin{figure}
  \centering
  \includegraphics[width=0.95\columnwidth]{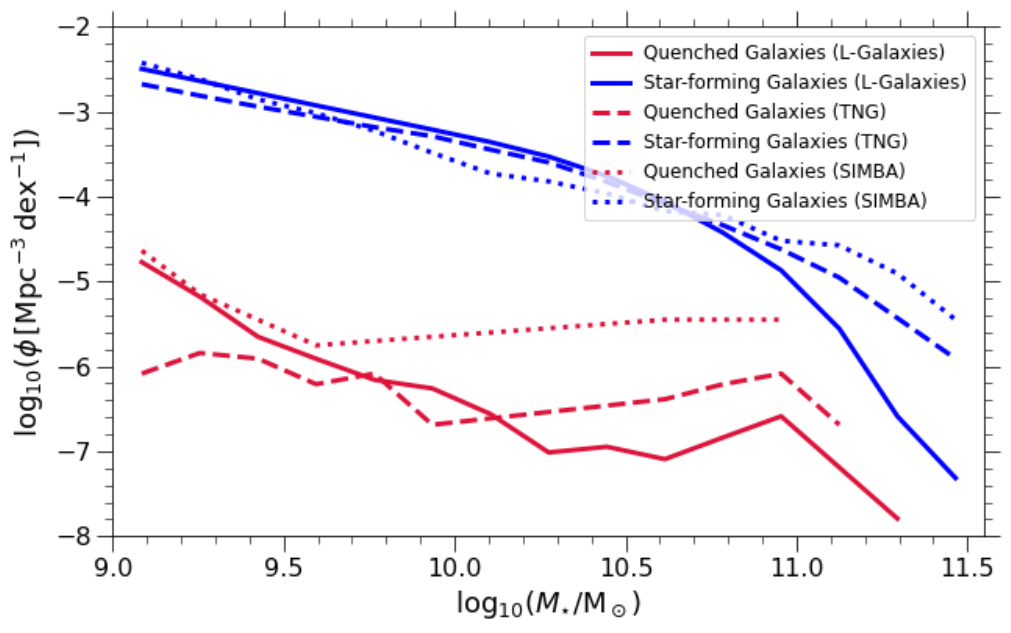} 
  \caption{SMF for quenched (red) and star-forming galaxies (blue) in \textsc{L-Galaxies} (solid), \textsc{TNG} (dashed), and \textsc{Simba} (dotted) at $z=4$.}
\label{fig:stellar_mass_function_quenched_vs_starforming}
\end{figure}

Figure~\ref{fig:stellar_mass_function_quenched_vs_starforming} shows the SMF at $z=4$ in the simulation, with the separated galaxies into star-forming and quenched populations. In all three simulations, the SMF is dominated by star-forming galaxies, especially at low stellar masses where their relative abundance is highest. Although quenched galaxies are less numerous overall, a non-negligible population is already present, and its relative contribution increases toward higher stellar masses ($\log_{10}({M_{\star}}/{\rm M_{\odot}}) \gtrsim 10.5$).

The abundance of the high-mass quenched galaxies ($\log_{10}({M_{\star}}/{\rm M_{\odot}}) \gtrsim 10.5$) differs significantly among the models. \textsc{Simba} predicts the highest number densities, whereas \textsc{L-Galaxies} yields the lowest, with \textsc{TNG} in between. This difference reflects the AGN-feedback implementations: \textsc{Simba}'s kinetic feedback efficiently ejects gas and quenches massive galaxies, while the AGN treatment in \textsc{L-Galaxies} primarily suppresses gas cooling without removing gas.

At the low-mass end, \textsc{L-Galaxies} and \textsc{Simba} predict a substantially higher abundance of quenched galaxies than \textsc{TNG}. The shape of the quenched SMF also differs among the models: \textsc{TNG} and \textsc{Simba} show an approximately flat trend for $\log_{10}({M_{\star}}/{\rm M_{\odot}}) \gtrsim 10$, whereas \textsc{L-Galaxies} declines toward higher stellar masses. By contrast, the star-forming SMFs are more similar across the models, with similar slopes and amplitudes and noticeable differences emerging only at the high-mass end. These differences highlight the sensitivity of quenched galaxy populations to the adopted feedback implementations.

\subsubsection{The fraction of quenched galaxies}
\label{sec:quenchedfraction}

\begin{figure*}
    \centering
    \includegraphics[width=0.48\textwidth]{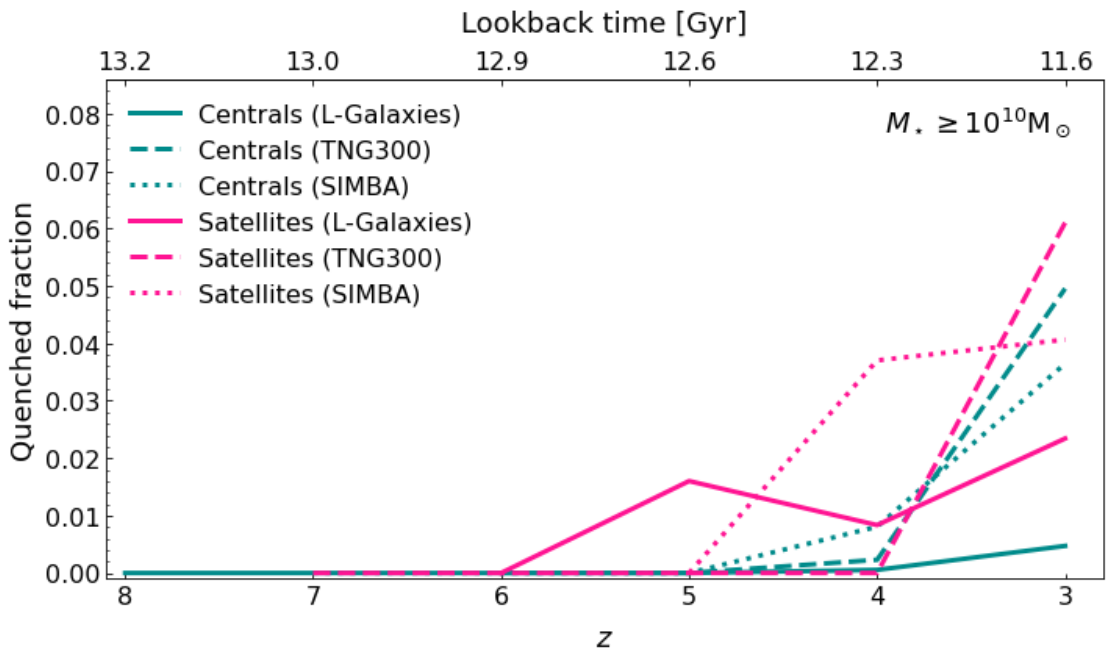}  
    \includegraphics[width=0.48\textwidth]{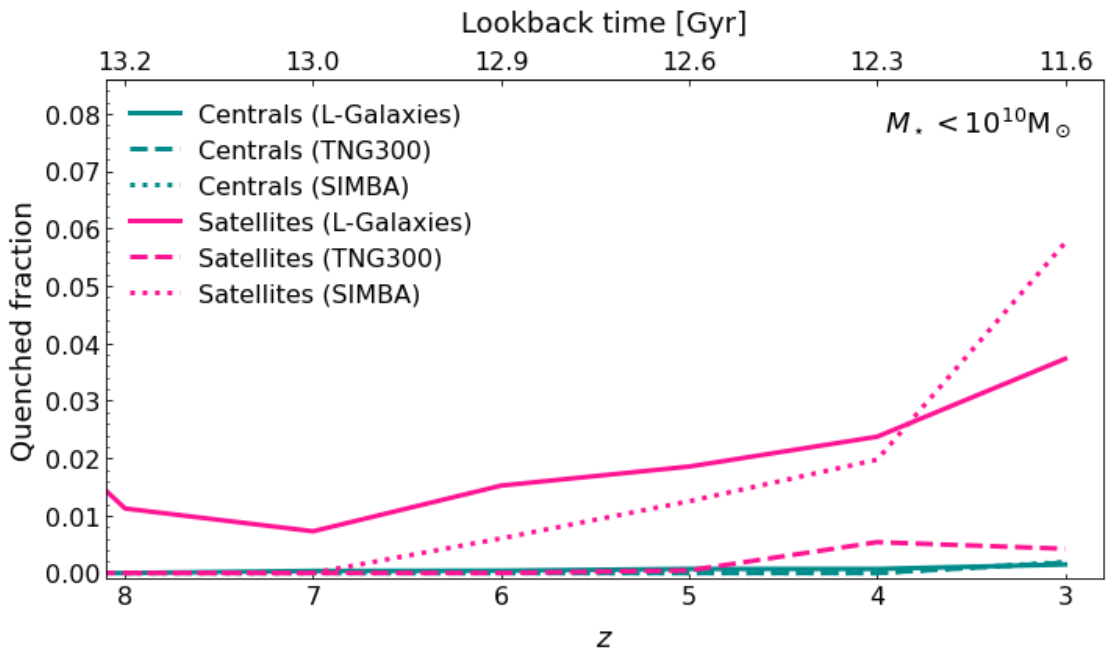}
    
    \includegraphics[width=0.48\textwidth]{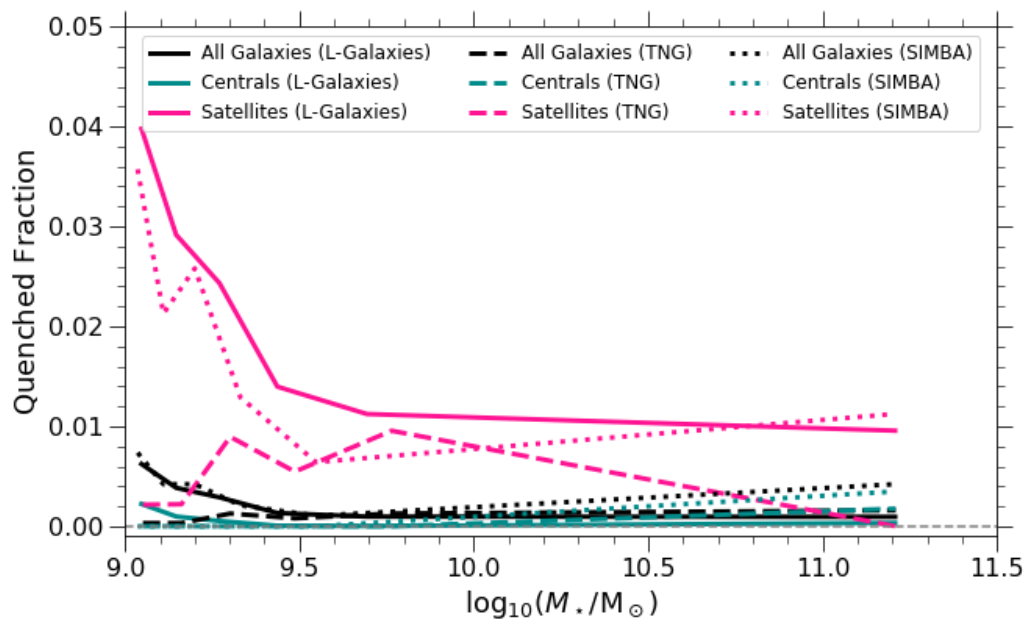} 
    \includegraphics[width=0.48\textwidth]{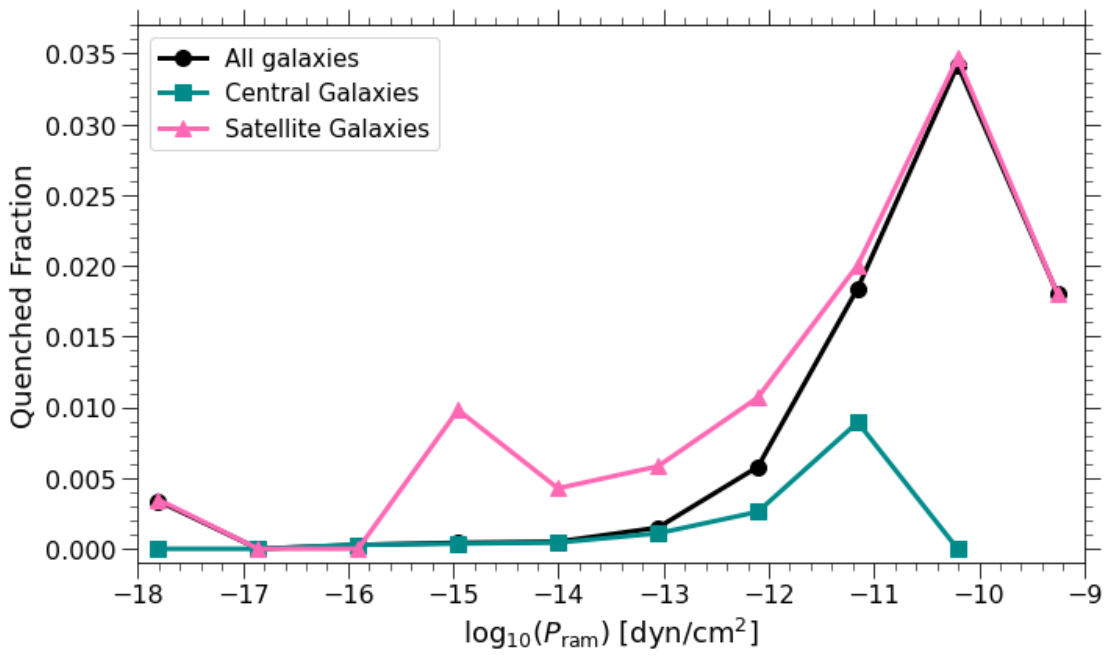} 
    \caption{\textit{Top}: Quenched fraction as a function of redshift for high- (\textit{left}) and low-mass (\textit{right}) galaxies. In all cases, the fraction of quenched galaxies increases with redshift. \textit{Bottom}: Quenched fraction at $z=4$ as a function of stellar mass (\textit{left}) and ram pressure (\textit{right}). Adaptive stellar-mass binning is used in the bottom-left panel to ensure robust statistics (see the top-left panel of Fig.~\ref{fig:QuenchedGalaxyFractionForCentralsandSatellites} for alternative binning). In both rows, line styles denote the simulations: \textsc{L-Galaxies} (solid), \textsc{TNG} (dashed), and \textsc{Simba} (dotted). Colors denote galaxy type: satellite (pink) and central (dark cyan). The bottom-right panel shows only \textsc{L-Galaxies}.}
    \label{fig:quenched_fraction_plots}
\end{figure*}

The top panel of Fig.~\ref{fig:quenched_fraction_plots} shows the evolution of the quenched fraction with redshift for central and satellite galaxies in two stellar mass bins across the three simulations. In the massive bin, the onset and subsequent growth of quenching differ among the models: quenched satellites appear by $z \sim 6$ in \textsc{L-Galaxies} and around $z \sim 5$ in \textsc{Simba}, while \textsc{TNG} shows the latest onset, remaining close to zero until $z \sim 4$. Despite this, \textsc{TNG} predicts the highest quenched fractions at $z \sim 3$. In all three simulations, satellites have systematically higher quenched fractions than centrals at fixed redshift, indicating that environmental processes contribute to the early build-up of the quenched population, while the timing and efficiency of quenching remain model-dependent.

The top-right panel of Fig.~\ref{fig:quenched_fraction_plots} shows that low-mass central galaxies remain predominantly star-forming across the full redshift range in all three simulations, with quenched fractions close to zero. By contrast, quenched satellites are already present. At $z \gtrsim 7$, only \textsc{L-Galaxies} predicts a nonzero quenched fraction for low-mass satellites, whereas neither \textsc{TNG} nor \textsc{Simba} show any quenched systems, which may partly reflect the larger volume of \textsc{L-Galaxies}. Toward lower redshifts, \textsc{Simba} shows a rapid rise in the quenched fraction of low-mass satellites and reaches the highest values among the three simulations by $z \sim 3$, whereas \textsc{TNG} remains close to zero. \textsc{L-Galaxies}, by contrast, exhibits a more gradual increase with decreasing redshift.

For reference, at $z \sim 4$ the quenched fraction of satellite galaxies is $\sim 2\%$ in \textsc{L-Galaxies} and \textsc{Simba}, and $<1\%$ in \textsc{TNG}. Although central galaxies constitute the vast majority of the population ($\sim 91$\% in \textsc{L-Galaxies}, $\sim 87$\% in \textsc{TNG}, and $\sim 84$\% in \textsc{Simba}), they account for only a small fraction of quenched systems. By contrast, satellite galaxies contribute disproportionately, representing the majority ($77$\% in \textsc{L-Galaxies}, $72$\% in \textsc{TNG}, and $86$\% in \textsc{Simba}) of all quenched systems (Table~\ref{tab:quenched_comparison}). This provides strong evidence that environmental quenching mechanisms are already effective at high redshifts ($z > 3$).

Next, we examined the quenched fraction as a function of stellar mass (Fig.~\ref{fig:quenched_fraction_plots}, bottom left). For central galaxies, the quenched fraction remains close to zero at $\log_{10}({M_{\star}}/{\rm M_{\odot}}) \lesssim 10.5$ in all simulations, with only a small number of low-mass quenched centrals in \textsc{L-Galaxies}. At higher masses, $\log_{10}({M_{\star}}/{\rm M_{\odot}}) \gtrsim 10.5$, the quenched fraction of centrals slightly increases with stellar mass. By contrast, quenched satellites are already present at the low-mass end in all models. In \textsc{L-Galaxies}, the quenched fraction decreases with stellar mass up to $\log_{10}({M_{\star}}/{\rm M_{\odot}}) \sim 9.8$ and then flattens at higher masses. \textsc{Simba} shows a similar trend, with a slight rise for $\log_{10}({M_{\star}}/{\rm M_{\odot}}) \gtrsim 9.6$, likely due to feedback combined with environment. In \textsc{TNG}, the quenched fraction of satellites rises with stellar mass up to $\log_{10}({M_{\star}}/{\rm M_{\odot}}) \sim 9.8$ and then declines, reaching zero at the high-mass end.

\begin{figure}
    \centering
    \includegraphics[width=0.95\columnwidth]{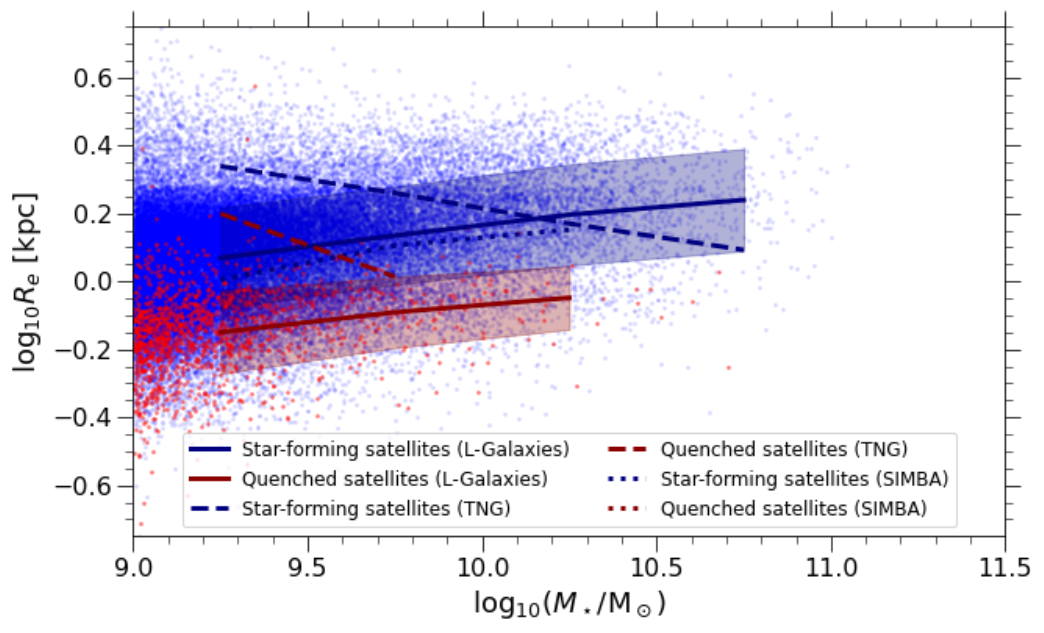}
    \caption{Stellar half-mass radius versus stellar mass for satellite galaxies at $z=4$ in \textsc{L-Galaxies} (solid), \textsc{TNG} (dashed), and \textsc{Simba} (dotted). Individual galaxies (points) and the standard deviation around the median (shaded regions) are shown for \textsc{L-Galaxies}, together with the median trends (lines) for all models. Quenched \textsc{Simba} satellites are not shown due to low statistics.}
    \label{fig:StellarHalfmassRadius}
\end{figure}

Finally, the bottom-right panel of Fig.~\ref{fig:quenched_fraction_plots} shows the quenched fraction as a function of ram pressure in \textsc{L-Galaxies}. Central galaxies have quenched fractions close to zero across the full ram-pressure range, whereas satellites show a strong positive correlation with ram pressure. This is expected because ram-pressure stripping is cumulative: centrals typically do not spend enough time in the dense intra-halo medium to be strongly affected, and they can continue to accrete gas from the intergalactic medium, unlike satellites. At fixed ram pressure, low-mass satellites reach higher quenched fractions owing to their shallower potential wells (see the top-right panel of Fig.~\ref{fig:QuenchedGalaxyFractionForCentralsandSatellites}). Because ram pressure is highest in the inner halo, where the LBE density is greatest (Sect.~\ref{sec:rampressureradial}), the increase in quenched fraction with ram pressure implies that quenching is most efficient in these dense regions (Sect.~\ref{sec:quenchedfraction}).

\subsubsection{Galaxy sizes across galaxy populations}
\label{sec:Stellarhalf-massradius}
\begin{figure}
    \centering
    \includegraphics[width=0.95\columnwidth]{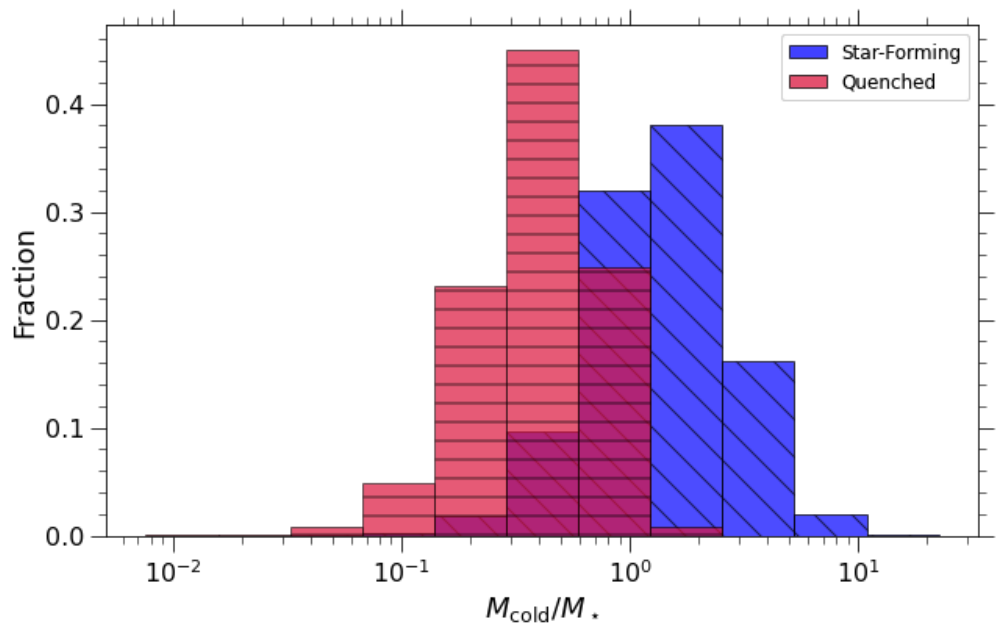}
    \includegraphics[width=0.95\columnwidth]{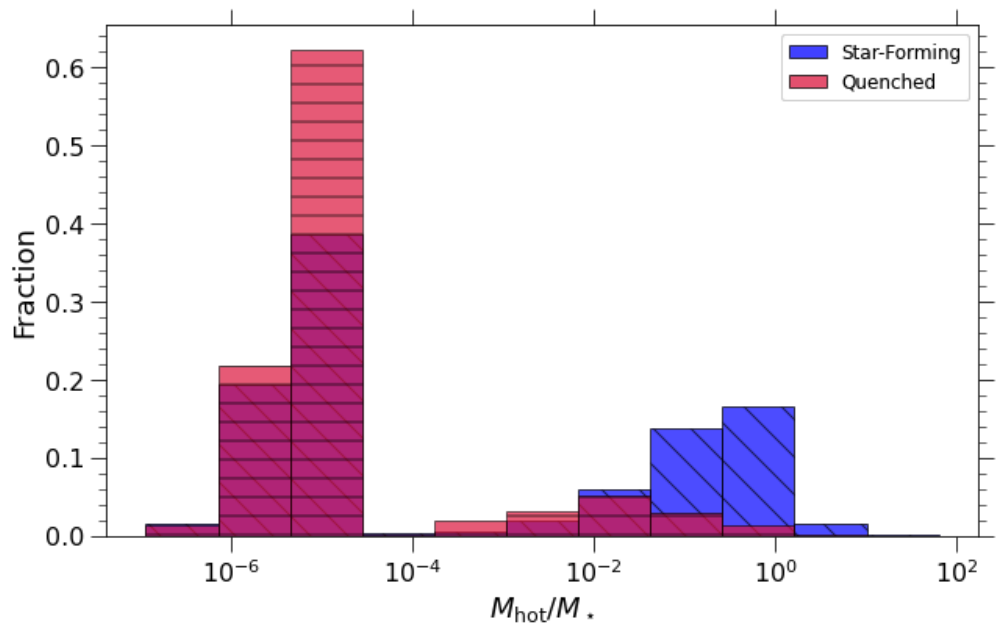}

    \caption{Normalized histograms of cold gas mass (\textit{top}) and hot gas mass (\textit{bottom}), divided by stellar mass, for satellite galaxies at $z=4$ in \textsc{L-Galaxies}. Galaxies are separated by type: star-forming (blue) and quenched (red). Hot gas masses below $10^4{\rm M_{\odot}}$ are set to this value. Since trends show only a weak dependence on halo mass, we do not separate the populations based on halo mass.}
    \label{fig:gas_histogram}
\end{figure}

Figure~\ref{fig:StellarHalfmassRadius} shows the relation between stellar half-mass radius and stellar mass for satellite galaxies at $z=4$ in the three simulations\footnote{Due to limited statistics in \textsc{Simba}, quenched satellites fall below the minimum-galaxy threshold and are therefore not shown.}. In both \textsc{L-Galaxies} and \textsc{Simba}, the half-mass radius increases with stellar mass, consistent with the usual size-mass relation in which more massive satellites tend to be larger. By contrast, \textsc{TNG} shows the opposite trend, with satellites becoming more compact at higher stellar masses, indicating a different structural evolution in this model. This compactification in \textsc{TNG} may be driven by rapid central mass assembly associated with gas-compaction processes at high redshifts \citep[e.g.,][]{Costantin2023expectations}. In this picture, high-mass satellites in \textsc{TNG} grow predominantly through dense, centrally concentrated star formation rather than extended disk growth, leading to smaller half-mass radii at higher stellar masses. Overall, high-mass satellites in \textsc{TNG} are more compact than \textsc{L-Galaxies} and \textsc{Simba}.

Satellite galaxies show a clear separation between quenched and star-forming populations in all models. At fixed stellar mass, quenched satellites are typically smaller than their star-forming counterparts. This size difference likely reflects their assembly histories: quenched satellites assemble most of their stellar mass earlier, during more dissipative phases with lower specific angular momentum, leading to more compact stellar distributions. By contrast, star-forming satellites continue to grow at later times through the accretion of higher-angular-momentum gas, resulting in more extended stellar components. Environmental processes after infall can further reduce the sizes of quenched satellites by suppressing late-time star formation and removing loosely bound outer material. This size contrast between quenched and star-forming galaxies, well established at lower redshifts (e.g., \citealt{vani2025probing} for \textsc{L-Galaxies} and \citealt{genel2018size} for \textsc{TNG}), is already clearly present at $z=4$.

\begin{figure}
    \centering
    \includegraphics[width=0.95\columnwidth]{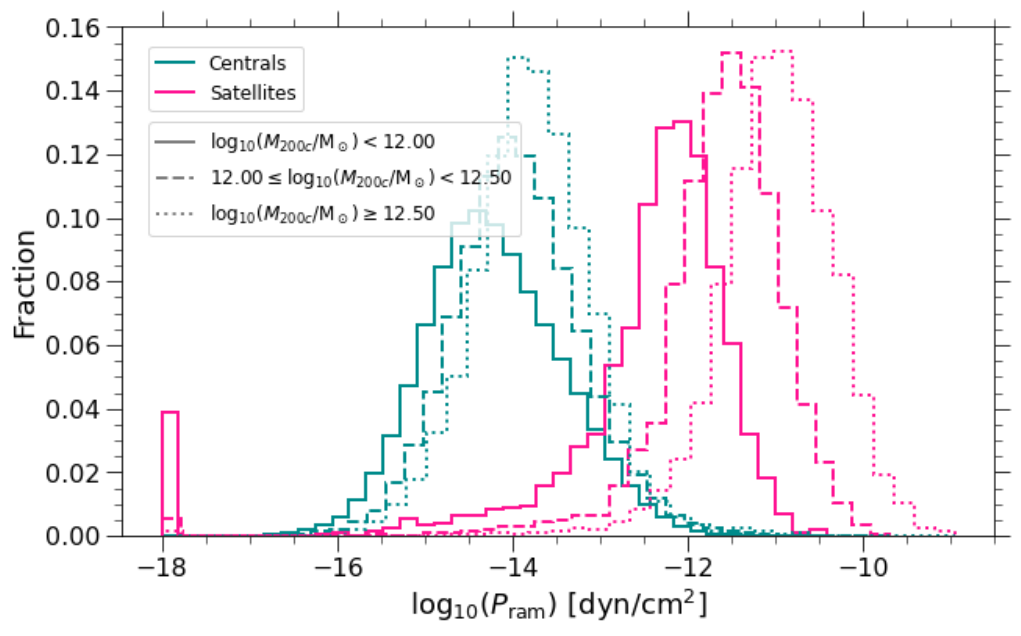}
    \includegraphics[width=0.95\columnwidth]{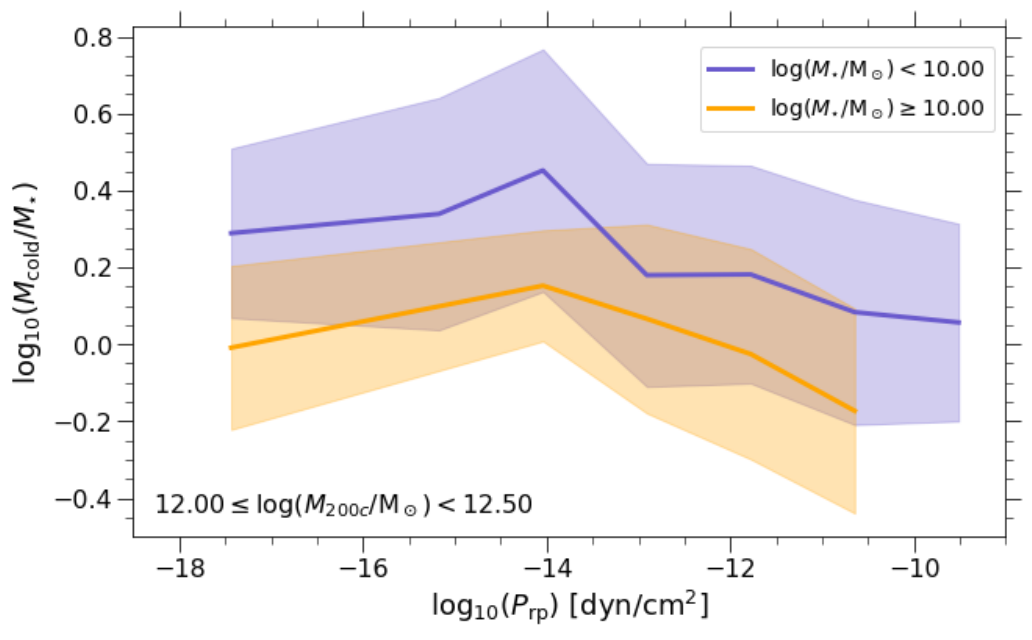}
    \includegraphics[width=0.95\columnwidth]{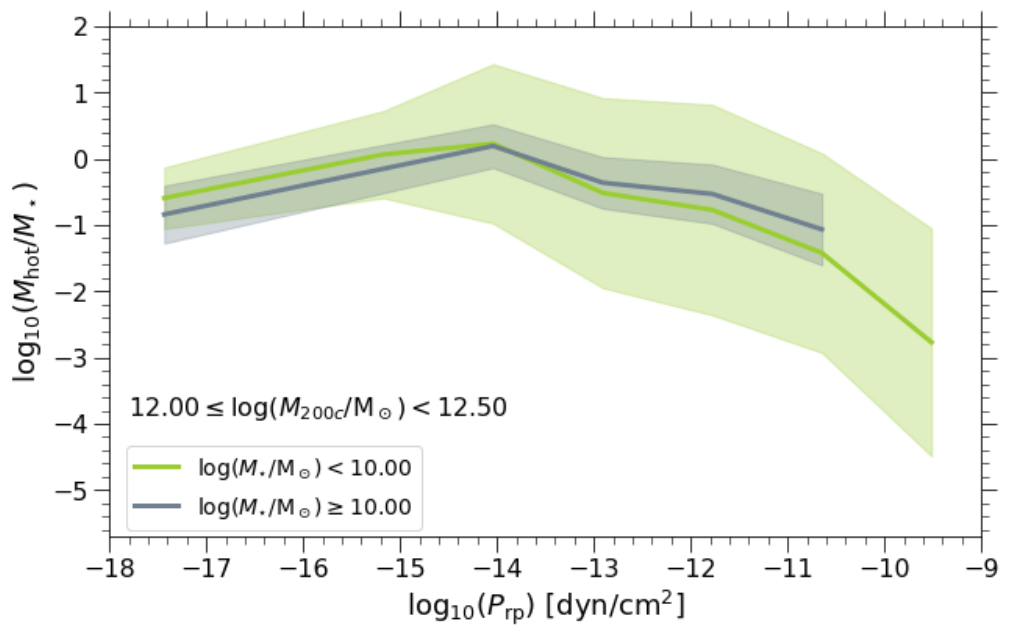}
    \caption{\textit{Top}: Normalized histogram of ram pressure for central and satellite galaxies in \textsc{L-Galaxies} at $z=4$ shown in three halo mass bins. Ram pressure values smaller than $10^{-18} \ \mathrm{dyn/cm^{-2}}$ are mapped to this value. \textit{Middle} and \textit{bottom}: Cold and hot gas to stellar mass ratio as a function of ram-pressure only for satellite galaxies in intermediate-mass halos in \textsc{L-Galaxies}. Shaded regions indicate the standard deviation around the median values.} 
    \label{fig:rampressure_hist}
\end{figure}

\begin{figure*}
    \centering
    \includegraphics[width=0.48\textwidth]{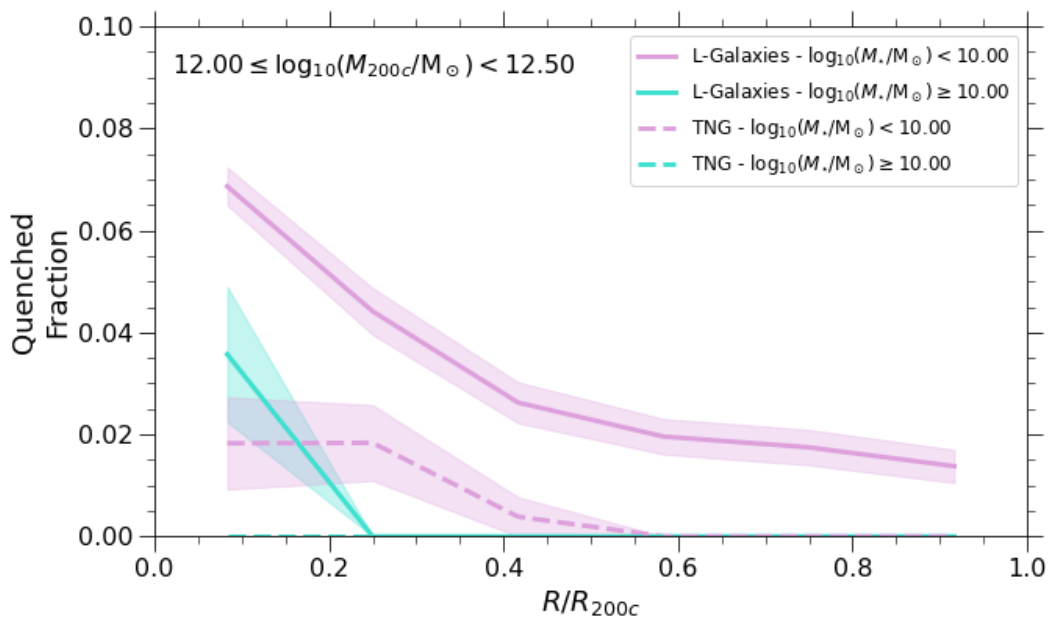}
    \includegraphics[width=0.48\textwidth]{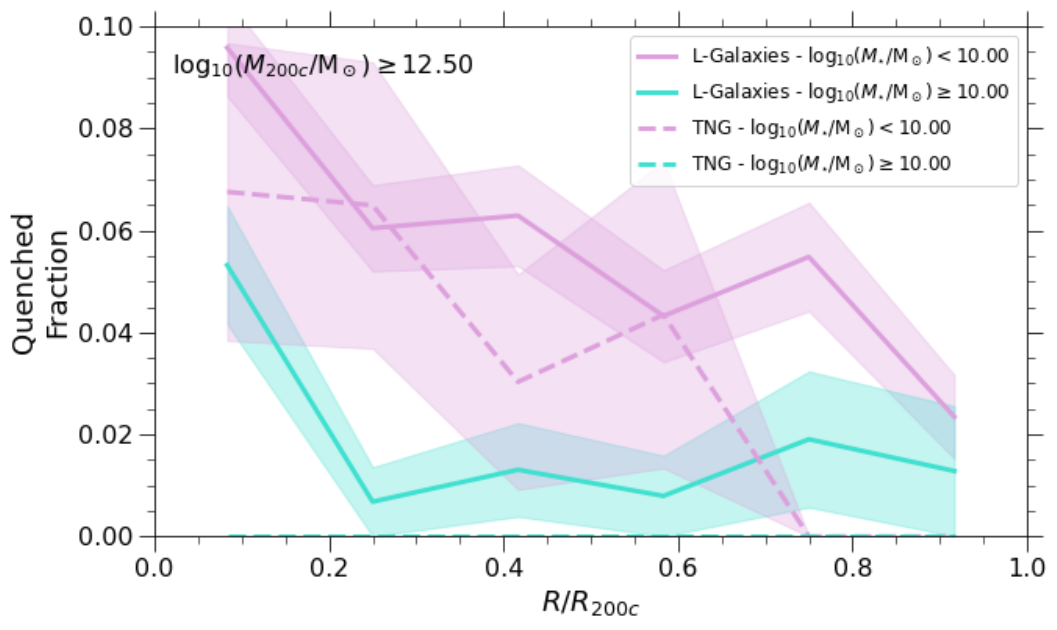}
    \caption{Quenched satellite galaxy fraction as a function of normalized halocentric distance ($R/R_{\rm 200c}$) for two simulations --- \textsc{L-Galaxies} (solid) and \textsc{TNG} (dashed) --- at $z=4$. The satellite population is divided into three host halo mass bins (panels) and stellar mass bins (colors). Shaded regions represent the standard deviation. \textsc{Simba} is excluded from the main analysis due to low-number statistics; however, it is shown for completeness in the bottom panel of Fig.~\ref{fig:QuenchedGalaxyFractionForCentralsandSatellites}, together with the lowest halo-mass bin.}
    \label{fig:quenched_fraction_profile_halo_MAIN}
    \end{figure*}

\subsubsection{Multiphase gas}
\label{sec:globalgascontent}
Figure~\ref{fig:gas_histogram} shows histograms of cold and hot gas mass for quenched and star-forming satellites  in \textsc{L-Galaxies}. The cold gas distributions show a clear separation between star-forming and quenched satellites. Quenched satellites have low cold gas fractions, whereas star-forming satellites extend to substantially higher values (often ${M_{cold}}\geq {M_{\star}}$).

The bottom panel shows the near-complete depletion of hot gas in quenched satellites, as well as a subset of star-forming satellites. This is expected because hot gas resides in the extended outer halo, where it is more weakly bound and therefore most easily stripped by environmental processes such as ram-pressure and tidal stripping \citep{bahe2013does, Wetzel2014Galaxy, ayromlou2019new}. By contrast, many star-forming satellites occupy a distinctly higher range of hot gas fractions, from ${M_{hot}}/{M_{\star}}\sim 10^{-3}$ up to $\sim 10$, depending on halo mass.

Overall, star-forming satellites typically retain substantial hot gas reservoirs, indicating that removal of the hot halo is a key step in high-$z$ quenching. By stripping the hot halo, ram pressure suppresses the supply of fresh cold gas, leading to gradual cold gas depletion and the eventual cessation of star formation.

\subsubsection{Ram pressure}
\label{sec:rampressurehistogram}

We present the distributions of ram pressure for galaxies in \textsc{L-Galaxies} in the top panel of Fig.~\ref{fig:rampressure_hist}. The ram-pressure values, taken from the \textsc{L-Galaxies} public data release, are measured using the LBE method based on the particle data of the simulation (see \citealt{ayromlou2019new,ayromlou2021galaxy} for details). The distributions reveal two main trends. First, ram pressure is systematically lower for central galaxies than for satellites, as expected because satellites orbit within the dense intra-halo medium and experience higher relative velocities and ambient gas densities. Second, ram pressure increases with host-halo mass, most strongly for satellites, reflecting the more intense environments of massive halos (higher densities and relative velocities). Thus, even at high redshifts, satellites can experience substantial ram pressure, particularly in massive halos, with important consequences for their gas content and star formation activity.

Next, we analyzed the relation between gas content and ram pressure (see the middle and bottom panels of Fig.~\ref{fig:rampressure_hist}), splitting satellites into low-mass ($\log_{10}({M_{\star}}/{\rm M_{\odot}}) < 10$) and high-mass ($\log_{10}({M_{\star}}/{\rm M_{\odot}}) \geq 10$) systems. The sample was further divided into three halo mass bins. For brevity, we show only the intermediate bin in Fig.~\ref{fig:rampressure_hist}; the remaining bins are presented in Fig.~\ref{fig:rampressure_coldgasAppendix}. For both gas phases, the median gas-to-stellar mass ratio decreases with increasing ram pressure, indicating progressive gas loss for satellites in denser environments. At fixed ram pressure, the hot gas fractions show little dependence on stellar mass, while low-mass satellites are more cold-gas-rich. This likely reflects cold gas removal in more massive systems by internal processes such as AGN feedback.

\subsection{Radial trends with halocentric distance}
\label{sec:environmentaldependenceofsatellitegalaxyquenching}

In this subsection we investigate how satellite properties vary with halocentric distance ($R/R_{\rm 200c}$), focusing on \textsc{L-Galaxies} and \textsc{TNG}, as they provide sufficient statistics.

\subsubsection{Quenched fraction across halocentric distance}
\label{sec:quenchedfractionradial}

Figure~\ref{fig:quenched_fraction_profile_halo_MAIN} shows the quenched fraction as a function of halocentric distance in \textsc{L-Galaxies}  and \textsc{TNG}. The strongest radial trends occur in the high-mass halo bin, where low-mass satellites show a clear increase in quenched fraction toward smaller radii in both simulations, indicating that quenching is most effective in the inner regions of massive halos. High-mass satellites in \textsc{L-Galaxies} show elevated quenched fractions at small radii, followed by a rapid decline and an approximately flat trend at larger distances, whereas in \textsc{TNG} high-mass satellites remain star-forming across the full radial range. In the intermediate halo mass bin, the radial trends persist but are noticeably weaker.

Differences between the stellar mass bins are also pronounced. Low-mass satellites show higher quenched fractions than high-mass satellites, consistent with environmental processes being more effective in systems with weaker self-gravity. Overall, these results indicate that although simulated galaxies at $z \gtrsim 4$ are predominantly star-forming, environmental quenching can already operate in dense halo environments, particularly for low-mass satellites in massive halos.

\begin{figure*}
    \centering

    \includegraphics[width=0.48\textwidth]{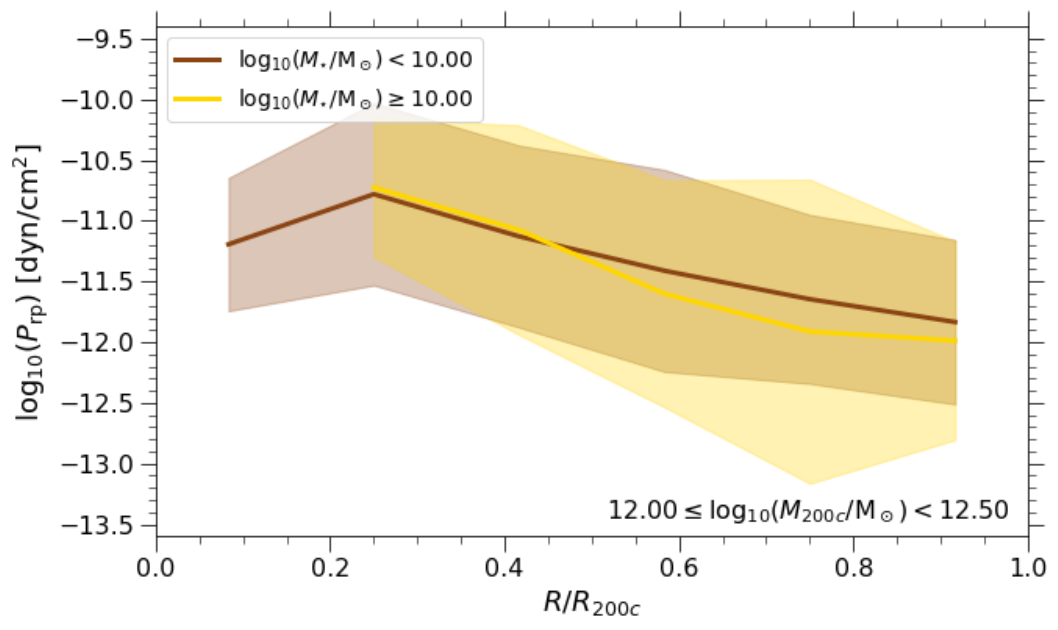}
    \includegraphics[width=0.48\textwidth]{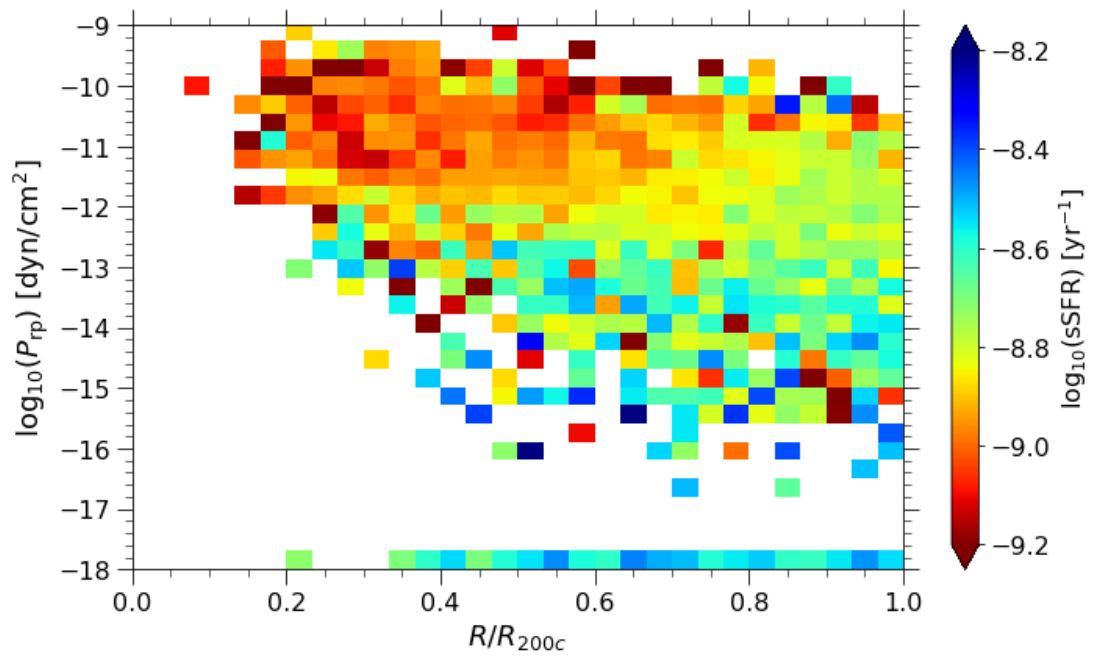}

    \includegraphics[width=0.48\textwidth]{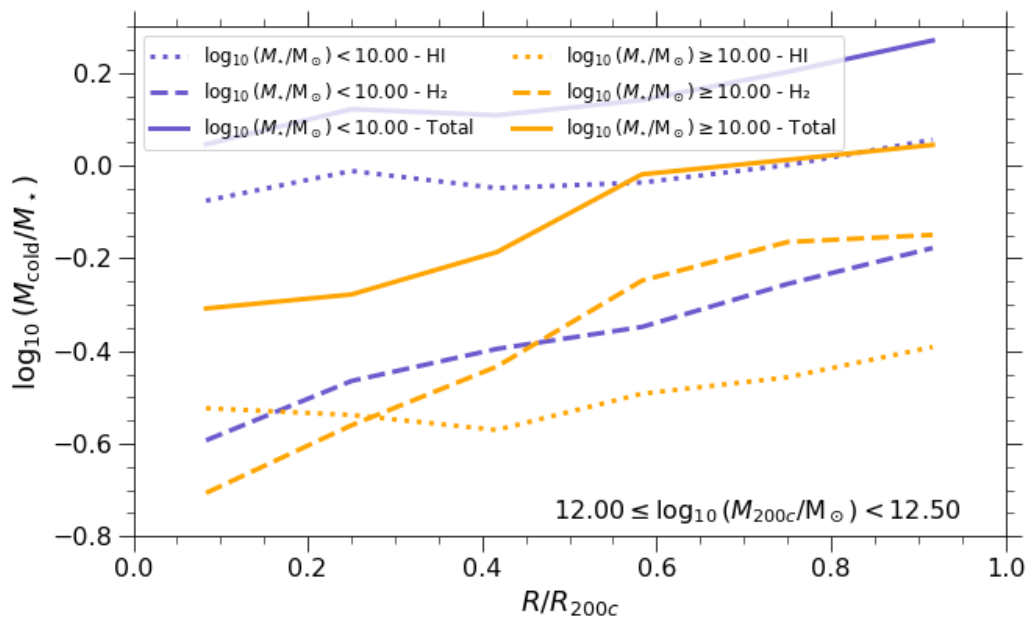}
    \includegraphics[width=0.48\textwidth]{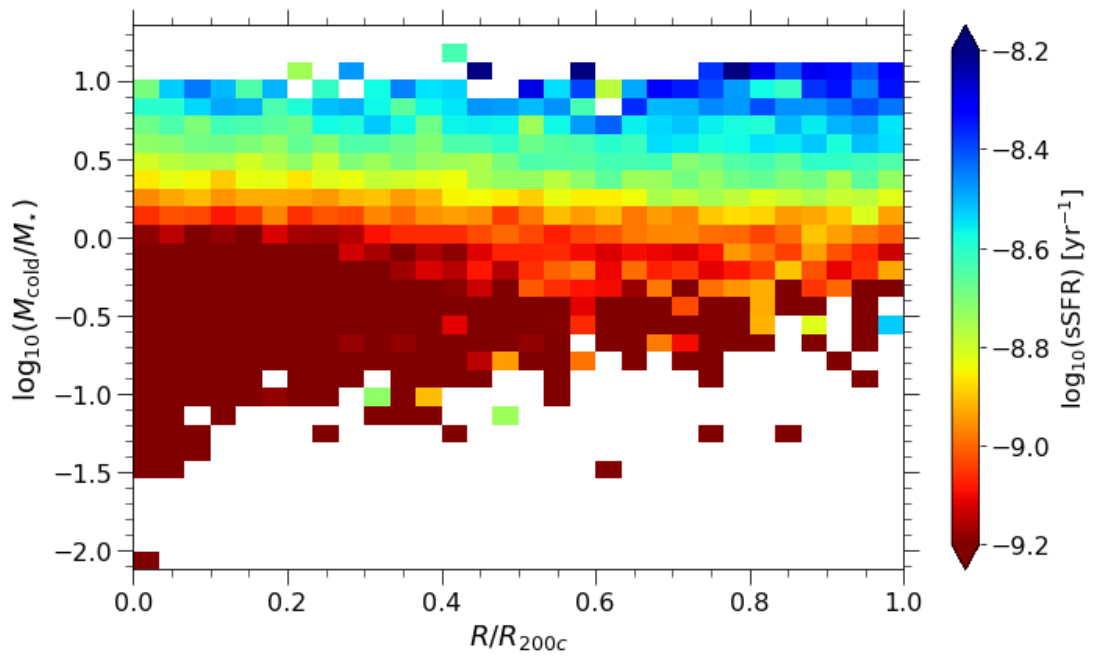}

    \includegraphics[width=0.48\textwidth]{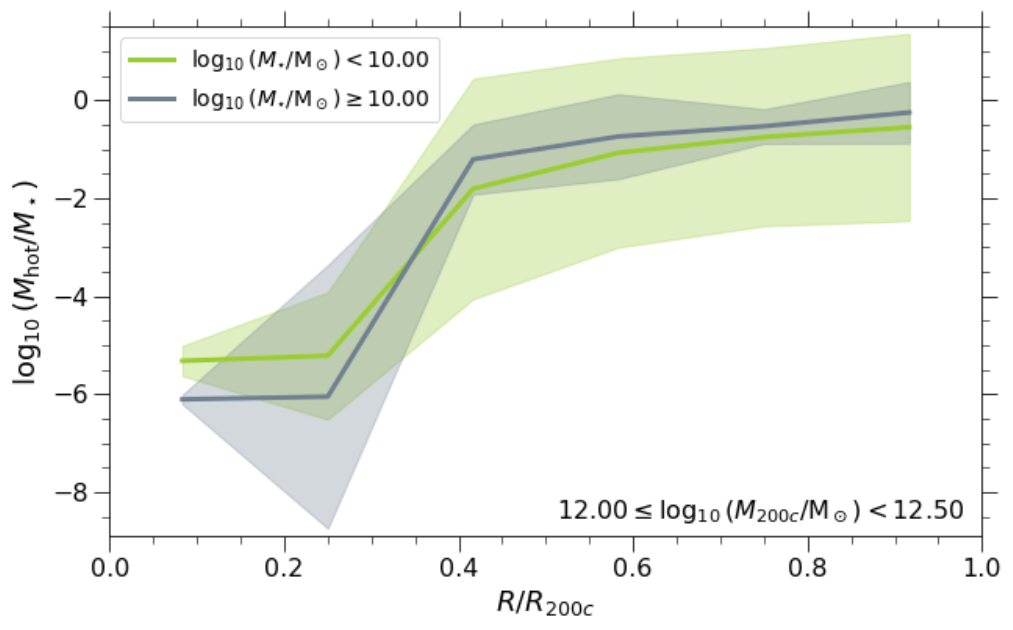}
    \includegraphics[width=0.48\textwidth]{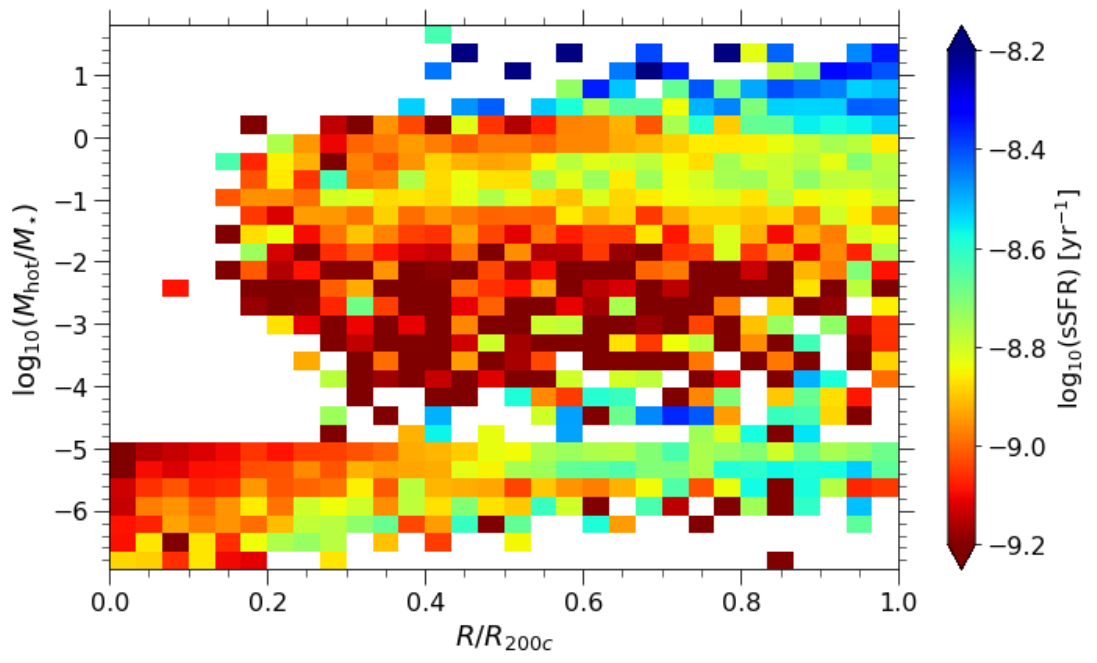}

    \caption{Radial trends in satellite galaxy properties at $z=4$ in \textsc{L-Galaxies}. From \textit{top to bottom}: ram pressure, cold-gas content, and hot-gas content. \textit{Left column}: Median profiles as a function of halocentric distance ($R/R_{\rm 200c}$). \textit{Right column}: 2D histograms of the same quantities, colored by sSFR. In the 2D histograms, bins are shown whenever they contain at least one galaxy. Shaded regions indicate the standard deviation around the median profiles. Zero hot-gas masses are mapped to $10^{4}{\rm M_{\odot}}$, and ram-pressure values below $10^{-18}\mathrm{dyn\,cm^{-2}}$ are set to this value. The intermediate halo-mass bin is shown here; the low- and high-mass bins, and, for completeness, trends for the half-mass radius are shown in Appendix~\ref{sec:appendixother}.}
    \label{fig:6panel_grid}
\end{figure*}

\subsubsection{Ram pressure across halocentric distance}
\label{sec:rampressureradial}

Figure~\ref{fig:6panel_grid} presents the relationship between satellite properties and halocentric distance in \textsc{L-Galaxies} for several quantities, including ram pressure and cold and hot gas masses. The first row reveals a steep radial gradient in ram pressure. Overall, ram pressure increases toward smaller halocentric distances. In the innermost regions ($R/R_{\rm 200c} \lesssim 0.2$), the median profile shows a slight decrease toward the center, which may reflect a resolution effect if satellites in the most intense environments fall below the resolution limit and are therefore missing from the sample. The radial ram-pressure trends show little dependence on stellar mass, and are primarily driven by environment. The 2D histogram shows a strong correlation between high ram pressure, low sSFR, and small halocentric distance: at fixed distance, satellites experiencing stronger ram pressure have systematically lower sSFRs. Moreover, at fixed ram pressure, satellites closer to the halo center tend to have lower sSFRs, consistent with longer exposure to the halo environment.

\subsubsection{Multiphase gas across halocentric distance}
\label{sec:multigas}
The middle and bottom rows of Fig.~\ref{fig:6panel_grid} illustrate the gas content of satellites as a function of halocentric distance. Satellites at larger halocentric distances ($R/R_{\rm 200c} \gtrsim 0.5$) retain substantially higher gas fractions than those near the halo center. The strong suppression of gas fractions at small radii indicates the efficiency of environmental processes in dense central regions.

The cold-gas fraction shows a pronounced dependence on stellar mass, with low-mass satellites retaining higher $ M_{\mathrm{cold}}/M_\star$ at fixed radius. This trend holds at all halocentric distances for the HI component. For the molecular component, low-mass satellites likewise exhibit higher $ M_{\mathrm{H_2}}/M_{\star}$ at smaller radii, but the trend reverses within $R/R_{\rm 200c} \gtrsim 0.45$, where higher-mass satellites retain slightly larger molecular gas reservoirs. In the H$_2$-based star-formation model employed by \textsc{L-Galaxies}, star formation is regulated by the H$_2$ reservoir. While both HI and H$_2$ decline toward the inner regions of halos, the depletion of H$_2$ becomes especially pronounced in the innermost regions, where the quenched fraction also reaches its maximum.

For the hot-gas component, $M_{\mathrm{hot}}/M_{\star}$ shows a sharp decrease from $R/R_{\rm 200c} \sim 0.4$ to $0.2$, indicating efficient hot gas stripping. Beyond this transition, the hot-gas fraction shows only a weak radial dependence and is nearly independent of stellar mass. The 2D histograms show a positive correlation between sSFR and gas at fixed halocentric distance. Conversely, at fixed gas fraction, satellites closer to the halo center show lower sSFR. Consistent with this, gas-rich, actively star-forming satellites are found in the outer halo, whereas gas-poor systems with low sSFR are more common near the halo center.

\section{Summary and conclusions}
\label{sec:Summary}

In this study we investigated the physical processes responsible for quenching the low-mass galaxies observed by JWST. To do so, we used three cosmological simulations (\textsc{L-Galaxies}, \textsc{TNG300}, and \textsc{Simba}), as well as the \textsc{TNG-Cluster} zoom-in simulation. In particular, we evaluated the efficiency of environmental quenching in the early Universe:

\begin{itemize}
  \item The simulations reproduce the JWST low-mass quenched population at $z > 4$ and suggest that these systems are predominantly satellites, based on their location in the SFR--stellar-mass plane (Fig.~\ref{fig:doubletroublesfr}).
  
  \item Most simulated galaxies quenched at $z \sim 4-5$ do not remain passive: $\sim 90\%$  have merged or undergone disruption by $z=0$, and only a small fraction remain continuously quenched across all stellar-mass bins (Fig.~\ref{fig:rejuv_plots}).

  \item The quenched fraction remains low at $z > 4$, but quenched systems are predominantly satellites in all simulations. Satellite quenching emerges earlier and reaches higher fractions than central quenching, consistent with efficient environmental quenching at early times (Fig.~\ref{fig:quenched_fraction_plots}).
  \item Ram-pressure quenching is effective for satellites at $z \sim 4$ but negligible for centrals, with satellite quenched fractions rising steeply with ram pressure (Fig.~\ref{fig:quenched_fraction_plots}, bottom right).
  \item Satellite sizes follow a common mass-size relation in \textsc{L-Galaxies} and \textsc{Simba}, whereas \textsc{TNG} predicts more compact star-forming satellites at the highest masses. In all models, quenched satellites are more compact than star-forming satellites at fixed stellar mass (Fig.~\ref{fig:StellarHalfmassRadius}).
  \item Quenched satellites have uniformly low cold- and hot-gas fractions, whereas star-forming satellites retain substantially larger gas reservoirs (Fig.~\ref{fig:gas_histogram}).
  \item Quenched satellites mostly reside at small halocentric distances, and the quenched fraction increases with host-halo mass. Controlling for the distance and host mass, quenching is more common among low-mass satellites (Fig.~\ref{fig:quenched_fraction_profile_halo_MAIN}).
  \item Cold- and hot-gas fractions show strong radial trends in satellites. Cold gas declines toward the halo center and closely tracks the sSFR, with gas-rich, high-sSFR satellites preferentially found in the outskirts (Fig.~\ref{fig:6panel_grid}). The hot gas fraction is highest at large radii, drops sharply within $R/R_{\rm 200c} \leq 0.4$, and shows little stellar-mass dependence, indicating that hot-halo stripping is primarily environment-driven (Fig.~\ref{fig:6panel_grid}).

\end{itemize}

Overall, the existence of simulated analogs to the JWST low-mass quenched systems suggests that environmental quenching can already operate efficiently in the early Universe, while differences among models highlight the sensitivity of these predictions to physical prescriptions. Future JWST samples with improved statistics and environmental characterization will enable decisive tests of our scenario.

\begin{acknowledgements}
We are grateful to the anonymous referee for the constructive comments, which have helped us improve the quality of our paper. We thank Romeel Dave for kindly assisting with the use of the \textsc{Simba} simulation data. MA is supported at the Argelander Institut für Astronomie through the Argelander Fellowship. We gratefully acknowledge the Collaborative Research Center 1601 (SFB 1601 sub-project C5) funded by the Deutsche Forschungsgemeinschaft (DFG, German Research Foundation) – 500700252. The computations and analyses presented in this work were performed using the AIfA computational cluster and the Marvin supercomputer at the University of Bonn. Data from the \textsc{L-Galaxies}, \textsc{IllustrisTNG}, \textsc{TNG-Cluster}, and \textsc{Simba} simulations are publicly available.
\end{acknowledgements}

\bibliographystyle{aa}
\bibliography{refbibtex}

\newpage
\begin{appendix}

\section{Observed quenched galaxies and their simulation analogs}
\label{sec:analogs}

\begin{table*}
\centering
\caption{Observed low-mass quenched galaxy candidates and their simulated analogs in \textsc{L-Galaxies}, \textsc{TNG-Cluster}, and \textsc{Simba}.}
\label{tab:pc_level_summary}
\begin{tabular}{lccccccc}
\toprule
Observation/Simulation & $\log_{10}({M_{\star}}/{\rm M_{\odot}})$ & SFR [${\rm M_{\odot}}$/yr] & $z$ & $\log_{10}({M_{\rm 200c}}/{\rm M_{\odot}})$ & U-V & V-J & Type \\
\midrule
\citetalias{baker2025doubleA} & $9.62^{+0.01}_{-0.01}$ & $0.01^{+0.12}_{-0.01}$ & $5.39^{+0.00}_{-0.00}$ & - & 0.97 & 0.52 &- \\ 
\midrule
\citetalias{baker2025doubleB}  & $9.55^{+0.12}_{-0.08}$ & $0.39^{+1.53}_{-0.35}$ & $5.11^{+0.01}_{-0.01}$ & - & 1.13 & 0.61 &- \\
\midrule
\citetalias{alberts2024highredshift} &  $9.89^{+0.03}_{-0.04}$ & $3.98\times 10^{-5}$  &  $5.33^{+0.16}_{-0.17}$ &  - & $1.26^{+0.04}_{-0.03}$ & $0.41^{+0.06}_{-0.07}$ & - \\
\midrule
\citetalias{delaVega25searchingA} & $9.86^{+0.07}_{-0.06}$ & 0.21 & $5.45^{+0.15}_{-0.28}$ &  - &  - & -  &- \\
\midrule
\citetalias{delaVega25searchingB} & $9.45^{+0.16}_{-0.15}$  & $2.04\times 10^{-4}$ & $5.52^{+0.27}_{-0.11}$ &  - & - & - & - \\
\midrule
\textsc{L-Galaxies}   & 9.41 & 0.23 & 5.46 & 12.00 & 1.03 & 0.70 & Satellite \\

& 9.30 & 0.03 & 5.03 & 12.04 & 1.13 & 0.79 & Satellite \\
 & 9.73 & 0.05 & 5.03 & 12.67 & 1.25 & 0.89 & Satellite \\
 & 9.33 & 0.17 & 5.03 & 12.17 & 1.17 & 0.83 & Satellite \\
 & 9.48 & 0.09 & 5.03 & 12.27 & 1.15 & 0.78 & Satellite \\
 & 9.37 & 0.03 & 5.03 & 12.46 & 1.17 & 0.87 & Satellite \\
 & 9.39 & 0.12 & 5.03 & 12.35 & 1.08 & 0.79 & Satellite \\
 & 9.48 & 0.16 & 5.03 & 12.63 & 1.18 & 0.84 & Satellite \\
 & 9.42 & 0.03 & 5.03 & 12.19 & 0.00 & 0.00 & Satellite \\
 & 9.62 & 0.29 & 5.03 & 12.03 & 1.13 & 0.73 & Satellite \\
 & 9.43 & 0.00 & 5.03 & 12.73 & 1.24 & 0.90 & Satellite \\
 & 9.82 & 0.09 & 5.03 & 12.20 & 1.12 & 0.72 & Satellite \\
 & 9.37 & 0.17 & 5.03 & 11.86 & 1.04 & 0.65 & Satellite \\
 & 9.36 & 0.12 & 5.03 & 12.41 & 1.10 & 0.79 & Satellite \\
 & 9.47 & 0.25 & 5.03 & 12.09 & 1.04 & 0.66 & Satellite \\
 & 9.56 & 0.36 & 5.03 & 12.04 & 0.94 & 0.43 & Satellite \\
 & 9.44 & 0.24 & 5.03 & 12.20 & 1.11 & 0.78 & Satellite \\
 & 9.66 & 0.34 & 5.03 & 12.15 & 0.00 & 0.00 & Satellite \\
 & 9.50 & 0.29 & 5.03 & 12.49 & 1.08 & 0.80 & Satellite \\
 & 9.38 & 0.13 & 5.03 & 11.95 & 1.01 & 0.62 & Satellite \\
 & 9.61 & 0.12 & 5.03 & 12.09 & 1.07 & 0.69 & Satellite \\
 & 9.47 & 0.04 & 5.03 & 12.39 & 1.22 & 0.85 & Satellite \\
 & 9.43 & 0.02 & 5.03 & 12.80 & 1.27 & 0.93 & Satellite \\
 & 9.55 & 0.06 & 5.03 & 12.19 & 1.14 & 0.81 & Satellite \\
 & 9.50 & 0.10 & 5.03 & 12.12 & 1.12 & 0.75 & Satellite \\
 & 9.33 & 0.14 & 5.03 & 12.00 & 1.15 & 0.84 & Satellite \\
 & 9.59 & 0.38 & 5.03 & 12.21 & 1.13 & 0.76 & Satellite \\
 & 9.41 & 0.06 & 5.03 & 12.60 & 1.24 & 0.89 & Satellite \\
 & 9.52 & 0.25 & 5.03 & 11.96 & 1.12 & 0.73 & Satellite \\
 & 9.36 & 0.11 & 5.03 & 12.09 & 1.16 & 0.81 & Satellite \\
 & 9.72 & 0.32 & 5.03 & 11.96 & 1.08 & 0.64 & Satellite \\
 & 9.83 & 0.10 & 5.03 & 12.19 & 1.16 & 0.83 & Satellite \\
 & 9.61 & 0.12 & 5.03 & 12.33 & 1.17 & 0.79 & Satellite \\
 & 9.48 & 0.21 & 5.03 & 12.03 & 1.07 & 0.80 & Satellite \\
 & 9.39 & 0.15 & 5.03 & 11.66 & 0.97 & 0.48 & Satellite \\
 & 9.39 & 0.21 & 5.03 & 12.88 & 1.05 & 0.69 & Satellite \\
 & 9.40 & 0.08 & 5.03 & 12.11 & 1.16 & 0.87 & Satellite \\
 & 9.35 & 0.07 & 5.03 & 11.83 & 1.11 & 0.73 & Satellite \\
 & 9.43 & 0.04 & 5.03 & 12.24 & 1.11 & 0.76 & Satellite \\
 & 9.79 & 0.38 & 5.46 & 12.38 & 1.06 & 0.64 & Satellite \\
 & 9.38 & 0.14 & 5.46 & 12.29 & 1.02 & 0.61 & Satellite \\
 & 9.41 & 0.13 & 5.46 & 12.19 & 1.10 & 0.76 & Satellite \\
 & 9.69 & 0.03 & 5.46 & 12.90 & 1.29 & 0.92 & Satellite \\
 & 9.52 & 0.25 & 5.46 & 12.51 & 1.20 & 0.87 & Satellite \\
 & 9.43 & 0.17 & 5.46 & 12.07 & 0.00 & 0.00 & Satellite \\
 & 9.44 & 0.04 & 5.46 & 12.18 & 1.11 & 0.75 & Satellite \\
 & 9.47 & 0.29 & 5.46 & 12.15 & 1.06 & 0.76 & Satellite \\
 & 9.37 & 0.07 & 5.46 & 12.12 & 1.03 & 0.60 & Satellite \\
 & 9.48 & 0.16 & 5.46 & 12.25 & 1.06 & 0.63 & Satellite \\
 & 9.88 & 0.50 & 5.46 & 12.43 & 1.12 & 0.71 & Satellite \\
 & 9.41 & 0.16 & 5.46 & 12.40 & 1.09 & 0.71 & Satellite \\
 & 9.35 & 0.05 & 5.46 & 12.25 & 1.05 & 0.66 & Satellite \\

\bottomrule
\end{tabular}
\end{table*}

\begin{table*}
\addtocounter{table}{-1}
\centering
\caption{Continued.}
\begin{tabular}{lccccccc}
\toprule
Observation/Simulation & $\log_{10}({M_{\star}}/{\rm M_{\odot}})$ & SFR [${\rm M_{\odot}}$/yr] & $z$ & $\log_{10}({M_{\rm 200c}}/{\rm M_{\odot}})$ & U-V & V-J & Type \\
\midrule
\textsc{L-Galaxies}   & 9.44 & 0.05 & 5.03 & 12.22 & 1.21 & 0.88 & Satellite \\
 & 9.48 & 0.24 & 5.46 & 12.35 & 1.10 & 0.78 & Satellite \\
        & 9.68 & 0.20 & 5.03 & 12.60 & 0.00 & 0.00 & Satellite \\
         
         & 9.59 & 0.38 & 5.03 & 12.32 & 1.17 & 0.87 & Satellite \\
         
     & 9.32 & 0.12 & 5.03 & 12.13 & 1.10 & 0.81 & Satellite \\
    
     & 9.36 & 0.07 & 5.03 & 11.98 & 1.16 & 0.80 & Satellite \\
    
     & 9.34 & 0.14 & 5.03 & 12.42 & 1.13 & 0.77 & Satellite \\
    
     & 9.35 & 0.17 & 5.03 & 12.50 & 1.12 & 0.83 & Satellite \\
    
     & 9.62 & 0.25 & 5.03 & 12.25 & 1.06 & 0.66 & Satellite \\
    
     & 9.78 & 0.16 & 5.03 & 12.44 & 1.20 & 0.85 & Satellite \\

             & 9.56 & 0.18 & 5.03 & 11.91 & 0.99 & 0.55 & Central \\
             & 9.39 & 0.10 & 5.03 & 11.66 & 1.02 & 0.73 & Central \\
             & 9.53 & 0.20 & 5.03 & 11.48 & 1.07 & 0.71 & Central \\
             & 9.31 & 0.20 & 5.03 & 11.31 & 1.00 & 0.58 & Central \\
             & 9.33 & 0.01 & 5.03 & 11.33 & 1.16 & 0.81 & Central \\
\midrule
\textsc{TNG-CLUSTER} & 9.58 & 0.00 & 5.0 & 12.52 & - & - &Satellite  \\
 & 9.59 &  0.30 & 5.0 & 12.94 & - & - &Satellite  \\
 & 9.6 & 0.00 & 5.0 & 12.52 & - & - &Satellite  \\
 & 9.62 &  0.00 & 5.0 & 12.78 & - & -  &Satellite  \\
 & 9.78 &  0.00 & 5.0 & 12.73 & - & - &Satellite  \\
 & 9.74 & 0.12  & 5.33 & 12.84 & - & - &Satellite  \\
\midrule
\textsc{Simba}  & 9.31 & 0.02 & 5.0 & 12.61 & 0.59 & 0.17 &Satellite  \\
 & 9.55 & 0.08 & 5.0 & 12.57 & 0.91 & 0.22 &Satellite \\
& 9.57 &  0.08 & 5.0 & 12.46 & 0.83 & 0.18 &Satellite  \\
\bottomrule
\end{tabular}
\end{table*}

\begin{figure*}

   \includegraphics[width=0.33\textwidth]{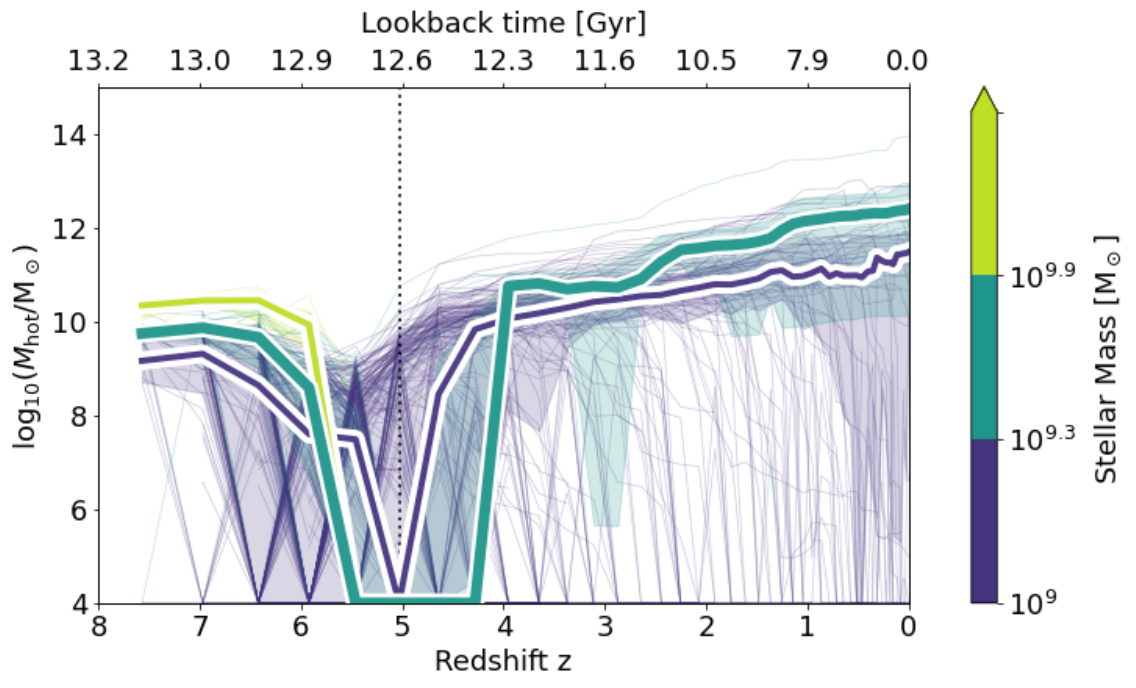}    \includegraphics[width=0.33\textwidth]{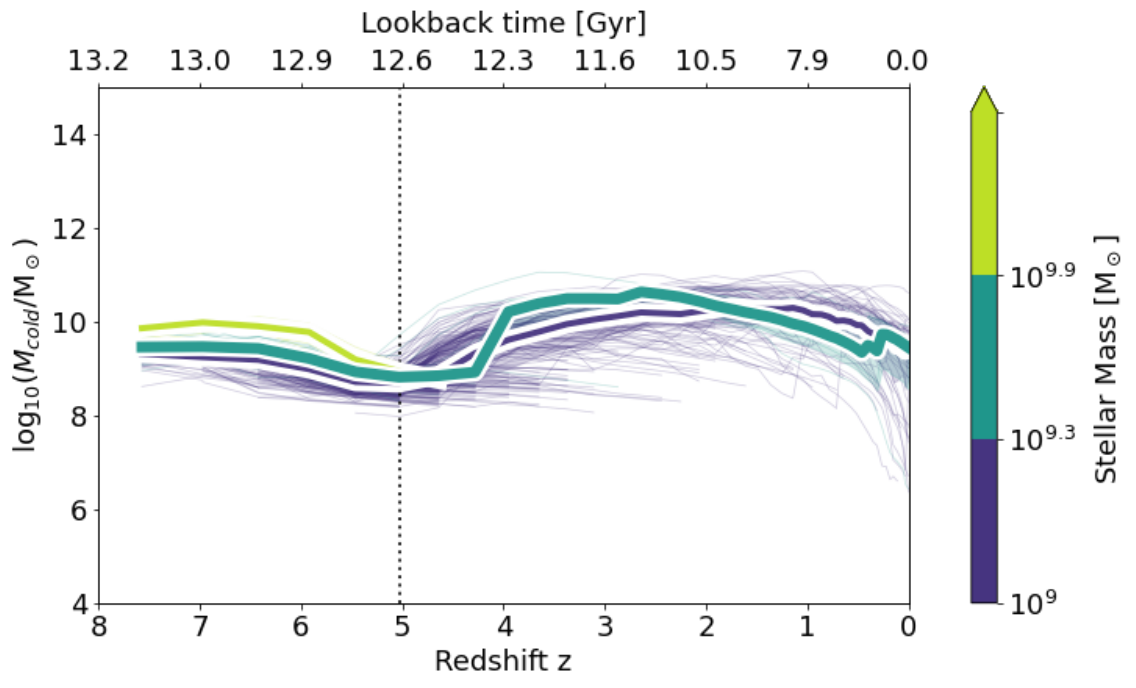} 
  \includegraphics[width=0.33\textwidth]{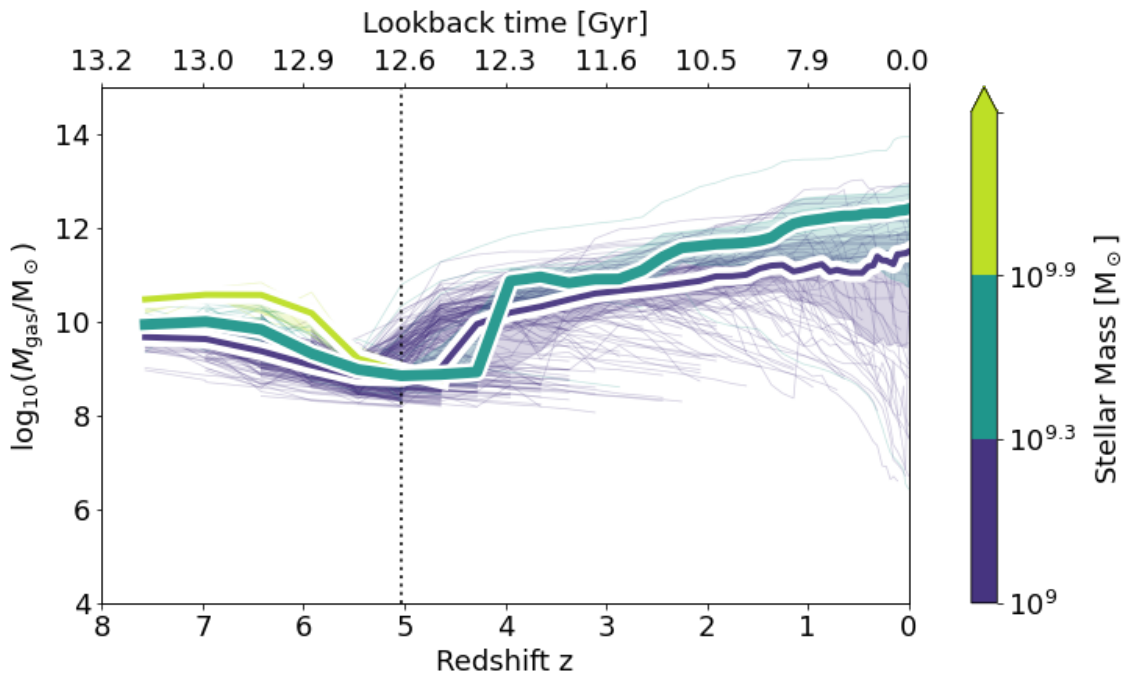}

     \includegraphics[width=0.48\textwidth]{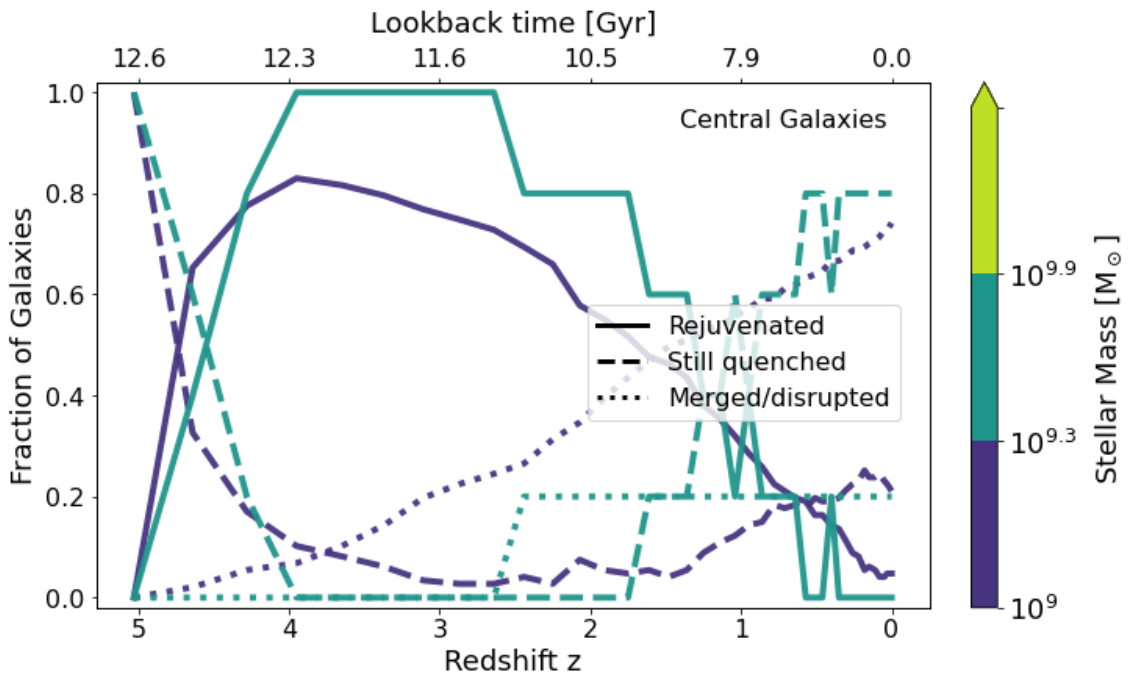}    \includegraphics[width=0.48\textwidth]{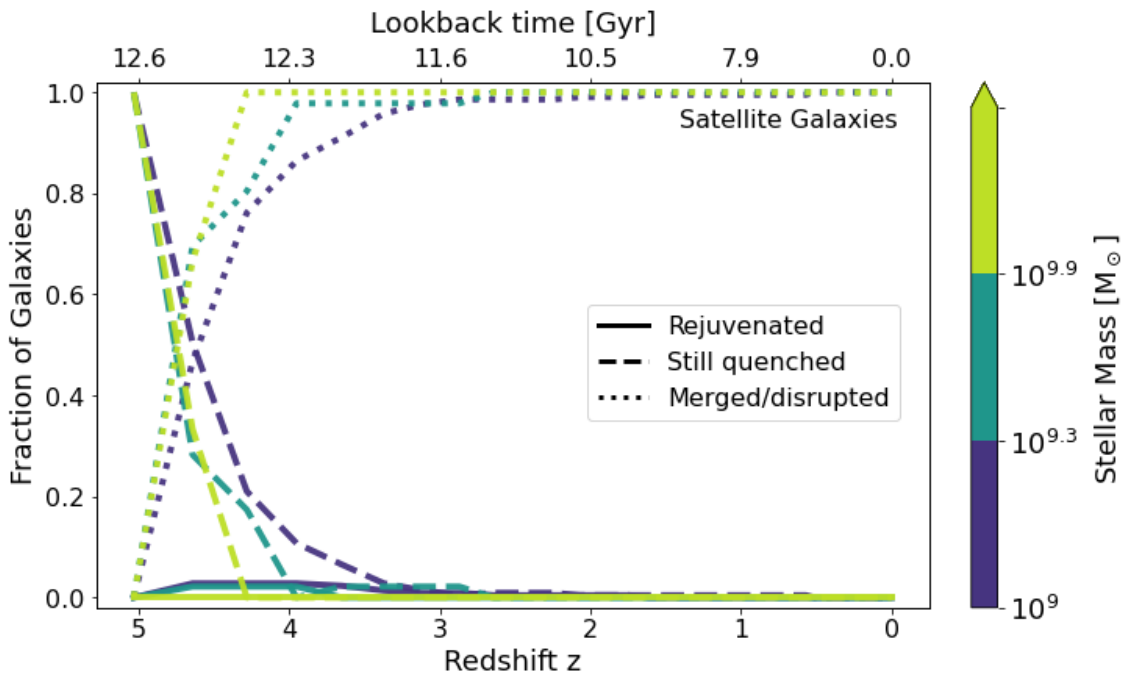} 
  \caption{\textit{Top}: Redshift evolution of galaxies quenched at $z=5$ in \textsc{L-Galaxies} in terms of hot (\textit{left}), cold (\textit{middle}), and total gas (\textit{right}). Colors indicate the stellar mass of each galaxy at $z=5$. The stellar mass range adopted in the observational analysis is highlighted with a thick green line. All zero hot gas masses are mapped to $10^4 \rm M_\odot$. \textit{Bottom}: Galaxy fraction as a function of redshift for three evolutionary classes for central (\textit{left}) and satellite (\textit{right}) galaxies. Different line styles indicate merged or disrupted systems (dotted), rejuvenated galaxies (solid), and still quenched galaxies (dashed), while colors denote different stellar mass bins.}
  \label{fig:gasfractionrejuvlgal5}
\end{figure*}

Here we provide additional analysis of the JWST analogs in \textsc{L-Galaxies} (see Fig.~\ref{fig:rejuv_plots}), focusing on systems that are quenched at $z=5$. The top panel of Fig.~\ref{fig:gasfractionrejuvlgal5} shows the evolution of the hot, cold, and total gas mass of galaxies as a function of redshift for both individual galaxy tracks and the median evolution in different stellar mass bins. At $z=5$, the galaxies were selected as quenched and were subsequently traced both backward and forward in time.

A similar trend is visible in all three panels: the gas content declines toward $z\sim5$, followed by a recovery at lower redshifts. This behavior is most pronounced for the hot gas component, where almost the entire hot gas reservoir is removed prior to quenching at $z\sim5$, followed by a subsequent rise. This indicates that the hot gas reservoir is replenished shortly after quenching. The cold gas component also decreases before quenching and increases after $z\sim5$, although this trend is much more gradual. At later times, the cold gas mass declines again toward lower redshifts. As a result, the total gas mass follows the combined evolution of the hot and cold gas phases, decreasing before quenching and increasing again after $z\sim5$. Clear differences between stellar-mass bins emerge only at later times, around $z\lesssim4$, with more massive galaxies showing higher median total gas masses.

It is important to note once more that the median curves shown here do not necessarily represent the full population of galaxies selected as quenched at $z=5$ at all later times. Rather, they trace only those galaxies that remain continuously identifiable in the merger tree at each snapshot. This explains, for example, why no median curve is present for the highest stellar mass bin at later times after $z=5$: the number of surviving galaxies in this category becomes too small, primarily because many have already merged or are no longer tracked as distinct systems. The figure therefore illustrates the evolution of the surviving subset of the original quenched population, rather than that of the complete $z=5$ sample at all redshifts.

The bottom panel of Fig.~\ref{fig:gasfractionrejuvlgal5} shows the fractions of galaxies in different evolutionary states, separately for centrals and satellites. Although some satellites do rejuvenate, this channel remains uncommon: fewer than $10\%$ exhibit renewed star formation even 200 Myr after their initial quenching. By contrast, the remaining quenched galaxies that are centrals are considerably more likely to rejuvenate than their satellite counterparts. Notably, the small number of galaxies that survive to the present day are exclusively centrals, as all quenched satellite galaxies have either merged or undergone disruption by $z=3$.

\begin{figure*}
    \centering  
    \includegraphics[width=0.48\textwidth]{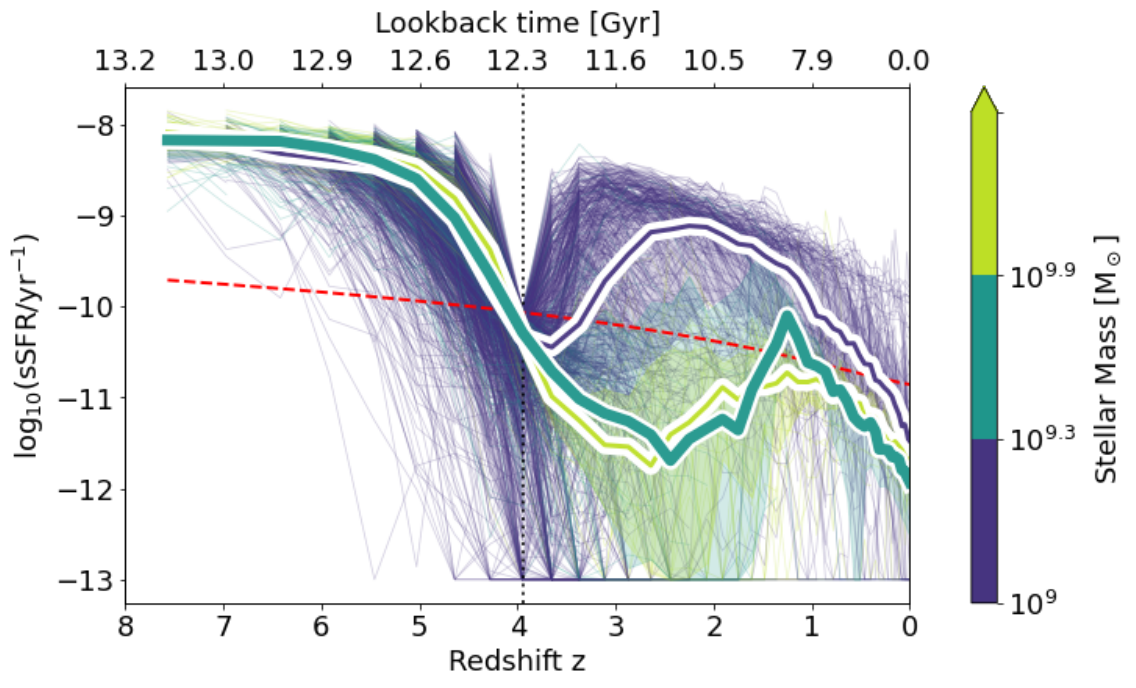}
    \includegraphics[width=0.48\textwidth]{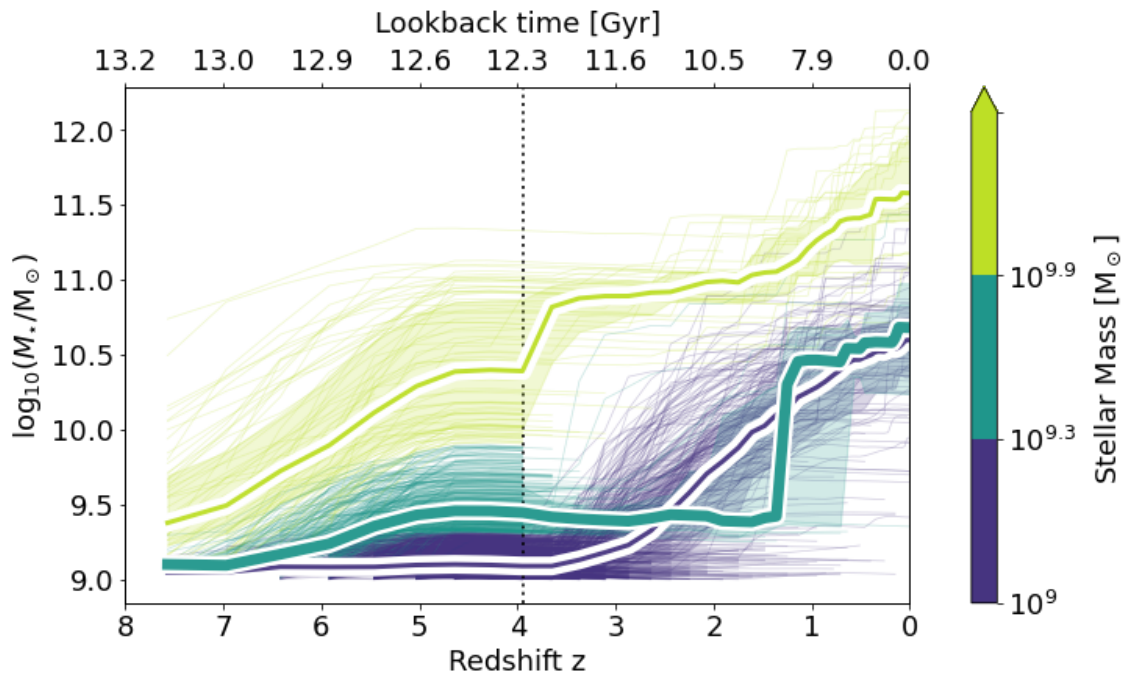}

     \includegraphics[width=0.33\textwidth]{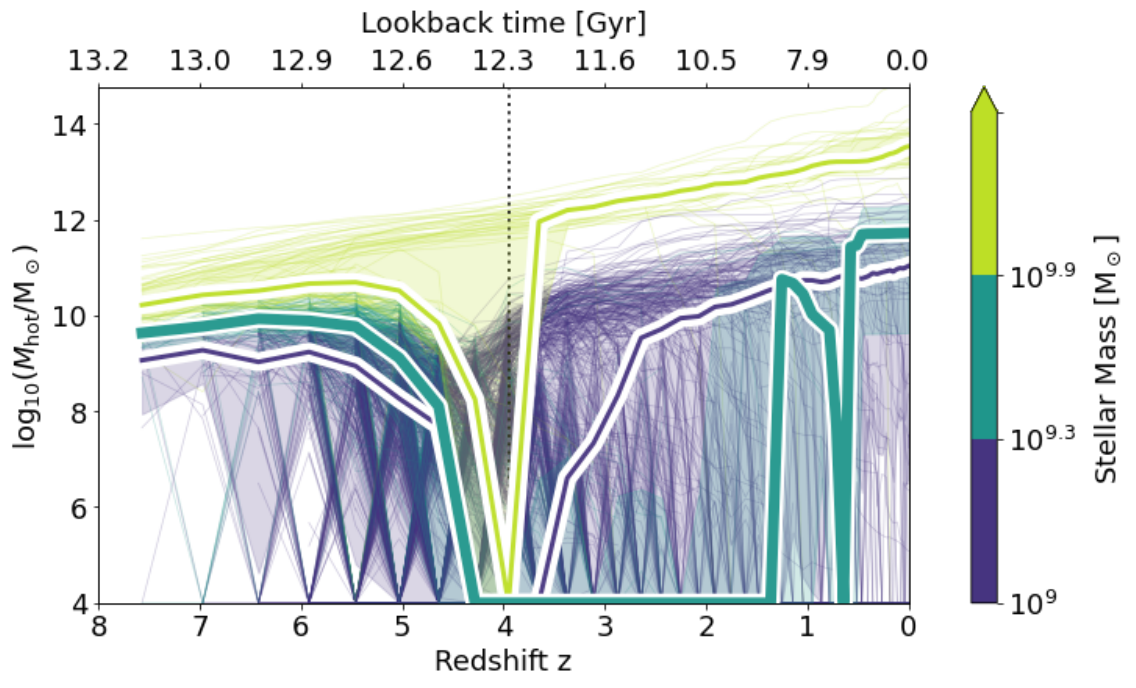}    \includegraphics[width=0.33\textwidth]{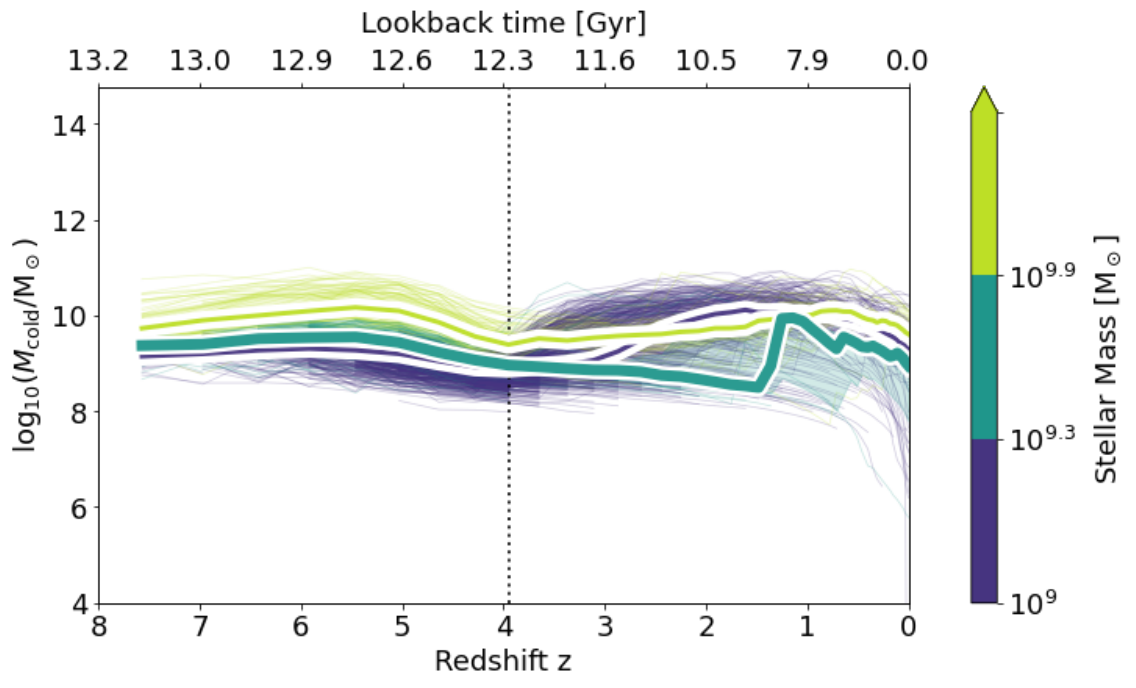} 
    \includegraphics[width=0.33\textwidth]{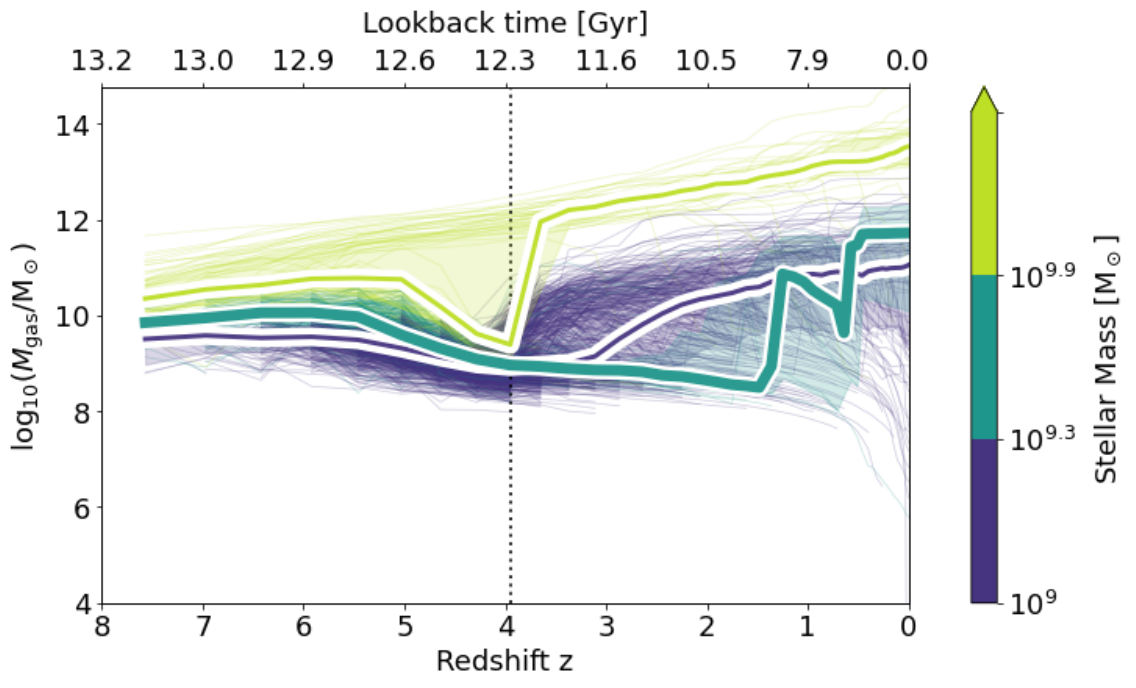}

    \includegraphics[width=0.33\textwidth]{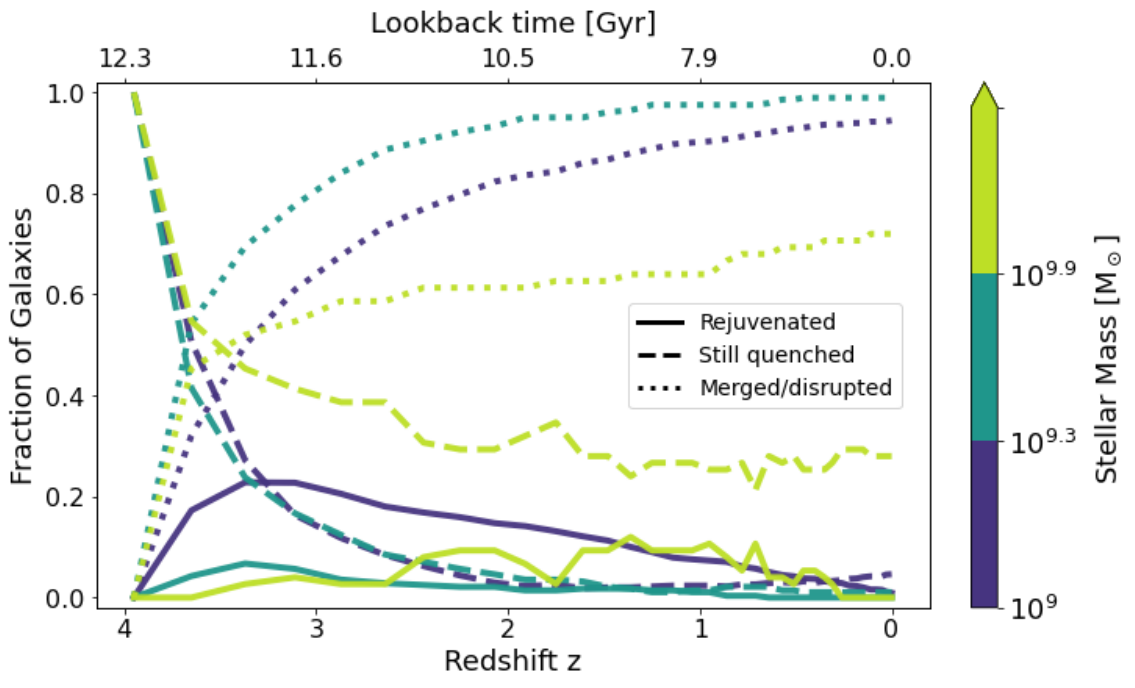}
    \includegraphics[width=0.33\textwidth]{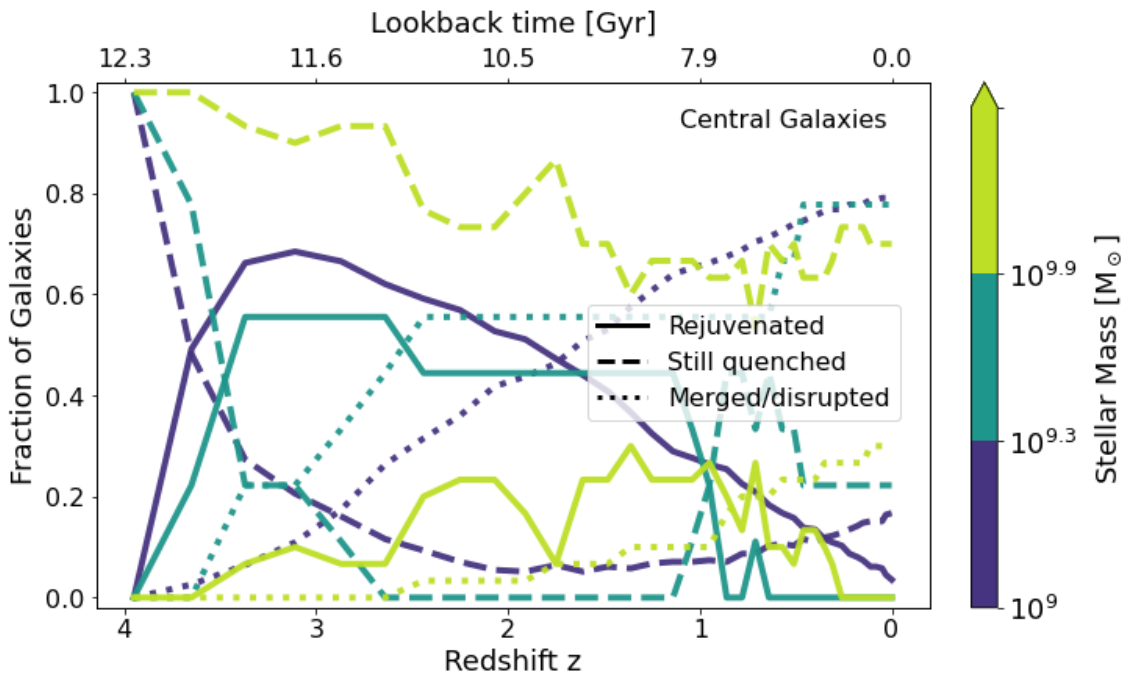}
    \includegraphics[width=0.33\textwidth]{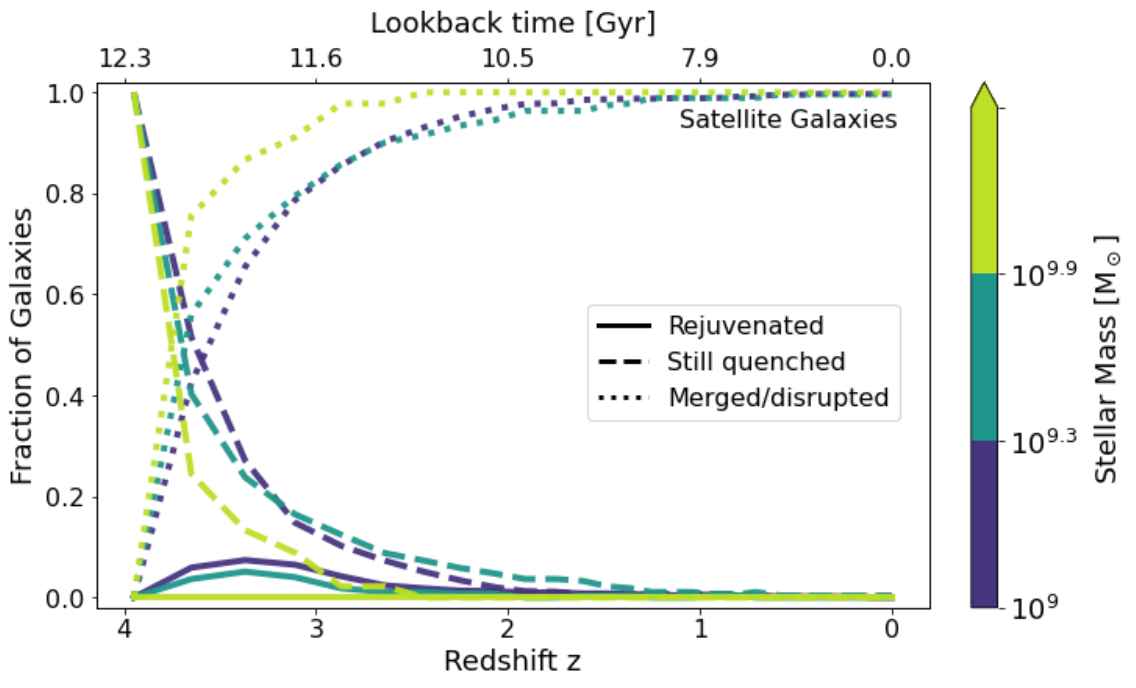}

    \caption{\textit{Top}: Redshift evolution of survivor galaxies quenched at $z=4$ in \textsc{L-Galaxies}, shown in terms of sSFR (\textit{left}) and stellar mass (\textit{right}). Colors indicate the stellar mass at $z=4$; the intermediate mass bin, which most closely matches the observed systems, is highlighted by the thick green line. sSFR values below $10^{-13}\mathrm{yr}^{-1}$ are set to this floor value. The quenching threshold, defined as $0.2/t_{\mathrm{H}}(z)$, is shown by the dashed red line. \textit{Middle}: Redshift evolution of galaxies quenched at $z=4$ in \textsc{L-Galaxies}, shown in terms of hot gas mass (\textit{left}), cold gas mass (\textit{middle}), and total gas mass (\textit{right}). All hot gas masses less than $10^4 \rm M_\odot$ are mapped to this floor value. \textit{Bottom}: Evolution of the fate of all (\textit{left}), central (\textit{middle}), and satellite (\textit{right}) galaxies quenched at $z=4$. Solid lines indicate the rejuvenated fraction, defined at each redshift as the fraction of the initially quenched population that is still present and star-forming at that time. Dashed lines show the fraction that remains quenched. Dotted lines represent the merged or disrupted fraction, defined as the fraction of galaxies that have disappeared from the merger tree by a given redshift. All fractions are normalized by the corresponding initially quenched population at $z=4$. }
    \label{fig:rejuv_plots4}
\end{figure*}

To complement the analysis of the $z=5$ quenched sample, we show the same trends for galaxies quenched at $z=4$ in Fig.~\ref{fig:rejuv_plots4}. The top panel presents the redshift evolution of sSFR and stellar mass for galaxies that have not merged. As in the $z=5$ case, the median lines shown here trace only the survivor galaxies that remain continuously identifiable within the merger tree at each successive snapshot. Unlike the $z=5$ sample, however, the most massive stellar mass bin ($\log_{10}({M_{\star}}/{\rm M_{\odot}}) > 9.9$) also retains a visible evolutionary track and does not disappear at later times. In the sSFR evolution, the high- and intermediate-stellar-mass bins show an overall decline in sSFR toward lower redshifts. By contrast, the low-stellar-mass bin exhibits a short-lived increase in median sSFR shortly after $z=4$, indicating rejuvenation episodes similar to those seen in the $z=5$ sample. This temporary rise is followed by a gradual decline toward lower redshifts, with many systems eventually returning to strongly passive states. This behavior suggests that low-mass galaxies can temporarily rejuvenate but may later be quenched again or merge into other, often more massive, systems.

The right panel provides additional insight into this behavior. After quenching at $z=4$, galaxies in the low- and high-stellar-mass bins continue to increase in stellar mass. This evolution likely reflects the combined effects of rejuvenated star formation, mergers, and merger-driven accretion. The intermediate-mass bin, however, follows a distinct pattern: its median stellar mass remains approximately flat until $z\sim1$, followed by a more rapid increase and then a return to a slower, plateau-like growth at later times.

The middle panel of Fig.~\ref{fig:rejuv_plots4} shows the evolution of the hot, cold, and total gas mass of galaxies as a function of redshift. For the hot gas component, all stellar mass bins show a decline toward $z=4$, with most galaxies losing almost all of their hot gas. The high-stellar-mass bin rapidly recovers its hot gas reservoir shortly after quenching, even reaching higher values than before, which suggests an earlier onset of hot gas replenishment. The low-stellar-mass bin also shows a recovery of hot gas, but this occurs more gradually. The most interesting behavior is seen in the intermediate-stellar-mass bin: it remains almost completely depleted of hot gas until $z\sim1.1$, after which the hot gas mass increases rapidly over a short period of time. This is followed by another sudden loss of hot gas and a subsequent rapid recovery.

The cold gas component evolves more smoothly up to the quenching epoch, with all stellar mass bins showing a gradual decline toward $z=4$. After quenching, however, the evolution differs between the bins. The highest stellar mass bin shows no strong feature and instead displays only a very mild increase toward lower redshifts. By contrast, the low-mass bin exhibits a small bump around $z\sim2$, after which the decline starts. The intermediate-mass bin shows a more pronounced peak around $z\sim1.1$; however, before reaching this peak, its cold gas mass continues to decline gradually after quenching, unlike the other stellar mass bins, which show a mild increase. This feature appears to coincide with the recovery of the hot gas reservoir in the same stellar mass bin.

The total gas mass reflects the combined contribution of these two phases. A mild decrease is visible between $z\sim5$ and $z=4$, after which the subsequent evolution differs by stellar mass bin. In the highest stellar mass bin, the total gas mass increases abruptly and then continues to grow more slowly toward lower redshifts. The low-mass bin shows a more modest and smoother increase, whereas the intermediate-mass bin initially declines, then undergoes a more sudden rise, followed by a more irregular evolution with alternating periods of decrease and increase.

The bottom panel of Fig.~\ref{fig:rejuv_plots4} shows the fractions of galaxies in different evolutionary states, separately for all, central, and satellite galaxies. While the satellite population follows broadly similar trends to the $z=5$ case, merging and disruption remain the dominant outcome. The central population, and consequently the full sample, shows a more differentiated evolution.

In particular, the rejuvenated fraction of centrals is lower than in the $z=5$ sample. For the intermediate stellar mass bin, it declines from nearly 1.0 at $z=3$ to approximately 0.55 at lower redshifts, while in the low-mass bin it decreases from about 0.8 to roughly 0.6. The high-stellar-mass bin, which remains visible in the $z=4$ sample, exhibits only a weak rejuvenation signal, with the rejuvenated fraction reaching at most $\sim0.3$. Instead, high-mass centrals show a clearly persistent still-quenched population: more than half remain passive once quenched. The same pattern is visible in the all-galaxy panel, where the still-quenched fraction in the highest mass bin remains nonzero and stays near $\sim0.3$ even at $z=0$. In contrast, the low- and intermediate-mass bins show still-quenched fractions that fall to zero, implying that all intermediate-mass systems and the large majority of low-mass systems ($>90\%$) have merged, been disrupted, or otherwise left the quenched population by the present epoch.

\section{Galaxy stellar mass function}
\label{sec:StellarMassFunctionofall}

\begin{figure}
    \centering
    \includegraphics[width=\columnwidth]{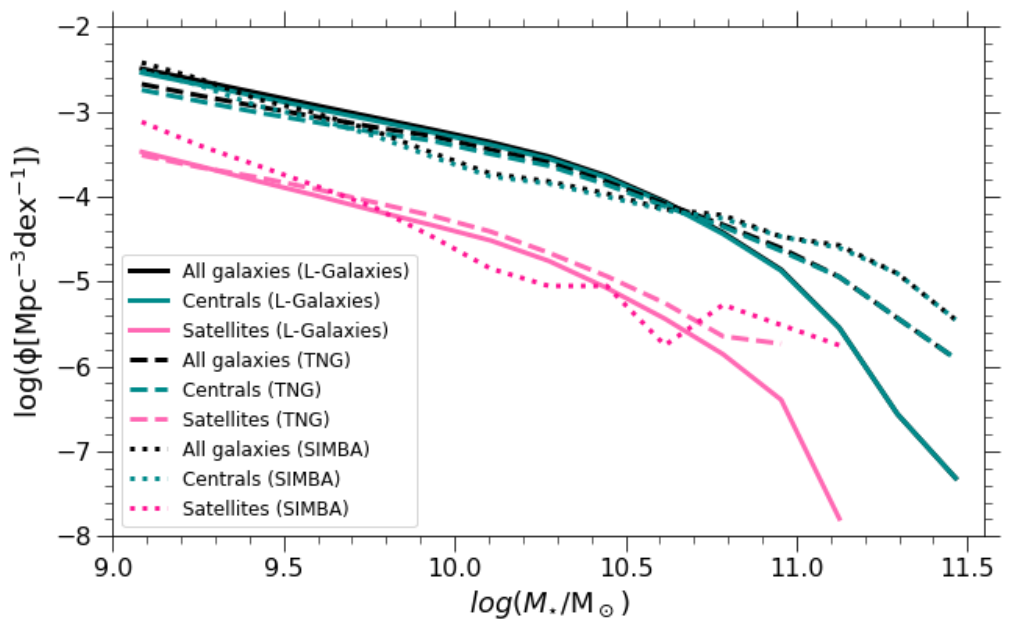} 
    \caption{SMFs of all galaxies (black), central galaxies (green), and satellite galaxies (pink) at redshift $z=4$, based on predictions from the \textsc{L-Galaxies} (solid line), \textsc{TNG} (dashed line), and \textsc{Simba} (dotted line) simulations. The x-axis shows the logarithm of stellar mass ($\log_{10}({M_{\star}}/{\rm M_{\odot}})$), and the y-axis displays the logarithmic number density of galaxies per dex in stellar mass per cubic megaparsec. This figure highlights the distinct contributions of central and satellite galaxies to the total SMF.}
    \label{fig:StellarMassFunctionofall}
\end{figure}

As a complementary analysis to Fig.~\ref{fig:stellar_mass_function_quenched_vs_starforming}, Fig.~\ref{fig:StellarMassFunctionofall} shows the total SMF derived from the \textsc{L-Galaxies}, \textsc{TNG}, and \textsc{Simba} at redshift $z=4$, with separate curves for all galaxies, centrals, and satellites. All simulations show the same trend. The SMF exhibits a steep decline at high stellar masses, as predicted by models and observed in high-redshift surveys, reflecting the rarity of high-mass galaxies in the early Universe \citep{vani2025probing}.  Central galaxies dominate the galaxy population across the full stellar mass range, particularly at the high-mass end (\( \log_{10}(M_{\star}/{\rm M_{\odot}}) \gtrsim 10.5 \)), where the satellite galaxies constitute only $\sim 3.2 \%  $ of the total galaxy population. At the low-mass end, satellites make up a growing fraction ($\sim 9.3 \%  $)  of the total galaxy population, even though centrals still dominate in absolute number.

The overall abundance of satellites is primarily driven by hierarchical structure formation, as galaxies fall into larger halos and become satellites. However, once they become satellites, environmental processes — such as ram-pressure stripping and strangulation — suppress their ability to accrete gas, effectively quenching or significantly reducing their star formation activity and thus limiting further stellar mass growth. In contrast, central galaxies, which are less affected by such environmental factors, continue to grow through sustained star formation. Consequently, satellites tend to remain at lower stellar masses, leading to a higher satellite fraction in the low-mass regime, even though their total number remains smaller than that of central galaxies \citep{cora2018semi}.

 Notably, the SMF does not exhibit a pronounced ``knee'' feature at this redshift for all of the simulations, typically observed around \( \log_{10}(M_{\star}/{\rm M_{\odot}}) \sim 10.5 \) at lower redshifts \citep{Moster2010Constraints, guo2010how}. This absence may indicate that the characteristic transition mass has not yet developed, consistent with the picture of a less evolved galaxy population at early times. Overall, the shape of the SMF at  $z=4$ reflects an early stage of galaxy assembly, where the galaxy population is still dominated by low-mass systems and high-mass galaxies are only beginning to emerge \citep{vani2025probing}.

\section{Spatial distribution of satellites}
\label{sec:Spatialdistributionofsatellites}

\begin{figure}

    \centering
\includegraphics[width=0.8\columnwidth]{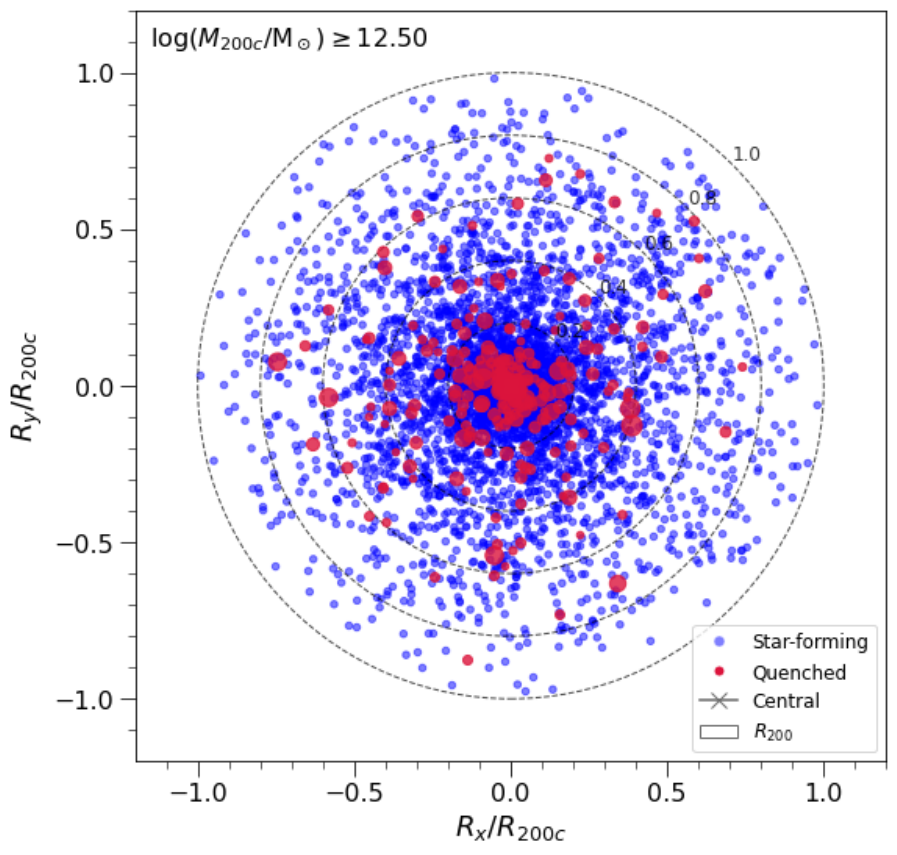}
\includegraphics[width=0.8\columnwidth]{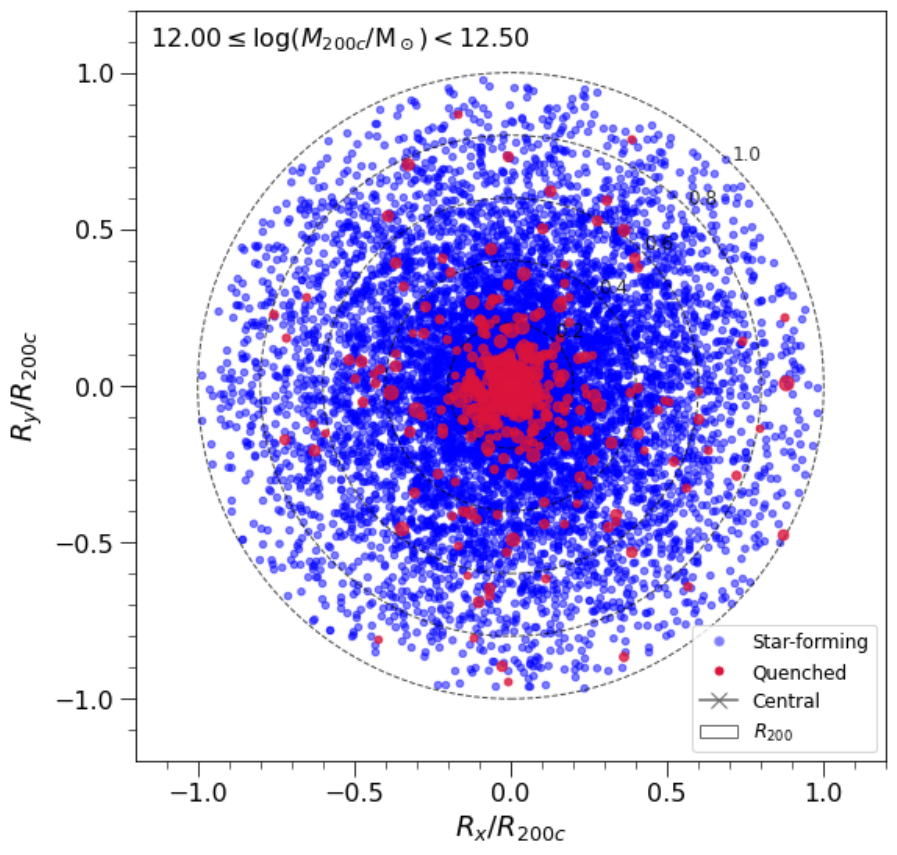}
\includegraphics[width=0.8\columnwidth]{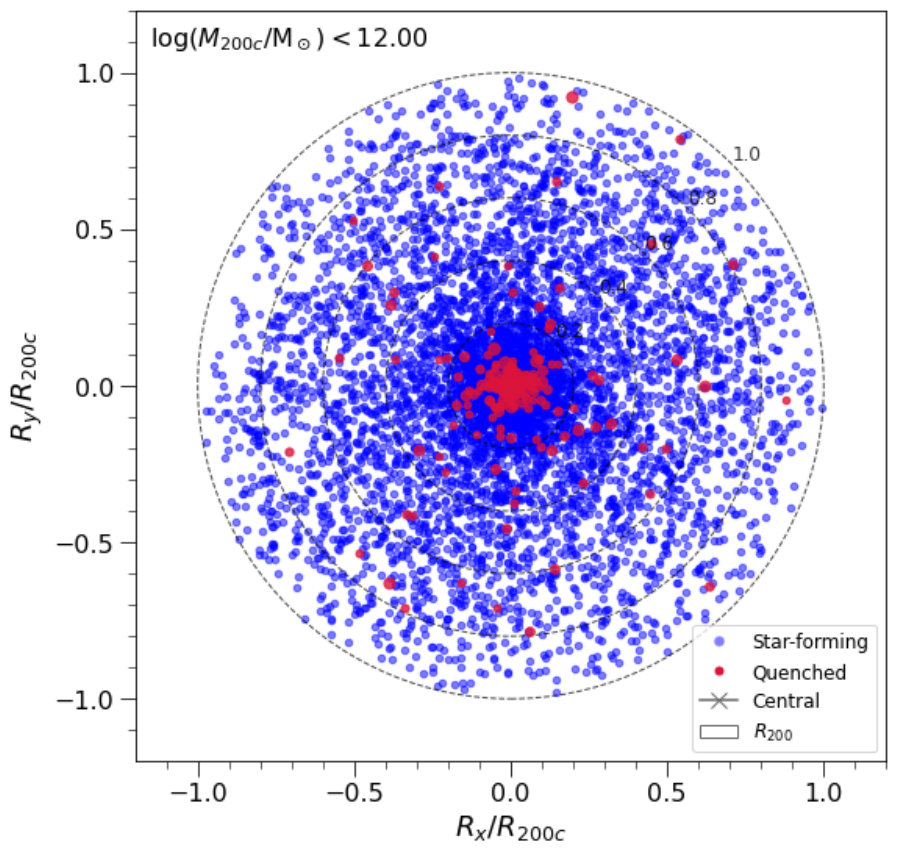}

\caption{Projected spatial distribution of satellite galaxies around central galaxies in \textsc{L-Galaxies} plotted in units of $R_{200c}$ and separated into three halo mass bins:
$\log_{10}(M_{\rm 200c}/{\rm M_{\odot}}) \geq 12.5$ (\textit{top}),
$12 \leq \log_{10}({M_{\rm 200c}}/{\rm M_{\odot}}) < 12.5$ (\textit{middle}),
and $\log_{10}({M_{\rm 200c}}/{\rm M_{\odot}}) < 12$ (\textit{bottom}). Satellite positions are normalized relative to their central galaxy and projected onto the $x$–$y$ plane. Satellites are classified as star-forming (blue) or quenched (red) using an sSFR threshold. The gray circles indicate the extent of $R_{\rm 200c}$. Marker sizes for quenched satellites are scaled according to their stellar masses within each halo mass bin, meaning that the largest markers in different bins do not necessarily correspond to the same absolute stellar mass. Star-forming satellites are displayed with uniform marker sizes for visualization clarity.}
\label{fig:satellite_distribution_grid}
\end{figure}

To further explore the environmental dependence of satellite quenching discussed in Sect.~\ref{sec:environmentaldependenceofsatellitegalaxyquenching}, we examined the projected spatial distribution of satellite galaxies around their central galaxies at $z=4$ in \textsc{L-Galaxies}.

Figure~\ref{fig:satellite_distribution_grid} shows the projected spatial distribution of satellite galaxies around central galaxies at redshift $z=4$ in \textsc{L-Galaxies}. Distances are normalized by the host halo virial radius, $R_{200c}$. Satellites are projected onto the $x$--$y$ plane. A line-of-sight depth cut is applied along the $z$-coordinate, restricting the projection to $\pm R_{200c}$ around each central galaxy.

Across all halo mass bins, star-forming satellites dominate in number and are distributed throughout the halo volume. In contrast, quenched satellites are less numerous, and their spatial and stellar-mass distributions show a dependence on host halo mass.

In the lowest halo mass bin, the quenched satellite population in \textsc{L-Galaxies} is relatively small and is strongly concentrated toward the inner halo, with most quenched satellites located within $\sim 0.2 R_{\rm 200c}$ of the central galaxy. This central concentration remains a persistent feature across the full halo mass range. As host halo mass increases, the number and fraction of quenched satellites also increase. In addition, the larger marker sizes indicate that quenched satellites in more massive halos tend to have higher stellar masses.

These trends indicate a clear halo-mass dependence of satellite quenching. Although quenched satellites remain subdominant relative to the star-forming population, they become more abundant, more massive, and more centrally concentrated in higher-mass halos. This behavior is consistent with a scenario in which environmental quenching mechanisms, such as ram-pressure stripping and strangulation, become increasingly efficient in denser and more massive halos at high redshifts \citep{bahe2015star}.

\section{Additional quenching trends}
\label{sec:Appendixqunechingtrends}

\begin{table*}
    \centering
    \caption{Comparison of the quenched galaxy statistics of the different simulation models.}
    \resizebox{\textwidth}{!}{%
    \begin{tabular}{lccccc}
        \hline
        Model & Galaxy type & Total population & Quenched population & Quenched fraction & Number of quenched galaxies \\
        \hline
        \textsc{L-Galaxies} & Central (607,808)  & 90.8\%   & 23.1\%  & 0.07\%  &  417 out of 607,808 \\
        \textsc{L-Galaxies} & Satellite (61,235)& 9.2\%    & 76.9\%  & 2.27\%  & 1391 out of 61,235 \\
        \textsc{TNG}        & Central (32,584)   & 86.5\% & 28.5\%  & 0.03\%  & 10 out of 32,584 \\
        \textsc{TNG}        & Satellite (5,070)  & 13.5\% & 71.5\%  & 0.5\%   & 25 out of 5,070 \\
        \textsc{Simba}       & Central (4,771)   & 83.6\% & 13.64\%  & 0.06\%  & 3 out of 4,771 \\
        \textsc{Simba}        & Satellite (939)  & 16.4\% & 86.36\%  & 2.02\%   & 19 out of 939 \\
        \hline
    \end{tabular}%
    }
    \label{tab:quenched_comparison}
\end{table*}

Table~\ref{tab:quenched_comparison} summarizes the galaxy statistics in \textsc{L-Galaxies}, \textsc{TNG}, and \textsc{Simba}. For each model, the total galaxy population is split into centrals and satellites, and the corresponding quenched fractions and absolute numbers of quenched systems are reported. Here, the quenched population refers to the distribution of quenched galaxies across centrals and satellites, while the quenched fraction quantifies the fraction of galaxies within a given population that are quenched. In all three simulations, quenched galaxies constitute only a small fraction of the total population, with quenched fractions remaining below a few percent. Satellites consistently dominate the quenched population despite representing a smaller fraction of the total sample, reflecting the strong association between quenching and satellite status. The low absolute numbers of quenched galaxies, particularly in \textsc{TNG} and \textsc{Simba}, highlight the role of limited number statistics and motivate the cautious interpretation of radial quenched-fraction profiles discussed in the main text.

\begin{figure*}
  \centering
  \includegraphics[width=0.48\textwidth]{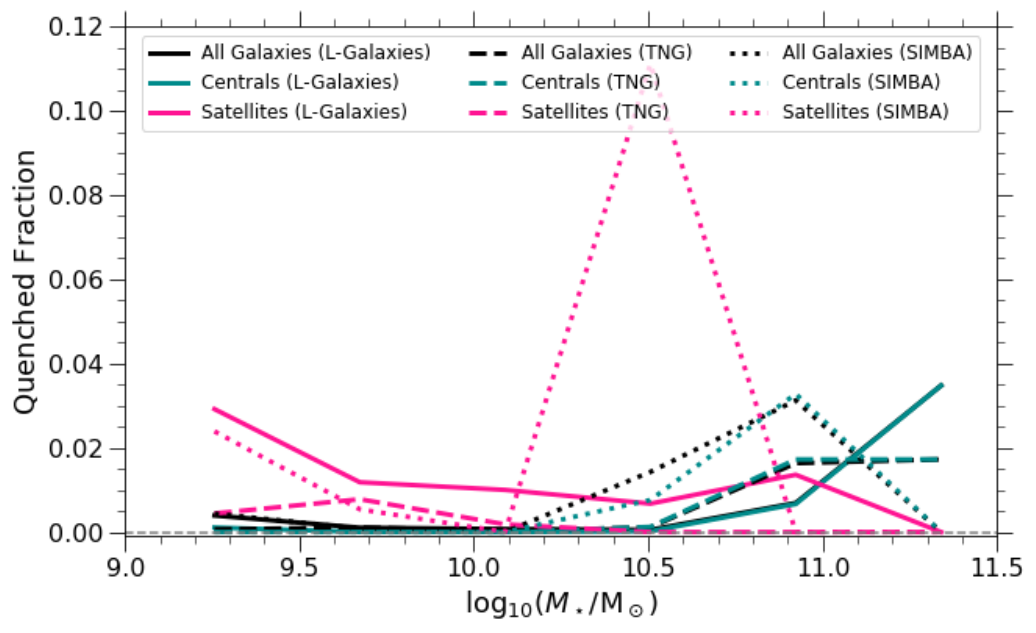} 
    \includegraphics[width=0.48\textwidth]{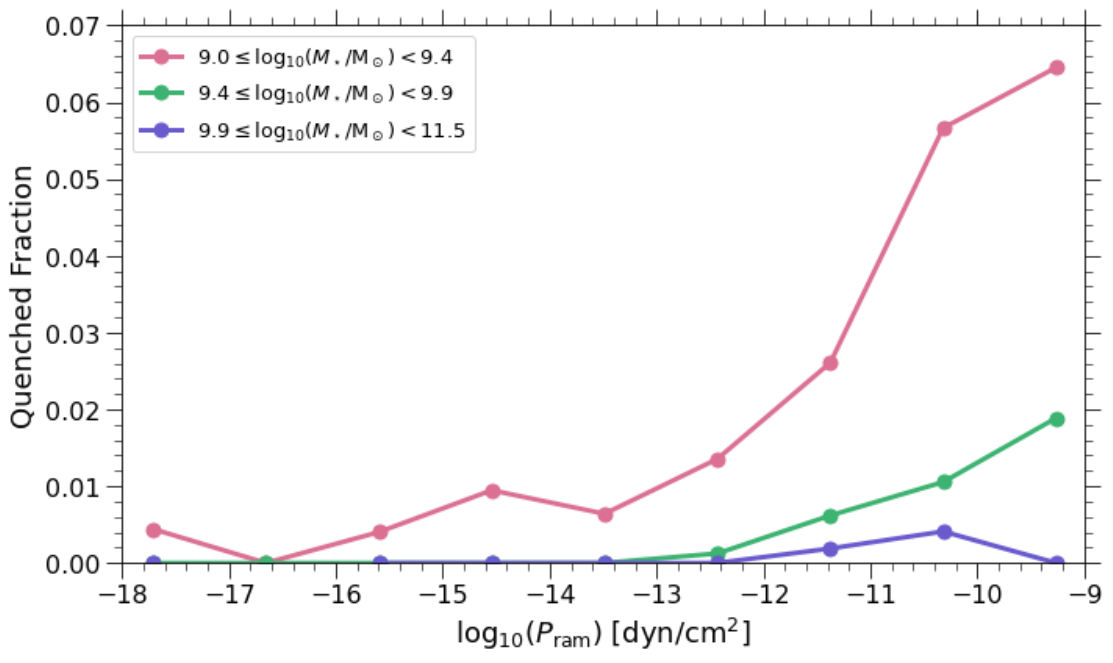}

        \includegraphics[width=0.33\textwidth]{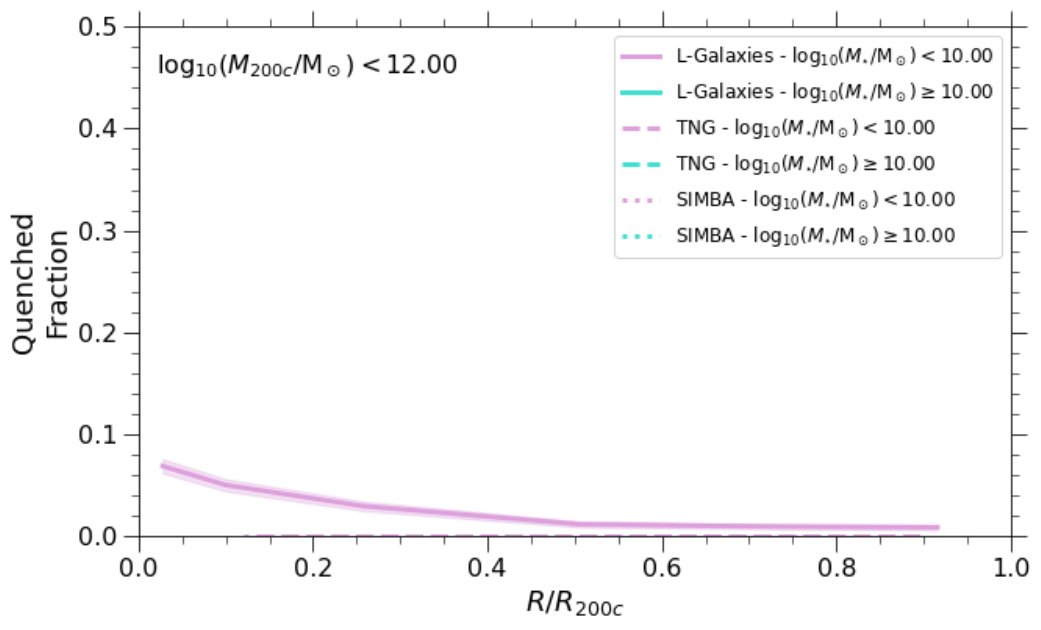}
    \includegraphics[width=0.33\textwidth]{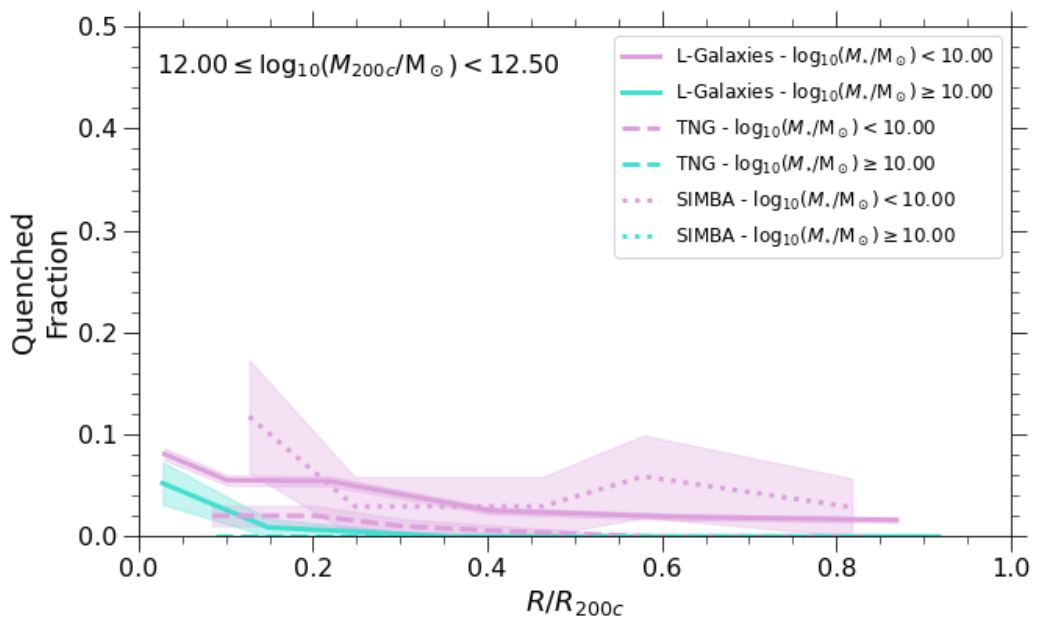}
    \includegraphics[width=0.33\textwidth]{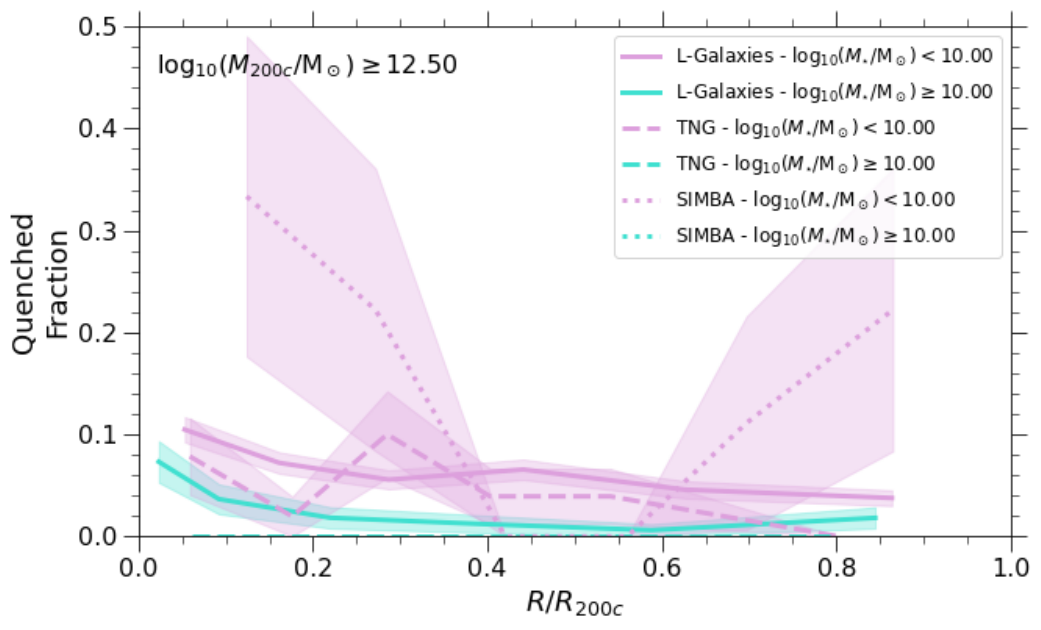}
  \caption{\textit{Top left}: Quenched fraction of galaxies as a function of stellar mass at $z=4$ computed from the \textsc{L-Galaxies} (solid line), \textsc{TNG} (dashed line), and \textsc{Simba} (dotted line) simulations. Galaxies are classified as quenched if their $\mathrm{sSFR} < 0.2/t_{\mathrm{H}}(z)$. The quenched fraction of all galaxies (black) is shown alongside the separate contributions from central galaxies (dark cyan) and satellite galaxies (pink). Fixed-width binning is used. \textit{Top right}: Quenched fraction of satellite galaxies as a function of ram pressure for different stellar mass bins. All ram-pressure values below $10^{-18}\mathrm{dyn\,cm^{-2}}$ are set to this value. \textit{Bottom}: Radial profiles of the quenched satellite galaxy fraction as a function of normalized halocentric distance ($R/R_{\rm 200c}$) for three simulations: \textsc{L-Galaxies} (solid lines),\textsc{TNG} (dashed lines), and \textsc{Simba} (dotted lines). Colored curves correspond to stellar mass subsamples: $\log_{10}({M_{\star}}/{\rm M_{\odot}}) < 10$ (purple) and $\log_{10}({M_{\star}}/{\rm M_{\odot}}) \geq 10$ (blue). Adaptive radial bins were used separately for each simulation and mass-bin combination to improve statistics in bins with few quenched satellites, especially for \textsc{Simba}.} 
\label{fig:QuenchedGalaxyFractionForCentralsandSatellites}
\end{figure*}

The top-left panel of Fig.~\ref{fig:QuenchedGalaxyFractionForCentralsandSatellites} shows the same relation as the bottom-left panel of Fig.~\ref{fig:quenched_fraction_plots} but using an alternative binning scheme, which shows that the detailed mass dependence of the quenched fraction is sensitive to the chosen binning strategy. In particular, the sharp feature observed for satellite galaxies in \textsc{Simba} around 
\( \log_{10}({M_{\star}}/{\rm M_{\odot}}) \approx 10.5 \) with fixed-width binning becomes less pronounced when using adaptive binning. This indicates that part of the apparent rise is driven by low-number statistics in sparsely populated high-mass bins. Adaptive binning, which ensures a more uniform number of galaxies per bin, smooths these fluctuations and reveals a more gradual, model-dependent trend. Nevertheless, the qualitative differences between the simulations persist under both binning schemes: central galaxy quenching remains confined to the high-mass regime, while satellite quenching is already present at low stellar masses, with \textsc{Simba} consistently showing a distinct behavior compared to \textsc{TNG} and \textsc{L-Galaxies}. This demonstrates that while binning affects the quantitative shape of the curves, the physical interpretation of early quenching and its model dependence is robust.

The top-right panel of Fig.~\ref{fig:QuenchedGalaxyFractionForCentralsandSatellites} shows the quenched fraction of satellite galaxies as a function of ram pressure for different stellar mass bins. At fixed ram pressure, low-mass satellites exhibit higher quenched fractions than their more massive counterparts, likely because their shallower gravitational potential wells make them more vulnerable to environmental gas removal. As a result, external processes such as ram-pressure stripping can more easily deplete their gas reservoirs and suppress star formation.

The bottom panel of Fig.~\ref{fig:QuenchedGalaxyFractionForCentralsandSatellites} shows the quenched fraction as a function of halocentric distance for all three cosmological simulations, including \textsc{Simba}, which was excluded from the main-text analysis because the number of satellite galaxies is too small for reliable statistical trends. As a result, the increased quenched fraction seen in the highest halo mass bin, which approaches 40\%, is unlikely to be statistically robust and should not be over-interpreted.

\section{Ram pressure trends}
\label{sec:additionalrampressuretrends}
\begin{figure*}
    \centering
    \includegraphics[width=0.48\textwidth]{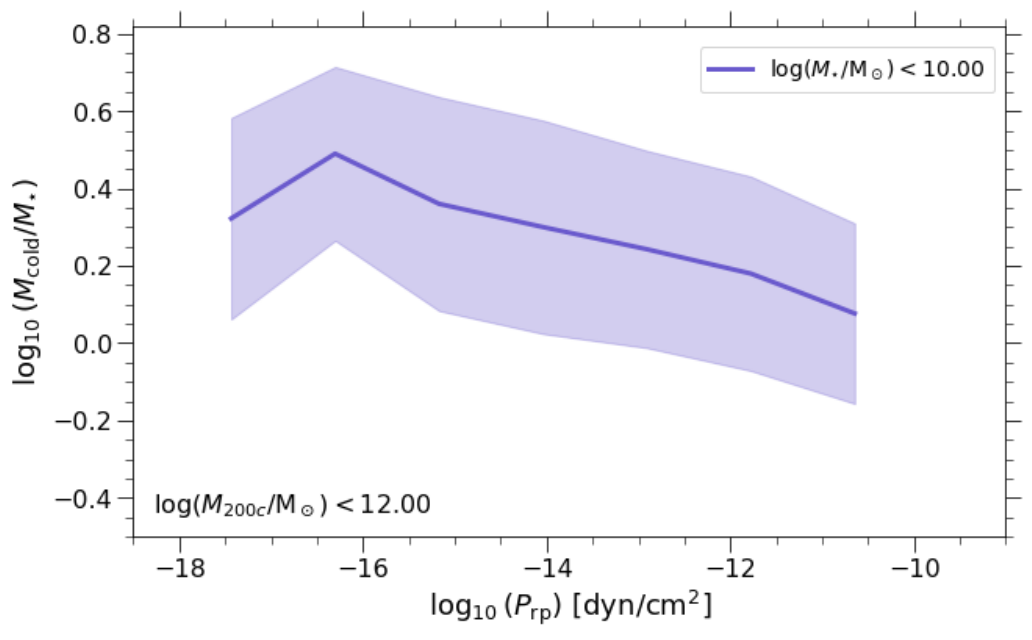}
    \includegraphics[width=0.48\textwidth]{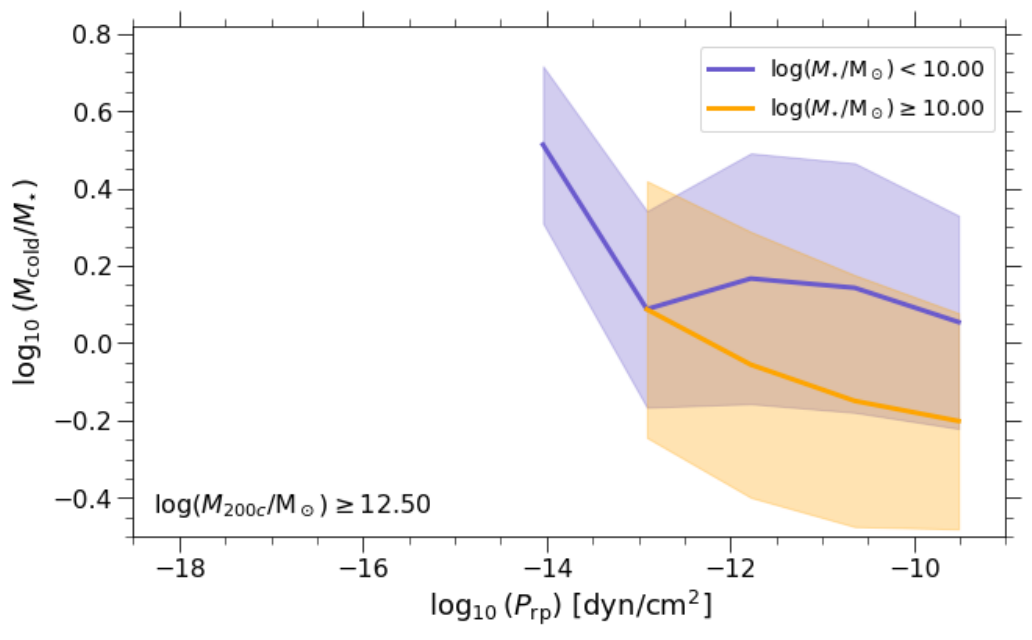}

    \includegraphics[width=0.48\textwidth]{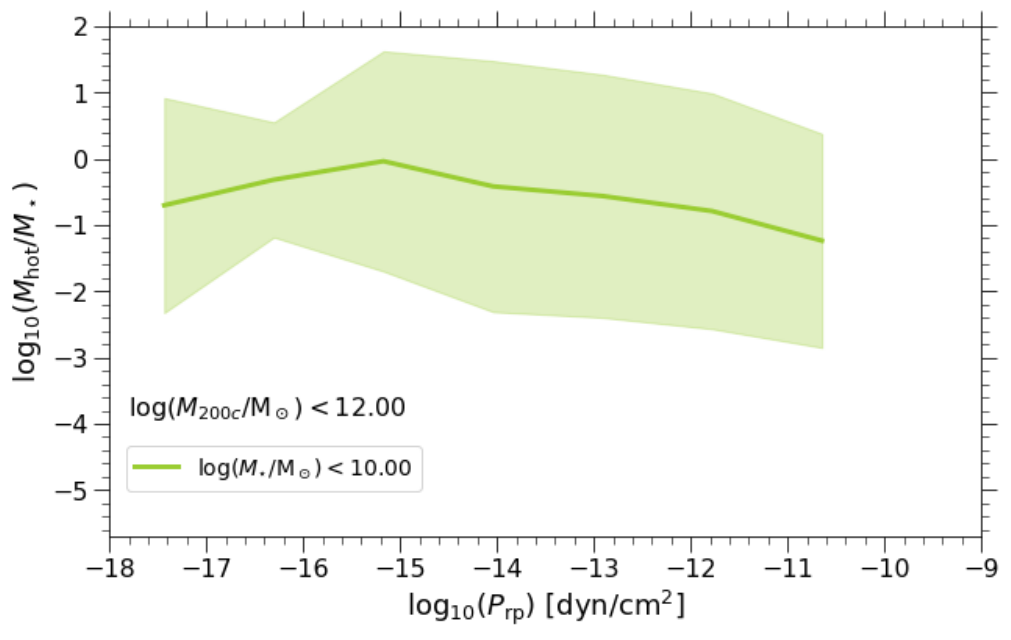}
    \includegraphics[width=0.48\textwidth]{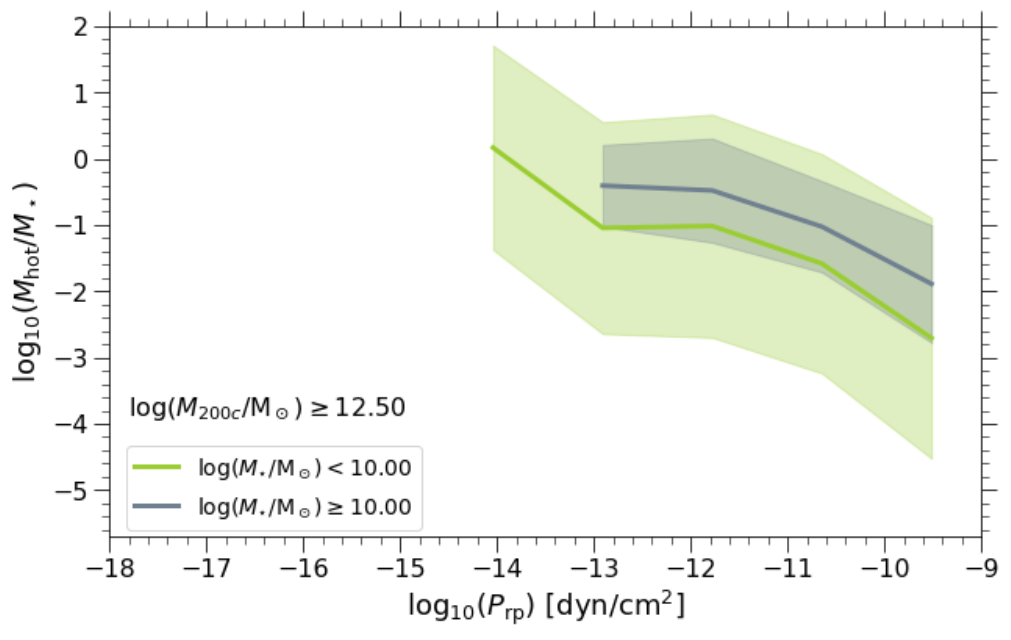}

    \caption{Cold (\textit{top}) and hot (\textit{bottom}) gas fraction as a function of ram pressure for the lowest ($\log_{10}({M_{\rm 200c}}/{\rm M_{\odot}}) < 12$) and highest ($\log_{10}({M_{\rm 200c}}/{\rm M_{\odot}}) \geq 12.5$) halo mass bins. Shaded regions indicate the standard deviation around the median profiles.}
    \label{fig:rampressure_coldgasAppendix}
\end{figure*}

To extend the ram-pressure analysis presented for the intermediate halo mass bin in Fig.~\ref{fig:rampressure_hist}, we show here the corresponding trends for the lowest and highest halo mass bins. The hot- and cold-gas-to-stellar-mass ratios as a function of ram pressure are shown in Fig.~\ref{fig:rampressure_coldgasAppendix} for the lowest and highest halo mass bins. The lowest halo mass bin is presented only for the lower stellar mass range, as the number of galaxies with \({M_{\star}} > 10^{10}{\rm M_{\odot}}\) is too small for a reliable analysis. In the highest halo mass bin, galaxies experience systematically higher ram pressure values than their counterparts in the intermediate halo mass bin.

The hot gas fraction shows no clear dependence on halo mass, as the trend remains consistent within the uncertainties. Moreover, the cold gas fraction exhibits a trend similar to that in the intermediate halo mass bin, with low-stellar-mass galaxies showing higher cold gas fractions across all ram pressure bins. As in the main text, the median gas-to-stellar mass ratio decreases with increasing ram pressure, indicating gas loss for satellites in denser environments.

\section{Halocentric trends}
\label{sec:appendixother}
\begin{figure*}
    \centering
    \includegraphics[width=0.35\textwidth]{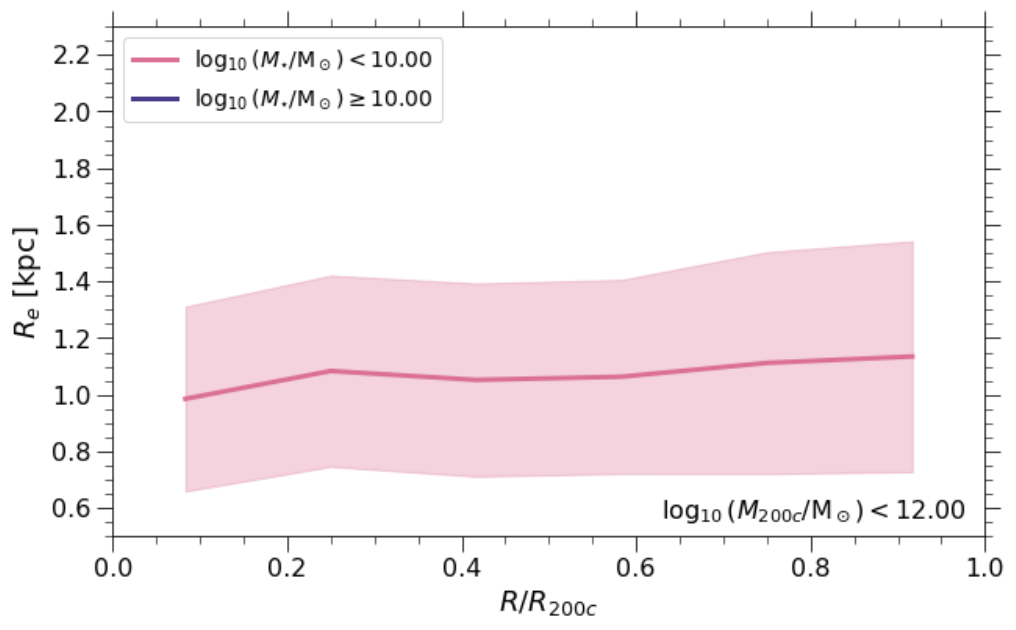}
    \includegraphics[width=0.35\textwidth]{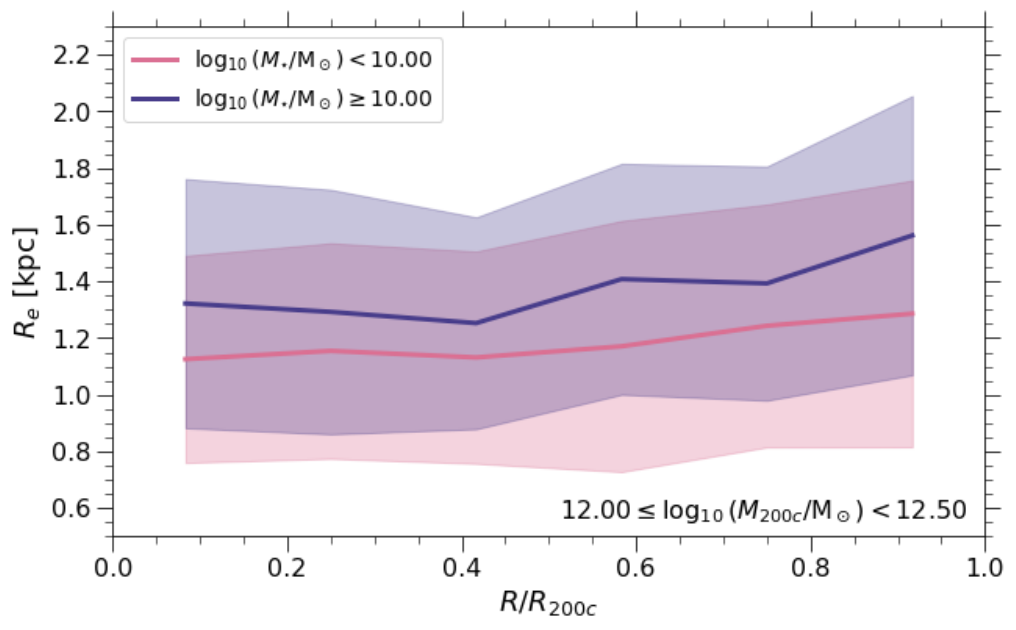}
   
    \includegraphics[width=0.35\textwidth]{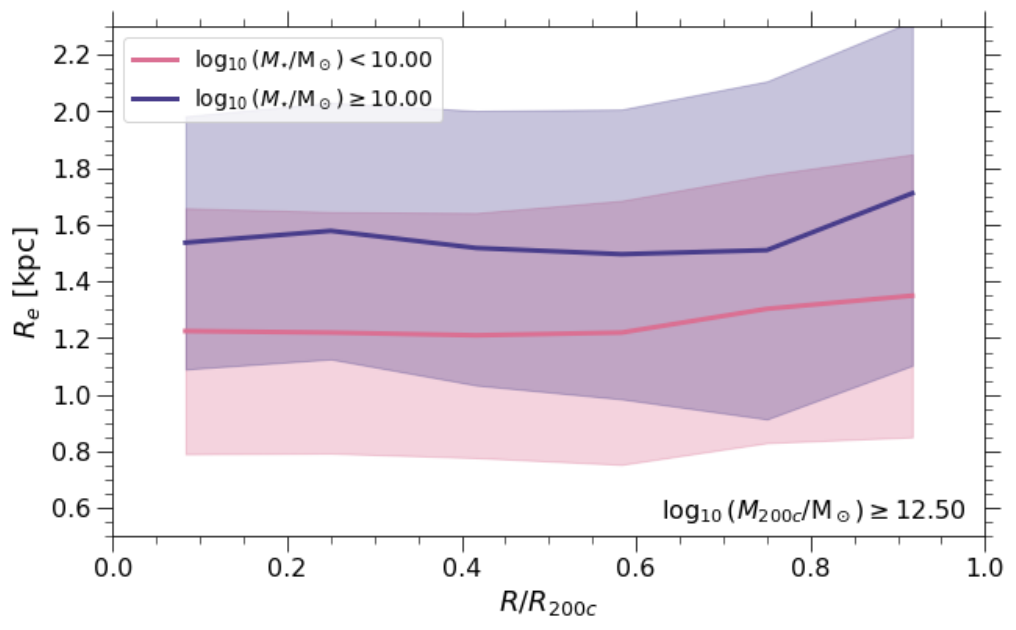}
    \includegraphics[width=0.35\textwidth]{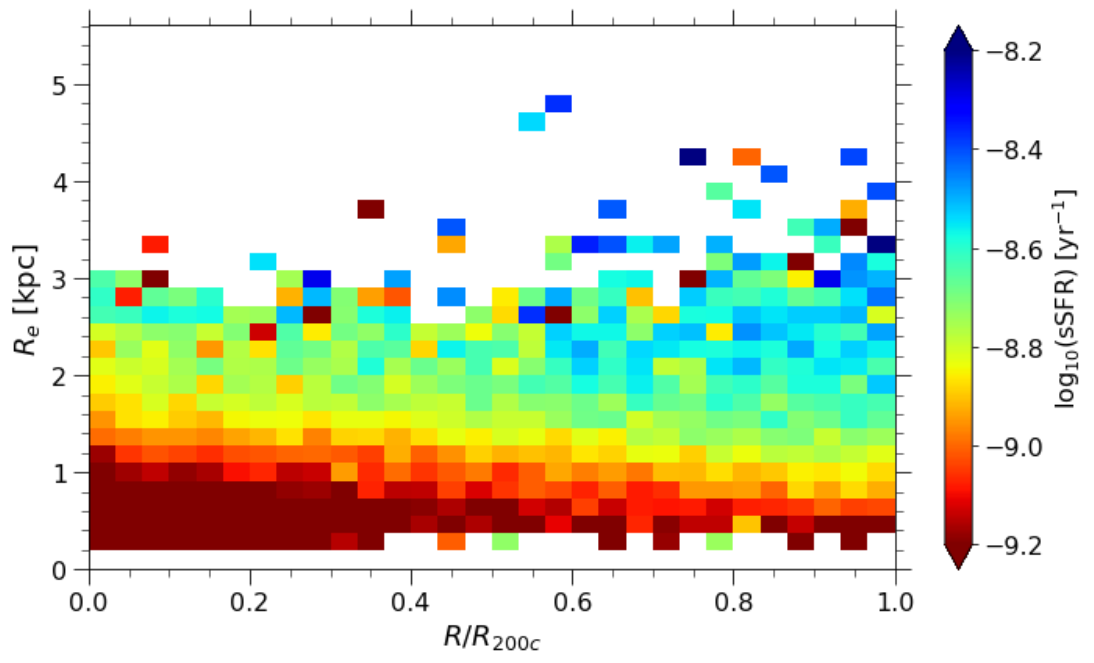}
    \caption{Median stellar half-mass radius as a function of halocentric distance in three halo mass bins: $\log_{10}({M_{\rm 200c}}/{\rm M_{\odot}}) < 12$ (\textit{top left}), $12 \leq \log_{10}({M_{\rm 200c}}/{\rm M_{\odot}}) < 12.5$ (\textit{top right}), and $\log_{10}({M_{\rm 200c}}/{\rm M_{\odot}}) \geq 12.5$ (\textit{bottom left}). In each panel, satellites are split into two stellar mass bins, $\log_{10}({M_{\star}}/{\rm M_{\odot}}) < 10$ (pink) and $\log_{10}({M_{\star}}/{\rm M_{\odot}}) \geq 10$ (blue). \textit{Bottom right}: Corresponding 2D histogram, color-coded by sSFR. Shaded regions indicate the standard deviation around the median profiles. In the 2D histogram, a minimum threshold of five galaxies was not used.}
    \label{fig:halfmassradiusappendix}
\end{figure*}
\begin{figure*}
    \centering

    \centering    
    \includegraphics[width=0.35\textwidth]{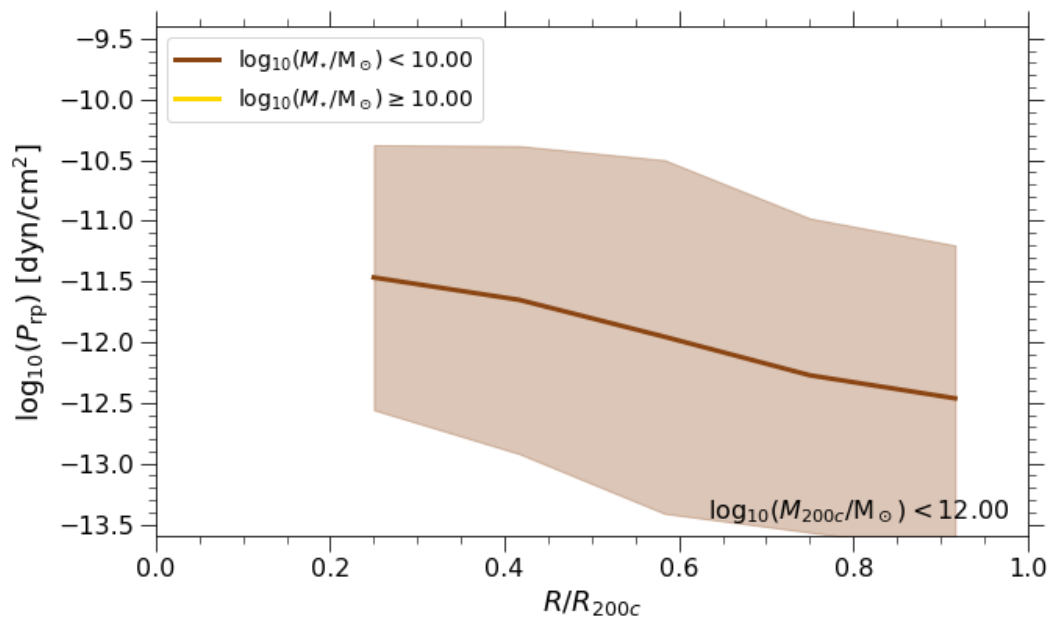}
    \includegraphics[width=0.35\textwidth]{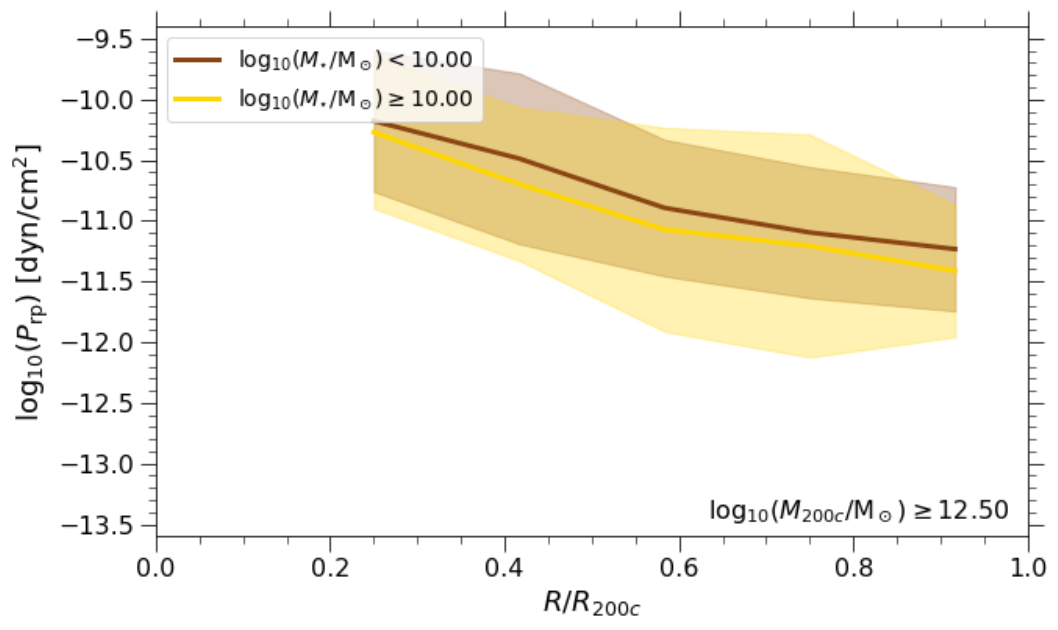}

    \centering
    \includegraphics[width=0.35\textwidth]{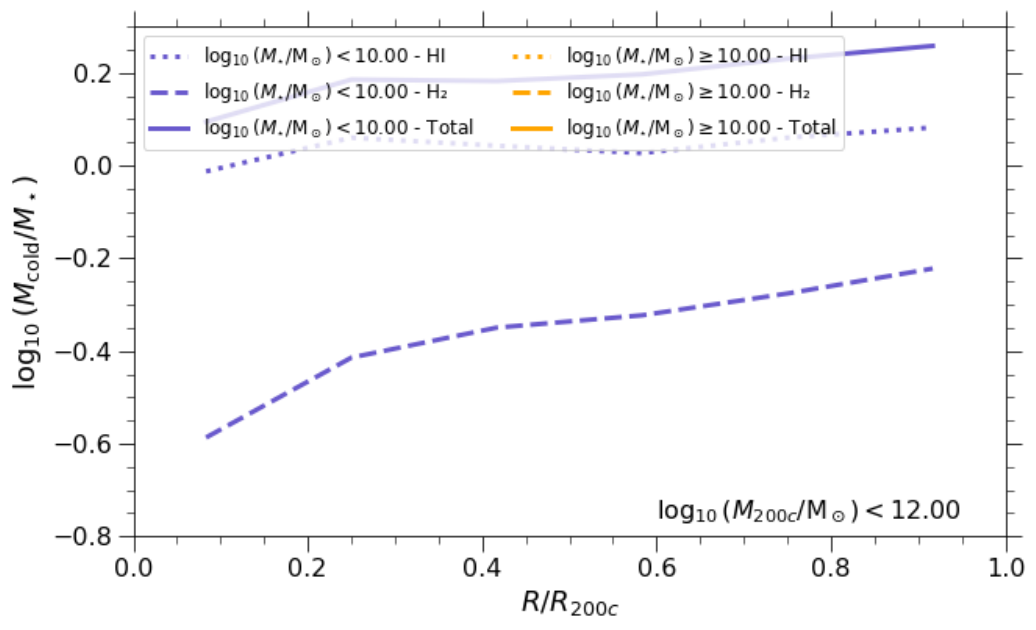}
    \includegraphics[width=0.35\textwidth]{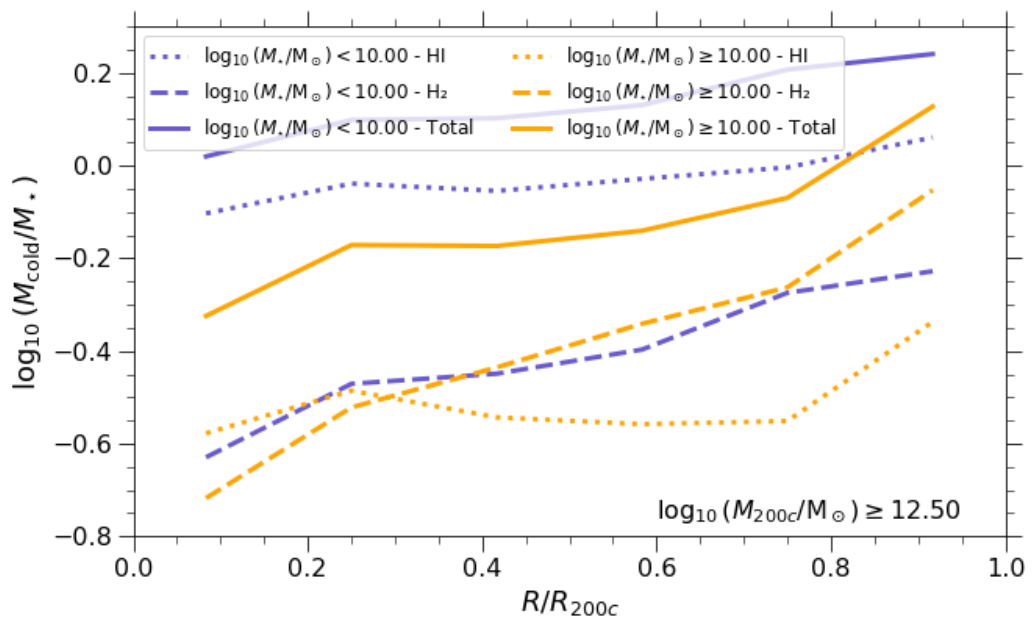}

    \centering
    \includegraphics[width=0.35\textwidth]{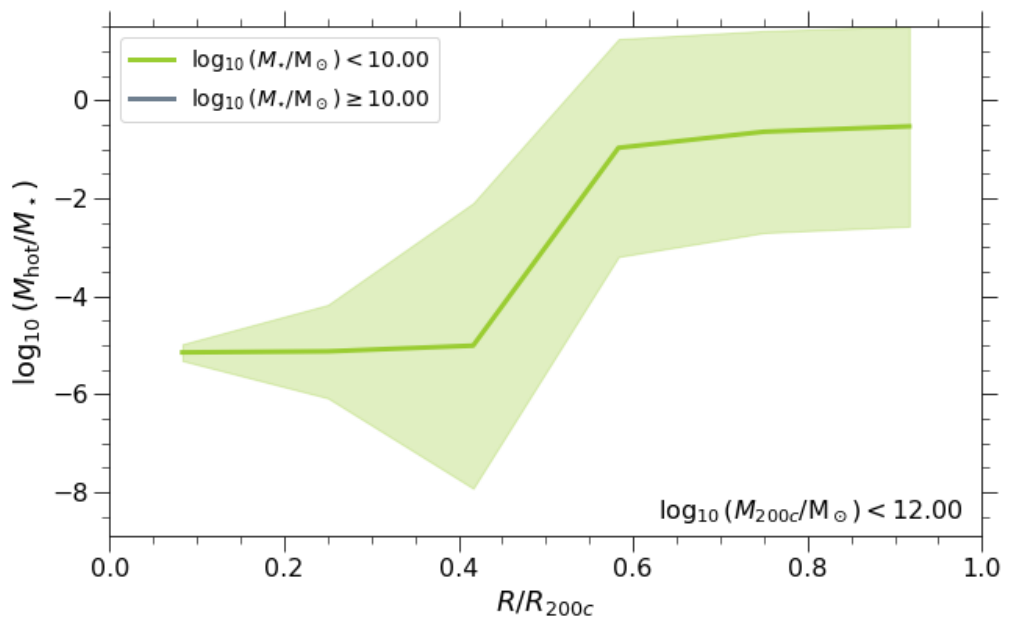}
    \includegraphics[width=0.35\textwidth]{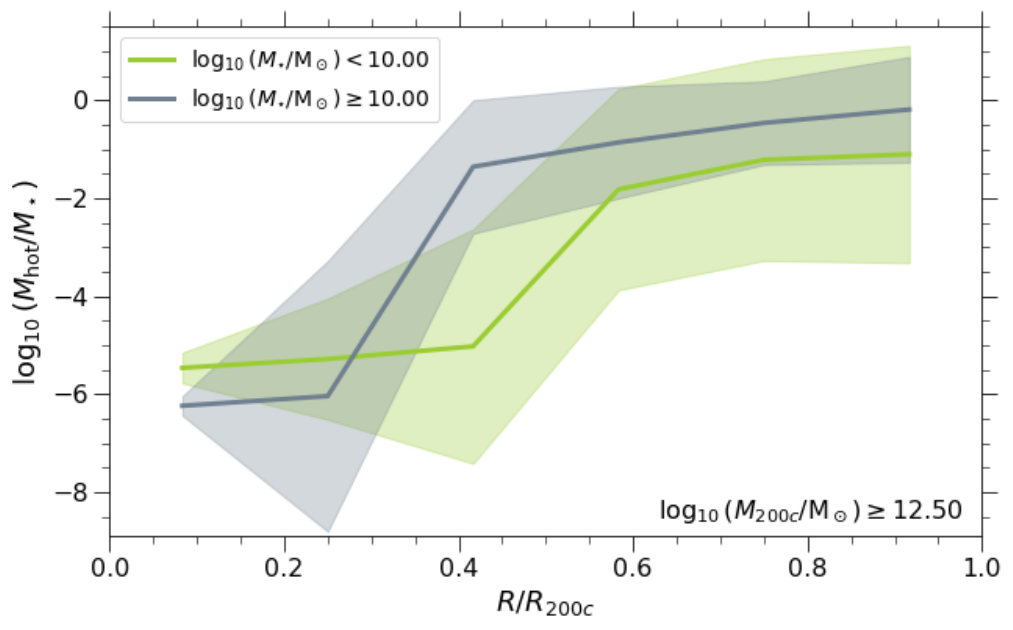}
    
    \caption{Median trends of ram pressure (\textit{top}), cold gas mass (\textit{middle}), and hot gas mass (\textit{bottom}) as a function of halocentric distance for the low-mass ($\log_{10}({M_{\rm 200c}}/{\rm M_{\odot}}) < 12$; \textit{left}) and high-mass ($\log_{10}({M_{\rm 200c}}/{\rm M_{\odot}}) \geq 12.5$; \textit{right}) halo mass bins. In each panel, satellites are separated into two stellar mass bins, $\log_{10}({M_{\star}}/{\rm M_{\odot}}) < 10$ and $\log_{10}({M_{\star}}/{\rm M_{\odot}}) \geq 10$. Shaded regions indicate the standard deviation around the median profiles. Zero hot-gas masses are mapped to $10^4 {\rm M_{\odot}}$, and ram-pressure values below $10^{-18}\mathrm{dyn\,cm^{-2}}$ are set to this value.}
    \label{fig:radialtrendsotherbins}
\end{figure*}

To further investigate the environmental dependence of satellite galaxy quenching discussed in Sect.~\ref{sec:environmentaldependenceofsatellitegalaxyquenching}, we examined the half-mass radii of satellite galaxies in \textsc{L-Galaxies} at $z=4$. Figure~\ref{fig:halfmassradiusappendix} presents the median half-mass radius as a function of halocentric distance, along with the corresponding 2D histogram color-coded by sSFR. Overall, the half-mass radius shows little dependence on halocentric distance, remaining nearly constant across the halo with only a slight increase toward larger radii. The 2D histogram highlights a clear relation between galaxy size, star formation activity, and environment: at fixed halocentric distance, larger satellites generally have higher sSFRs, while more compact satellites are more strongly quenched. Conversely, at fixed half-mass radius, satellites closer to the halo center exhibit systematically lower sSFRs than those at greater distances.

Figure \ref{fig:radialtrendsotherbins} provides the counterpart to Fig.~\ref{fig:6panel_grid} for the low- and high-mass halo bins. In the low-mass halo bin, no trend is visible for the higher stellar mass bin, $\log_{10}({M_{\star}}/{\rm M_{\odot}}) \geq 10$, owing to the very small number of satellite galaxies in this regime. The behavior in both halo mass bins is otherwise broadly consistent with that seen in the intermediate-mass bin in the main text.

\end{appendix}
\end{document}